\documentclass[traditabstract]{aa}
\usepackage{amsmath}
\usepackage[varg]{txfonts}
\usepackage{amssymb}
\usepackage{natbib}
\bibpunct{(}{)}{;}{a}{}{,} 
\usepackage{mathrsfs}
\usepackage{graphicx}
\usepackage[draft]{hyperref}
\usepackage{afterpage,array,longtable}

\usepackage{microtype}
\usepackage{wasysym}
\usepackage{colortbl}
\usepackage[usenames,dvipsnames,svgnames,table]{xcolor}

\usepackage[normalem]{ulem} 

\def\asec{\ifmmode ^{\prime\prime}\else$^{\prime\prime}$\fi}

\def\degs{\ifmmode ^{\circ}\else$^{\circ}$\fi}
\def\amin{\ifmmode ^{\prime}\else$^{\prime}$\fi}

\def\la{\mathrel{\mathchoice {\vcenter{\offinterlineskip\halign{\hfil
$\displaystyle##$\hfil\cr<\cr\sim\cr}}}
{\vcenter{\offinterlineskip\halign{\hfil$\textstyle##$\hfil\cr
<\cr\sim\cr}}}
{\vcenter{\offinterlineskip\halign{\hfil$\scriptstyle##$\hfil\cr
<\cr\sim\cr}}}
{\vcenter{\offinterlineskip\halign{\hfil$\scriptscriptstyle##$\hfil\cr
<\cr\sim\cr}}}}}
\def\ga{\mathrel{\mathchoice {\vcenter{\offinterlineskip\halign{\hfil
$\displaystyle##$\hfil\cr>\cr\sim\cr}}}
{\vcenter{\offinterlineskip\halign{\hfil$\textstyle##$\hfil\cr
>\cr\sim\cr}}}
{\vcenter{\offinterlineskip\halign{\hfil$\scriptstyle##$\hfil\cr
>\cr\sim\cr}}}
{\vcenter{\offinterlineskip\halign{\hfil$\scriptscriptstyle##$\hfil\cr
>\cr\sim\cr}}}}}
\def\arcmin{\hbox{$^\prime$}}

\def\arcsec{\hbox{$^{\prime\prime}$}}

\def\degr{\hbox{$^\circ$}}
\def\magg{\ensuremath{\text{mag}\,\text{arcsec}^{-2}}}

\def\ere{\ensuremath{r_{\text{e}}}}
\def\mue{\ensuremath{\mu_{\text{e}}}}
\def\pz{\ensuremath{\phantom{0}}}
\def\pzz{\ensuremath{\phantom{00}}}
\def\sdash{\ensuremath{\phantom{-}}}
\def\sdashx{$\phantom{-0}$}
\def\sd{$\phantom{0}$-}
\def\ssd{$\phantom{00}$-}
\def\snul{$\phantom{0}$}

\def\hst{{\sl HST\/}}

\def\rS{Paper\,I}
\def\Abe{ABC99}
\def\BO{B{\"O}02}
\def\Cairos{C01}
\def\Caon{C05}
\def\Fry{F99}
\def\Jong{J08}
\def\Kormendy{K09}
\def\KrSe{KS81a}
\def\Michard{M02}
\def\Michevaa{M13a}
\def\Michevab{M13b}
\def\Michevac{M10}
\def\Papaderos{P96}
\def\Wu{W02}
\def\BZC{BZC10}
\def\EPB{EPB08}
\def\DB{DB00}
\def\YD{YD06}
\def\sersic{S{\'e}rsic}

\def\PSFA {PSF$_{\text{A99}}$}
\def\PSFF {PSF$_{\text{F99}}$}
\def\PSFK {PSF$_{\text{K71}}$}

\def\PSFVa{PSF$_{V,\text{0m}}$}
\def\PSFVb{PSF$_{V,\text{3m}}$}
\def\PSFIa{PSF$_{i,\text{0m}}$}
\def\PSFIb{PSF$_{i,\text{3m}}$}
\def\PSFMBH{PSF$_{\text{MBH}}$}
\def\PSFMBHn{PSF$_{\text{MBH}}^{\text{new}}$}
\def\PSFMich{PSF$_{\text{M13}}$}

\begin{document}

\title{The influence of diffuse scattered light\thanks{Figures \ref{fig:haro11_x} and \ref{fig:VCC0001} are only available in electronic form via http://www.edpsciences.org}}
\subtitle{II. Observations of galaxy haloes and thick discs and hosts of BCGs}

\authorrunning{C.\ Sandin}

\author{Christer~Sandin\thanks{E-mail: CSandin@aip.de}}
\institute{Leibniz-Institut f\"ur Astrophysik Potsdam (AIP), An der Sternwarte 16, 144 82 Potsdam, Germany}

\date{Received 16 October 2014 / Accepted 25 February 2015}

\abstract{Studies of deep photometry of galaxies have presented discoveries of excess light in surface-brightness and colour profiles at large radii in the form of diffuse faint haloes and thick discs. In a majority of the cases, it has seemed necessary to use exotic stellar populations or alternative physical solutions to explain the excess. Few studies have carefully scrutinized the role of scattered light in this context. I explore the influence of scattered light on ground-based observations of haloes and thick discs around edge-on galaxies, haloes around face-on disc galaxies, host galaxies around blue compact galaxies (BCGs), and haloes around elliptical galaxies. Surface-brightness structures of all considered types of galaxies are modelled and analysed to compare scattered-light haloes and thick discs with measurements. I simulate the influence of scattered light and accurate sky subtraction on simplified {\sersic}-type and face-on disc galaxy models. All galaxy models are convolved with both lower-limit and brighter point spread functions (PSFs); for a few galaxies it was possible to use dedicated PSFs. The results show bright scattered-light haloes and high amounts of red excess at large radii and faint surface brightnesses for nearly all types of galaxies; exceptions are the largest elliptical-type galaxies where the influence of scattered light is smaller. Studies have underestimated the role of scattered light to explain their surface-brightness profiles. My analysis shows surface-brightness profiles that include scattered light that are very similar to and overlap measurements at all radii. The derivation of physical properties of haloes, thick discs, and BCG hosts from diffuse data is misleading since accurate and radially extended PSFs are non-existent. Significantly improved analyses that include new measurements of PSFs are required to study diffuse haloes further.}


\maketitle

\section{Introduction}\label{sec:introduction}
Diffuse excess light has been discovered around objects as diverse as galaxies, supernova remnants, planetary nebulae, and stars. For galaxies, it has been found around, for example, edge-on and face-on disc galaxies, low surface-brightness galaxies (LSBGs), blue compact galaxies (BCGs) where it is thought to be part of the host component, and elliptical galaxies. Excess light seen atop a single-component intensity profile near the centre region of the galaxy is referred to as a thick disc or a halo at larger radii. Stellar populations of haloes are determined through studies of radial colour gradients, which often show strong discrepant red excess; the shape of the surface-brightness profile and the halo thereby provide valuable information about the galaxy structure and evolution. Haloes are also seen around galaxies in observations of resolved stars, but such observations are by necessity limited to more nearby galaxies. Haloes are typically very faint, and therefore difficult to observe; because of this, the discoveries are for some types of objects limited to a small number. However, the number of new discoveries of haloes can be expected to increase as the number of deep observations increase \citep[for example,][]{DoAbMe:14,SoKaWaVe:14,DuCuKa.:15,ZhThHe.:15}.

Few haloes have been found around edge-on disc galaxies. Excess light has more often been observed closer to the centre of the disc in so-called thick discs, which presence is even reported as ubiquitous \citep[hereafter {\YD}]{YoDa:06}. Surface-brightness structures around face-on disc galaxies sometimes show flatter intensities at larger radii, beginning at a so-called anti-truncation radius; such profiles have even been assigned their own class, type III \citep[hereafter {\EPB}]{ErBePo:05,PoTr:06,ErPoBe:08}, extending the two previous types first defined by \citet{Fr:70}. In BCGs, the occurrence of star formation in the centre regions is determined by properties of the surrounding stellar host. Structural properties of hosts of BCGs give clues to the gravitational potential where starbursts take place. Although, the host galaxy is typically much fainter than the starbursts and is therefore difficult to observe.

Observations of haloes and thick discs are challenging and require careful assessments of data-analysis issues such as the accuracy of flat fields, sky subtraction, and other data-reduction properties. Accurate sky subtraction is essential with faint galaxies where haloes are visible at intensities a fraction of the intensity of the night sky. The influence of scattered light from the instrument, the telescope, and the sky has not been checked as carefully as the sky subtraction. Most studies merely conclude that their intensity profiles are shallower than the point spread function (PSF), which defines and measures the extent of scattered light, and that their data therefore are unaffected by it.

In \citet[hereafter {\rS}]{Sa:14}, I overview the existing sample of radially extended PSFs, and conclude that the PSF has rarely been measured out to large radii (angles) and is mostly inaccurate. The paper describes how the PSF can be measured more easily at large radii, and that it is necessary to account for variations of the PSF with time and wavelength, as well as spatially away from the optical axis. Example models of edge-on disc galaxies illustrate that it is necessary to analyse observations of edge-on galaxies with PSFs at least 1.5 times as extended as the measurements, or the influence of scattered light is underestimated. An alternative solution regarding the origin of the halo around NGC 5907 is also presented. My re-assessment of the accuracies of the sky and the PSF suggests that the halo observations of NGC 5907 are mainly explained as scattered light.

The mathematically correct procedure to remove scattered light is to deconvolve the measured image with the corresponding PSF. This is difficult, however, as a deconvolution requires a high accuracy at all intensities, both in the data and in the PSFs. Another approach is to probe the influence of scattered light by modelling surface-brightness profiles that are at first convolved with the PSF and thereafter compared with the measurements. Notably, such tests have been made for elliptical galaxies \citep[for example,][hereafter \Michard]{CaVa:83,Mi:02} and edge-on disc galaxies \citep[hereafter \Jong]{Jo:08}. The studies on elliptical galaxies find that effects of scattered light are small. Some 20--80 per cent of the excess light in data of the Sloan Digital Sky Survey (SDSS) has, initially, been found to be scattered light (\Jong).

The question is how scattered light more generally affects the faint outer parts of surface-brightness profiles of galaxies observed in diffuse light. If scattered light is a key effect, it implies that conclusions of observations need to be reconsidered with a more careful analysis that involves accurately measured PSFs. My simulation-based study scrutinizes haloes and thick discs in galaxies of the different kinds mentioned above to show that scattered light indeed is such a key effect. I largely use the same PSFs as in {\rS}, which span lower-limit to larger effects for all kinds of instruments and telescopes. The modelled profiles are compared with available measurements, and excess is studied in one colour for each galaxy.

As in {\rS}, this study focuses on ground-based observations in the visual wavelength range, 300--900nm. The focus is also on symmetric structures, and I ignore much fainter structures that show parts of shells or tidal tails and streams as are seen in, for example, \citet{MaGaCr.:10} and \citet{DuCuKa.:15}. The analysis method used to model all intensity profiles is described in Sect.~\ref{sec:methods}.

I scrutinize analyses of observations of different types of galaxies in separate sections. The case studies are ordered according to how strongly observations of the specific types are affected by scattered light, beginning with the most affected objects; most studies have delimited themselves to one galaxy-structure type following the same scheme. Three additional haloes around edge-on galaxies are studied in Sect.~\ref{sec:edgeon}, which also contains a discussion of the thick discs around edge-on disc galaxies of {\YD}, and a critical study of the finding of \citet[hereafter {\BZC}]{BeZaCa:10}, who conclude that red excess around edge-on LSBGs is not explained by scattered light. Five face-on disc galaxies are studied in Sect.~\ref{sec:faceon}, with a focus on the type III profile. Numerous surface-brightness profiles of galaxies described with a {\sersic} function are discussed in Sect.~\ref{sec:sersic}, including five BCGs and elliptical galaxies. Each section begins with an overview of studies of the respective type of galaxy structure to indicate the extent and potential impact of this work. Several of the studied galaxies are discussed in appendices to corroborate the results in the main text. Section~\ref{sec:discussion} contains a brief discussion including consequences for future observations, and the paper is closed with conclusions in Sect.~\ref{sec:conclusions}.

\section{Method and its application to example models}\label{sec:methods}
The method of the simulations and the analysis of each galaxy is split in three parts, as in {\rS}. In Sect.~\ref{sec:mPSF}, I select a set of PSFs used to estimate varying scattered-light effects. A set of models of the surface-brightness structure is configured in Sect.~\ref{sec:mSBS}. Section~\ref{sec:mMOD} describes how the PSFs are applied to the model structures and the outcome is analysed. In addition, I illustrate how the radial extent of the PSFs and the sky-subtraction accuracy affect models of the considered types of galaxies in Sect.~\ref{sec:example}. 

\subsection{Choosing representative PSFs}\label{sec:mPSF}
The PSFs used in this study should describe both temporal variations and the red-halo effect in the $i$ band \citep[as described by][]{SiClHa.:98}, and they should extend out to $r\simeq900\arcsec$. As in {\rS}, I chose to use the only published extended PSFs of {\Michard}, which were measured with both the (Cousins) $V$ and the (Gunn) $i$ bands, at two distinct occasions separated by three months: {\PSFVa} and {\PSFIa} were measured three months before {\PSFVb} and {\PSFIb}. The difference with radius of {\PSFVb}$-${\PSFVa} on average is higher than that of {\PSFIb}$-${\PSFIa} (see Fig.~\ref{fig:p2psf}). The radially extended {\PSFK} of \citet{Ki:71}, which was measured (mostly) in the $B$ band is used as a general lower-limit indicator of scattered light (cf.\ {\rS}). The comparison of the PSFs in figs.~1 and 2 in {\rS} demonstrate that {\PSFVa} is $\la0.5\,\magg$ brighter at intermediate radii ($20\la r\la60\arcsec$) than the other treated PSFs. {\PSFVb}, meanwhile, is another $0.5\,\magg$ brighter. In {\rS}, I reassessed {\PSFMBH} that was measured by \citet{MoBoHa:94} in their study of NGC 5907; the new {\PSFMBHn} lies between {\PSFVa} and {\PSFVb}. {\PSFIa} appears to represent the average $i$-band SDSS PSFs well where $r\ga10\arcsec$. Some PSFs show more light at large radii than the $r^{-2}$ power-law of {\PSFK}; the predictive ability of models of larger objects is weaker, as the far wings of the PSF are generally poorly determined. The complementary PSFs discussed in this paper, which all extend beyond $r\simeq10\arcsec$, are shown in Fig.~\ref{fig:p2psf}.

In the simulations, I assumed that the two $V$-band PSFs, as well as {\PSFK}, are the same in the $B$, $g$, $R$, and $r$ bands, as well as the photographic $J$ band (that is measured using a IIIa-J emulsion and a Wratten-2C filter), and that the two $i$-band PSFs are the same in the $I$ band; the colour predictability is hereby delimited to one colour, for example, $V-i$. I do not differentiate between Cousins, Johnson, or other photometric systems. All five PSFs were measured at a seeing of several arc seconds, which is why the spatial resolution in the brightest centre region is poor. Here, faint diffuse emission is studied, where the spatial resolution and resulting lower intensities in the centre regions are of minor importance.

\subsection{Configuring sets of model surface-brightness structures}\label{sec:mSBS}
I discriminate between galaxies, which surface-brightness structures are fitted well with models that only vary with the radius, and disc galaxies viewed at low (face-on) and high (edge-on) inclination angles. The relation $\mu=-2.5\log_{10}(I)$ is used to convert between magnitudes $\mu$ and intensities $I$.

The common {\sersic} function \citep{Se:68} is used to model surface-brightness structures of, for example, elliptical and lenticular galaxies, cD galaxies, bulges of disc galaxies, and BCGs,
\begin{eqnarray}
I(r)=I_{\text{e}}\exp\left(-b_{n}\left[\left(\frac{r}{r_{\text{e}}}\right)^{\frac{1}{n}}-1\right]\right),\label{eq:sersic}
\end{eqnarray}
where $r$ is the radius, $n$ the {\sersic}-profile index, $r_{\text{e}}$ an effective radius that encloses half of the total light of the profile, $I_{\text{e}}$ an effective surface brightness at $r_{\text{e}}$, and $b_{n}\simeq2n-0.327$ \citep[for example,][]{Ca:89,CaCaOn:93}. Also, $I(0)=I_{\text{e}}\exp\left(b_{n}\right)$. For exponential profiles ($n=1.0$), the function can be replaced with a representation that instead uses a scale length $h$,
\begin{eqnarray}
I(r)=I_{0}\exp\left(-\frac{r}{h}\right),
\end{eqnarray}
where $I_0=I_{\text{e}}\exp\left(b_1\right)$ is the surface brightness at the centre, and $h=r_{\text{e}}/b_{1}\simeq0.598r_{\text{e}}$. Moreover, when modelling flattened spheroidals, it is assumed that the minor-to-major axis intensity ratio equals the $c/a$ ratio. The ellipticity (flattening) is defined as $e=1-c/a$. Otherwise, any azimuthal dependence of measured profiles is ignored. {\sersic}-profile surface brightnesses reach very high values at sub-arc second angles for higher values on $n$. To scale the models reasonably, each model {\sersic} profile is normalized with the model centre intensity that results after convolving the model with a 1{\arcsec} seeing profile; I make a note of resulting non-zero offsets in figures of affected profiles.

Disc galaxies are suitably described in cylindrical coordinates; assuming an isothermal disc, the space-luminosity density can be described by (\citealt{Kr:88}; \citealt{KrSe:81a}, hereafter {\KrSe})
\begin{eqnarray}
I^{\prime}(r',z')=I^{\prime}_{0,0}\exp\left(-\frac{r'}{h_{\text{r}}}\right)\times2^{-2/n_{\text{s}}}\text{sech}^{2/n_{\text{s}}}\left(\frac{n_{\text{s}}z'}{z_{0}}\right),
\end{eqnarray}
where $I^{\prime}_{0,0}$ is the centre intensity, $r'$ the radius, $h_{\text{r}}$ the scale length, $z'$ the vertical distance from the centre, $z_{0}$ the vertical scale height, and $n_{\text{s}}$ is set to $1$, $2$, or $\infty$. The intensity drops to zero at a galaxy-specific truncation radius. When a disc galaxy is projected edge on, there is a(n optional) {\sersic}-profile bulge, and the truncation radius and dust extinction are ignored, the surface-brightness structure becomes
\begin{eqnarray}
I(r,z)=I(r)+I_{0,0}\frac{r}{h_{\text{r}}}K_{1}\left(\frac{r}{h_{\text{r}}}\right)\times2^{-2/n_{\text{s}}}\text{sech}^{2/n_{\text{s}}}\left(\frac{n_{\text{s}}z}{z_{0}}\right),\label{eq:edgeon}
\end{eqnarray}
where $I_{0,0}=2h_{\text{r}}I^{\prime}_{0,0}$ is the centre intensity, $r$ the major-axis radius, $z$ the vertical-axis distance from the major axis, and $K_{1}$ the modified Bessel function of the second kind. The intensity is slightly lower at larger radii when the truncation radius is finite and taken into account; considering how uncertain the PSFs are, this effect is ignored here. The vertical structure can be assumed as `isothermal' ($n_{\text{s}}=1$), exponential ($n_{\text{s}}=\infty$), or in between ($n_{\text{s}}=2$). When $n_{\text{s}}=\infty$, the $z$-dependent function reduces to $\exp(-z/z_{0})$; papers that discuss exponential structures often use the notation $h_{\text{z}}=z_{0}$. No attempt is made to make a perfect fit of the centre regions of edge-on galaxies.

Finally, with disc galaxies that are instead viewed face on, a surface-brightness structure is used that accounts for radial breaks in the intensity structure as well as a {\sersic}-profile bulge. Following the approach of \citet{BaTr:12},
\begin{eqnarray}
I(r,z)=I(r)&+&I_{0,\text{i}}\exp\left(-\frac{r}{h_{\text{i}}}\right)\Theta\left(r\right)\nonumber\\
             &+&I_{0,\text{o}}\exp\left(-\frac{r}{h_{\text{o}}}\right)\left(1-\Theta\left(r\right)\right),
\end{eqnarray}
where $I_{0}=2z_{0}I^{\prime}_{0,0}$ is the centre intensity, $I_{0,\text{i}}$ and $I_{0,\text{o}}$ ($h_{\text{i}}$ and $h_{\text{o}}$) are the centre intensities (scale lengths) of the inner and the outer discs, respectively, and the step function $\Theta\left(r\right)=1$ if $r\le r_{\text{brk}}$ ($r_{\text{brk}}$ is the break radius), and otherwise zero.

\subsection{Models and measurements analysis procedure}\label{sec:mMOD}
Each two-dimensional $V$-band (as well as $B$-, $g$-, $R$-, $r$-, and $J$-band) model image is convolved individually with the three resampled and normalized two-dimensional $V$-band {\PSFVa} and {\PSFVb}, and the $B$-band {\PSFK}. This is repeated for each $i$-band (as well as $I$-band) model image with {\PSFIa} and {\PSFIb}. Unless specified otherwise, the $i$-band models use the same model parameters as the $V$-band model, with two exceptions; $r_{\text{e},i}=0.95r_{\text{e},V}$ ($h_{i}=0.95h_{V}$) and $z_{0,i}=0.95z_{0,V}$, which result in negative slopes in $V-i$ with increasing radius. Thereby, any red-excess haloes in convolved models are induced by the PSF. Except the LSBG models in Sect.~\ref{sec:stack}, I made no attempt to fit observed $V-i$ colours, but $\mu_{\text{e},V}-\mu_{\text{e},i}$, $\mu_{0,V}-\mu_{0,i}$, and $\mu_{0,0,V}-\mu_{0,0,i}$ are set to values that result in a very rough general agreement between models and measurements; for example, I used $\mu_{V}-\mu_{i}=1.3\,\magg$ with all edge-on disc galaxy models. The used PSF image is always twice as large as the model image, to avoid PSF truncation effects in convolved images (see the outcome of the toy models for edge-on galaxies in {\rS} and for {\sersic}-type and face-on disc galaxies here, in Appendix~\ref{sec:toy}). The PSF and the model images are resampled to use the same pitch and about 100--200 pixels on the side, typically, to keep calculation times short. All convolutions are made by direct integration.

Convolved surface-brightness profiles are plotted together with model surface-brightness profiles for a cut in the image that extends from the centre and outwards. For edge-on galaxies, this cut is always positioned on the minor axis, directed outwards from the centre on the same (vertical) axis. In cases where deep enough $B$-, $V$-, $g$-, $R$-, $r$-, or $J$-band measurements exist in the literature, they are plotted as well; where tables were missing, I used \textsc{dexter}\footnote{\textsc{dexter} is available at: \href{http://dexter.sourceforge.net}{http://dexter.sourceforge.net}.} \citep{DeAcEi.:01} to extract the data. The three resulting $V-i$ colour profiles are shown in a separate lower panel: the model, convolved models that use the earlier {\PSFVa} and {\PSFIa}, or the later {\PSFVb} and {\PSFIb}.

The PSF induces scattered light, where surface-brightness profiles of input models and convolved models differ. Larger differences between profiles of convolved models that use {\PSFVa}, {\PSFVb}, and {\PSFK} in the $V$-band, or {\PSFIa} and {\PSFIb} in the $i$-band, are indicative of stronger time dependence in the PSFs. The opposite applies when such differences are smaller. Measured values due to scattered light should fall atop profiles of a model convolved with the PSF at the time of the observations. For some telescope and instrument setups, the PSF contains even more energy at large radii than {\PSFVb}, and in this case it is necessary to convolve affected models using an even brighter PSF. The PSF is mostly not measured or published, and in this case measurements expected to be scattered light plausibly fall between profiles of convolved models that assume the lowest and highest amount of such scattered light ({\PSFK} and {\PSFVb}). The predictive power is weaker without a published PSF, but models nevertheless provide valuable hints of the PSF dependence of specific model parameter setups. Finally, measurements that are brighter than the convolved model that uses the correct PSF are real and are not (only) due to scattered light, and measured values that are fainter than the same convolved model are non-physical.

A scattered-light halo radius $r_{110}$ is defined as the lower limiting radius where the convolved model intensity at all larger radii is $\ge10$ per cent higher than the model intensity. Note that $r_{110}$ depends on both object parameters and the PSF. The usefulness of $r_{110}$ at larger radii ($r\ga100\arcsec$) is at the moment doubtful in comparisons with real galaxies, since PSFs are nearly unknown there.

\subsection{Using example models to test the influence of the radial extent of PSFs and the accuracy of the sky subtraction}\label{sec:example}
I demonstrate effects of the use of truncated PSFs in the analysis of observations of edge-on galaxies in {\rS}. In Appendix~\ref{sec:toy}, I extend the study with completely analogous models for face-on disc galaxies and galaxy structures described with {\sersic}-type profiles (where $n=1$). The results of the demonstration are unambiguous -- surface-brightness structures of these galaxies are also strongly affected by scattered light, and effects are only somewhat dependent on the size of the galaxy. These new models indicate that the required extent of the PSFs for these structures is some $1.1$ times the radius of the measurements. Depending on the brightness of the PSF, the halo is up to $2.5\,\magg$ brighter around the edge-on galaxy models than around the {\sersic}-type galaxy models. The face-on galaxies are an intermediate case where the brightness of the halo depends on the properties and extent of the galaxy inside a (possible) break radius. I show in Sect.~\ref{sec:sersic} that effects are smaller in elliptical galaxies ($n=4$) and cD galaxies ($n\ga8$).

\begin{figure*}
\centering
\includegraphics{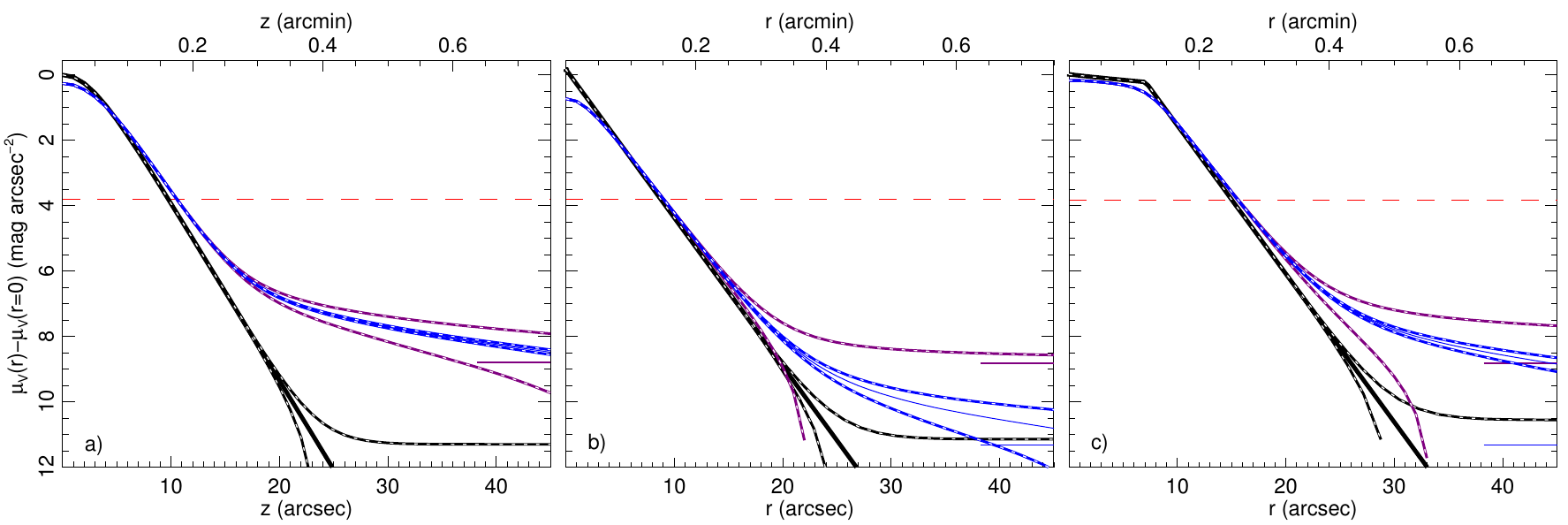}
\caption{Surface-brightness profiles of the three intermediate-size example models, which illustrate effects of the accuracy of the sky subtraction using {\PSFVa}. The three panels show the: \textbf{a}) edge-on disc galaxy (intensities are drawn versus the vertical distance $z$ instead of the radius $r$), \textbf{b}) {\sersic}-type galaxy, and \textbf{c}) face-on disc galaxy. In each panel, the galaxy model profile is drawn with a thick black solid line, and the convolved model profile is drawn with a thin blue line. The horizontal dashed red lines indicate the used background level of the sky, and the purple and blue short horizontal lines on the right-hand side of each panel 0.01 and 0.001 times the sky level. The thick lines overplotted with white dotted lines indicate the resulting profiles when the sky is under or over subtracted by 0.1 per cent (black and blue lines) and 1 per cent (purple lines).}
\label{fig:toysky}
\end{figure*}

\begin{table}
\caption{Results of the models of the sky, using the intermediate-size configurations of table~3 in {\rS} and Table~\ref{tab:toy}. Values in rows printed in boldface mark input models and remaining values models convolved with {\PSFVa}.}
\centering
\label{tab:toysky}
\tabcolsep=3.1pt
\begin{tabular}{lccllcllcl}\hline\hline\\[-1.8ex]
Model & $\delta_{\text{s}}$ & \multicolumn{2}{c}{$\delta=0.05$} && \multicolumn{2}{c}{$\delta=0.10$} && \multicolumn{2}{c}{$\delta=0.15$}\\\cline{3-4}\cline{6-7}\cline{9-10}\\[-1.7ex]
&& \multicolumn{1}{c}{$r_{\text{lim}}$} & \multicolumn{1}{c}{$\Delta$sky} &&
    \multicolumn{1}{c}{$r_{\text{lim}}$} & \multicolumn{1}{c}{$\Delta$sky} &&
    \multicolumn{1}{c}{$r_{\text{lim}}$} & \multicolumn{1}{c}{$\Delta$sky}\\\hline\\[-1.8ex]
edge-on & $\mathbf{7.5}$ & $\mathbf{18.0}$ & $\mathbf{4.4}$--$\mathbf{4.5}$ && $\mathbf{19.0}$ & $\mathbf{4.9}$--$\mathbf{5.1}$ && $\mathbf{20.0}$ & $\mathbf{5.4}$--$\mathbf{5.8}$\\
        & $7.5$ & $36.0$ & $4.3$ && - & - && - & -\\
        & $5.0$ & $15.0$ & $1.8$--$1.9$ && $18.0$ & $2.5$--$2.8$ && $20.0$ & $2.9$--$3.2$\\
{\sersic} & $\mathbf{7.5}$ & $\mathbf{18.2}$ & $\mathbf{4.2}$--$\mathbf{4.4}$ && $\mathbf{19.3}$ & $\mathbf{4.9}$--$\mathbf{5.1}$ && $\mathbf{20.4}$ & $\mathbf{5.1}$--$\mathbf{5.5}$\\
        & $7.5$ & $20.0$ & $4.2$--$4.3$ && $22.0$ & $4.9$--$4.9$ && $24.0$ & $5.2$--$5.4$\\
        & $5.0$ & $13.0$ & $1.6$--$1.7$ && $15.0$ & $2.4$--$2.6$ && $16.0$ & $2.7$--$3.0$\\
face-on & $\mathbf{7.5}$ & $\mathbf{22.9}$ & $\mathbf{3.5}$--$\mathbf{3.6}$ && $\mathbf{24.2}$ & $\mathbf{4.3}$--$\mathbf{4.7}$ && $\mathbf{25.7}$ & $\mathbf{4.7}$--$\mathbf{5.0}$\\
        & $7.5$ & $27.0$ & $3.5$--$3.6$ && $32.0$ & $4.2$--$4.3$ && $38.0$ & $4.6$--$4.8$\\
        & $5.0$ & $18.0$ & $1.0$--$1.3$ && $20.0$ & $1.8$--$1.9$ && $22.0$ & $2.2$--$2.6$\\[0.3ex]
\hline
\end{tabular}
\tablefoot{Column 1, intermediate-size galaxy model; Col.~2, error of added and subtracted sky (\magg); Cols.~3 and 4, radius (arcsec) and magnitude range (\magg) at the innermost radius where the difference between the model and the model with added and subtracted sky is $\delta\simeq0.05\,\magg$; Cols.~5 and 6 (Cols.~7 and 8), same as Cols.~3 and 4 for $\delta\simeq0.10\,\magg$ ($\delta\simeq0.15\,\magg$).}
\end{table}

Accurate subtraction of the sky is imperative, considering that faint haloes and thick discs are observed around already faint galaxies. It is beyond the scope of this paper to provide a general study of the influence of sky subtraction for all possible configurations of galaxy models and levels of the sky background. Instead, I show three examples of the already modelled galaxies to illustrate the interplay between sky subtraction and scattered light. The three models are the intermediate-size models of an edge-on disc (\rS), a {\sersic}-type, and a face-on disc galaxy.

I simulated the influence of the sky by adding a constant value to each modelled two-dimensional structure (arbitrarily) $3.8\,\magg$ fainter than the centre intensity. Thereafter, I subtracted two constant values in separate models offset from the added value by $\pm1$ per cent ($\delta_{\text{s}}=5.0\,\magg$) and $\pm0.1$ per cent ($\delta_{\text{s}}=7.5\,\magg$), respectively. The added-and-subtracted sky models were then convolved with {\PSFVa}. Resulting profiles are shown for all three galaxy types in Fig.~\ref{fig:toysky}. I measured minimum radii $r_{\text{lim}}$ and corresponding magnitudes where the sky-subtracted models deviate from the input model by $\delta>0.05$, $0.10$, and $0.15\,\magg$; these values are presented in Table~\ref{tab:toysky}.

The plots clearly show that the accuracy of the sky-affected input models are several magnitudes lower than the added perturbation of $\delta_{\text{s}}=7.5\,\magg$ when the allowed deviation $\delta\le0.15\,\magg$. The limiting magnitudes for the edge-on ({\sersic}-type) galaxy are then $4.4$--$4.5$ ($5.4$--$5.8$) {\magg} fainter than the sky when $\delta=0.05,\,0.15$ {\magg}; the corresponding radii are $r_{\text{lim}}=18,\,20\arcsec$. These values correspond to a lowering of the accuracy of the sky by $3.0$--$3.1$ ($3.1$--$3.3$) {\magg} when $\delta=0.05\,\magg$ and by $1.7$--$2.1$ ($2.0$--$2.4$) {\magg} when $\delta=0.15\,\magg$. The values are slightly lower for the more extended face-on disc galaxy model, where the limiting magnitude is $3.5$--$3.6$ ($4.7$--$5.0$) {\magg} for $\delta=0.05$ ($0.15$) {\magg}, at the radius $r_{\text{lim}}=23\arcsec$ ($r_{\text{lim}}=26\arcsec$); in this case, the accuracy of the sky is lowered by $3.9$--$4.0$ ($2.5$--$2.8$) {\magg}. Surface-brightnesses measured at larger radii can be compared with models to show if they may be affected by scattered light. However, it is only possible to recover an exponential profile, as in this case, out to a maximum radius determined by the accuracy of the sky.

Limiting magnitudes are only slightly lower in the convolved models than in the input models; the tabulated values show the range $0.0$--$0.4\,\magg$, considering all three values of $\delta$ for all three types of galaxies. The convolved sky-affected models that used a perturbation of $\delta_{\text{s}}=5.0\,\magg$ deviate from the convolved models at smaller radii and much brighter intensities; the values are very close to $2.5\,\magg$ lower than the models that use a $\delta_{\text{s}}=7.5\,\magg$ perturbation. For the edge-on galaxy model, and compared to the model that used a $7.5\,\magg$ perturbation, the magnitude is $2.5\,\magg$ lower with $\delta=0.05$, whilst the limiting radius is $r_{\text{lim}}=15\arcsec$. The corresponding values of the {\sersic}-type (face-on disc) galaxy model are $2.5$--$2.6$ ($2.4$--$2.5$) {\magg} lower and $r_{\text{lim}}=13$--$16\arcsec$ ($18$--$22\arcsec$) for all three values on $\delta$.

\section{About analyses of observations of faint structures around edge-on galaxies}\label{sec:edgeon}
Thick discs and haloes are observed in resolved star counts of nearby galaxies and in photometry of both near and far galaxies; here, I only consider diffuse photometric measurements. Resolved star counts are always real, regardless of if stars are attributed to a thick disc or an extended thin disc \citep{BoRiHo:12} -- discrete counts are unaffected by scattered-light effects.

\citet{Ts:77,Ts:79} presents the first detection of diffuse excess light around three bright and nearby lenticular S0 galaxies. She uses observations with photographic plates in the $B$ and $V$ bands. \citet{Bu:79}, who also uses photographic plates in the $B$ band, coins the expression `thick disc' in his study of five S0 galaxies that all show excess light. Following studies switch from photographic plates to CCD measurements and report on both positive and negative identifications of diffuse thick discs in a number of galaxies and different bands. The selection of objects for this study is made in Sect.~\ref{sec:cedgeon}, which also outlines the remaining parts of this section.

\subsection{On scattered light and details of the object selection}\label{sec:cedgeon}
Few of the numerous studies of detections in diffuse light test the influence of scattered light on their observations. In general, studies of ground-based observations merely note that their objects are unaffected by scattered light. Neither \citet{ElEl:06} nor \citet{CoElKn.:11} address the effects of the extended PSF in the analyses of their space-based observations. {\Jong} reports that edge-on galaxy observations are affected by scattered light, but does not specifically mention thick discs. In view of the results of {\rS}, it seems that {\Jong} underestimates the influence of the scattered light as he both neglects temporal variations of the PSF at large radii and extrapolates it with a too steep slope.

Here, I extend the example of NGC 5907 in {\rS} with three corroborating examples of diffuse haloes. An extended PSF was measured for the observations of the comparatively distant galaxy IC 5249, but its influence has not been checked. The thick disc and halo of NGC 4565 was measured multiple times; the resulting surface-brightness structures of the halo are suspiciously similar. No halo is found in analyses of diffuse observations of NGC 4244, but resolved-star counts do show a halo. These three galaxy haloes are scrutinized in the context of scattered light in Appendix~\ref{sec:aedgeon}. To demonstrate that scattered light can give rise to diffuse thick discs in small-to-intermediate size galaxies, I show two examples with the size-limiting galaxies FGC 310 and FGC 1285 in Sect.~\ref{sec:fgc}. Finally, as a general attribute, the influence of the PSF has been dismissed in a study of stacked SDSS images of edge-on LSBGs (\BZC). I show in Sect.~\ref{sec:stack} how this dismissal can be explained when accounting for temporally varying PSFs, and therefore scattered light cannot be excluded as an explanation to the surface-brightness structures for these galaxies either. Details of the observations and existing and new model parameters of all six considered galaxies are collected in Tables~\ref{tab:obs} and \ref{tab:edgeon}, respectively.

\subsection{On the origin of the observed thick discs around the galaxies FGC 310 and FGC 1285}\label{sec:fgc}
\citet[hereafter {\DB}]{DaBe:00} and \citet{DaBe:02} select a sample of 47 edge-on galaxies from the Flat Galaxy Catalog \citep[FGC;][]{KaKaPa:93}. The authors carefully consider issues of calibration uncertainties, the influence of field stars, and sky subtraction. They measured PSFs for each of the observed bandpasses in the radial range $0\le r\la6\arcsec$, which they compare with the observed profiles, concluding that the PSF is too narrow to have any effect.

\begin{figure*}
\sidecaption
\includegraphics{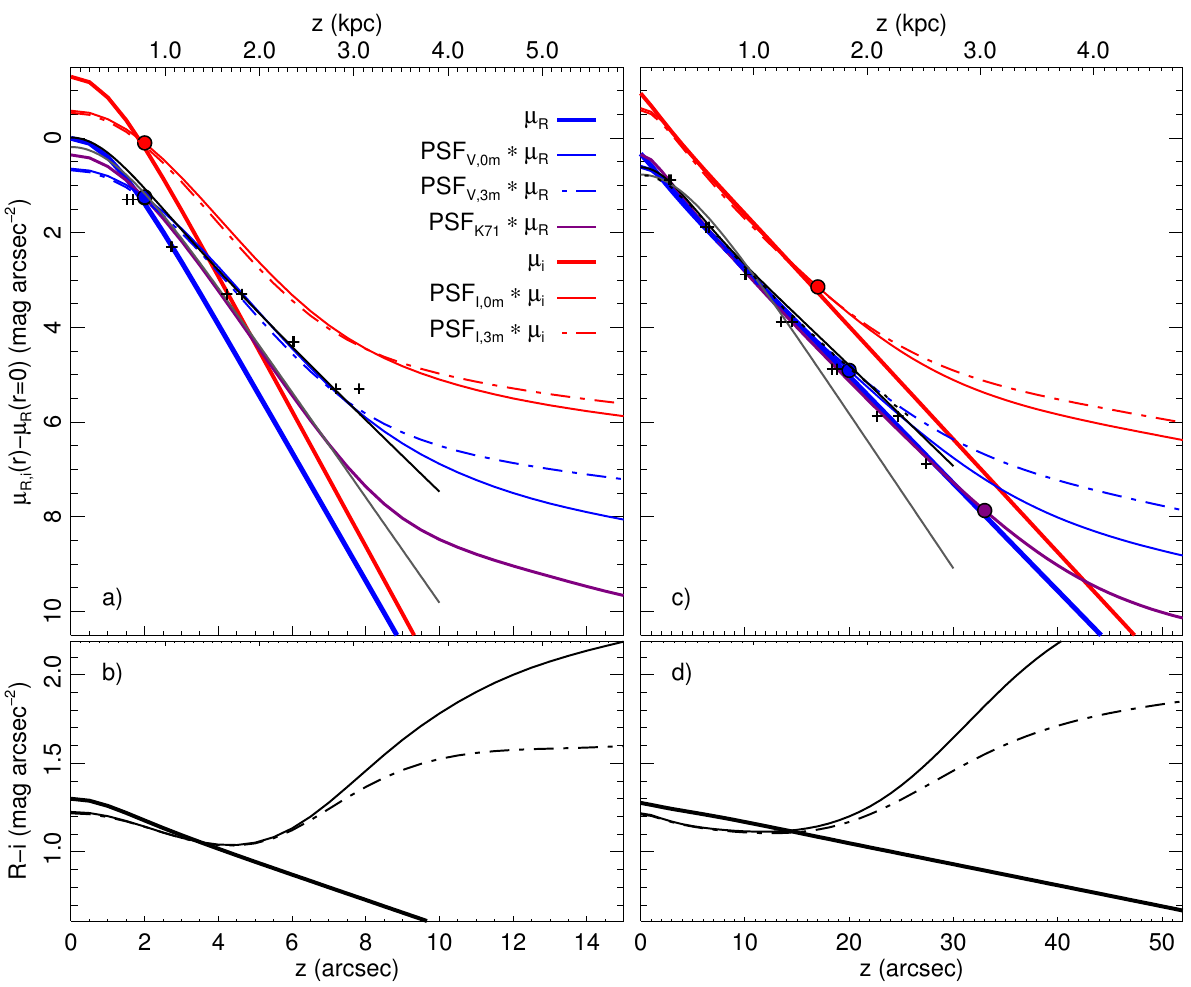}
\caption{Minor-axis surface-brightness profiles versus the vertical distance $z$ for two edge-on disc galaxies. \textbf{a}) and \textbf{b}) FGC 310, and \textbf{c}) and \textbf{d}) FGC 1285. Blue and purple lines in panels \textbf{a}) and \textbf{c}) show $R$-band profiles, and red lines $i$-band profiles. Model profiles are drawn with thick solid lines. Solid (dash-dotted) lines are profiles of convolved models using {\PSFVa} and {\PSFIa} ({\PSFVb} and \PSFIb), the purple line used \PSFK. The black (grey) line shows the respective fitted combined thin-and-thick-disc model (single-disc model; of \YD). Crosses indicate $R$-band measurements that were extracted from fig.~3 in {\DB}. The lower limiting radius $r_{110}$ -- where the convolved models using {\PSFVa}, {\PSFIa}, and {\PSFK} lie $\ge10$ per cent above the input model -- is marked with a coloured bullet that is surrounded by a black ring. \textbf{b}) Three colour profiles $R-i$ are shown for: the model (thick solid line), the convolved model using {\PSFVa} and {\PSFIa} (solid line), and the convolved model using {\PSFVb} and {\PSFIb} (dash-dotted line).}
\label{fig:thickdisc}
\end{figure*}

These observations are worth a closer examination, since the authors omitted a calculation of the integrated light of the galaxies. {\DB} present contour diagrams for each galaxy, where contours are drawn at a $1\,$mag resolution. Later, \citet{DaBe:02} present minor-axis surface-brightness profiles, which, however, are not as easily interpreted (they appear to differ from the data of {\DB} and {\YD}). {\YD} present two model fits for each galaxy: one with a single disc and one where a thick disc is superposed on a thin disc. It is worth noting that the authors weight their two-dimensional fits to the pixels with the lowest signal-to-noise ratio; this approach could perhaps explain differences between the model profile fits and the profiles that can be extracted from the contour diagrams of {\DB}. I chose one of their smallest and the largest appearing galaxies FGC 310 and FGC 1285, respectively. The centre parts of the smaller galaxy are poorly modelled here since the seeing values of the PSFs I used are large (about $4\farcs8$, which results in somewhat lower centre intensities).

I modelled FGC 310 with one disc where $h_{\text{r}}=8\farcs4$, $z_{0}=1\farcs6$, and $n_{\text{s}}=1$, Fig.~\ref{fig:thickdisc}a. The input model convolved with {\PSFK} agrees with the single-disc model, and when it was instead convolved with {\PSFVa} it agrees with the superposed thin-and-thick-disc (two-disc) model. By chance, the input model agrees with the thin disc of {\YD}. The minor-axis profile that I extracted from the contour diagram in {\DB} is more similar to the two-disc model, at least where $r\ga5\arcsec$. The colour profiles in Fig.~\ref{fig:thickdisc}b show red excess for $z\ga4\arcsec$, where $\Delta\left(R-i\right)\simeq0.44$--$0.50\,\magg$ at $z=7\arcsec$ and about $0.92$--$1.2\,\magg$ at $z=10\arcsec$. The lower (higher) values again used the latter {\PSFVb} and {\PSFIb} (earlier {\PSFVa} and {\PSFIa}). Notably, the effects of scattered light are larger for the smallest appearing galaxy in the study of {\YD}, FGC 1063, because the model parameters are even lower than in FGC 310.

I modelled the larger galaxy FGC 1285 with a bulge that is superposed on a disc. The parameters of the bulge are $n=1.0$, $r_{\text{e}}=4\farcs4$, and $\mu_{\text{e}}=19.3\,\magg$, whilst the disc parameters are $h_{\text{r}}=17\farcs3$, $z_{0}=9\farcs6$, $\mu_{0,0}=19.5\,\magg$, and $n_{\text{s}}=1$. Minor-axis surface-brightness profiles are shown in Fig.~\ref{fig:thickdisc}c. For smaller radii $r\la20\arcsec$, there are small differences between profiles of the input models and the convolved models. The observations of this object are less strongly affected by scattered light than FGC 310. The single-disc profile of {\YD} agrees poorly with the two-disc profile at all radii. In this case, when using {\PSFVa}, the scattered-light halo only becomes visible at an intensity that is some $4.8\,\magg$ fainter than the centre intensity; with {\PSFK} this value is about $7.8\,\magg$. The colour profiles in Fig.~\ref{fig:thickdisc}d show red excess for $z\ga15\arcsec$, where $\Delta\left(R-i\right)\simeq0.22$--$0.28\,\magg$ at $z=23\arcsec$ and $\Delta\left(R-i\right)\simeq0.53$--$0.69\,\magg$ at $z=30\arcsec$.

The small sizes of all galaxies of {\YD} imply that surface-brightness profiles and colours are highly sensitive to the PSF shape -- and thereby to temporal variations of the PSF. The exact values of the model parameters that are fitted here are unimportant. It should be clear that it appears to be of little value to fit a thin disc and a thick disc unless the data are first corrected for scattered light. Currently, the fitted components do not reveal any information about the real structure in the faint parts. To recover the correct structure out to $z=10\arcsec$, it is necessary to deconvolve observed structures with PSFs that extend to at least $r=15\arcsec$ in the considered bandpass ($1.5\times10\arcsec$, cf.\ \rS). For FGC 310 and using {\PSFK} (\PSFVa), the convolved structure at $z=8\arcsec$ is $1.9$ ($3.3$) {\magg} brighter than the model structure, which corresponds to a required S/N$=5.8$ ($21$). The corresponding values at $z=11\arcsec$ are $4.4$ ($5.8$) {\magg} and S/N$=58$ ($210$), and at $z=14\arcsec$ the values are $7.6$ ($9.0$) {\magg} and S/N$=1100$ ($4000$), and the values increase for larger radii. The requirements are lower for FGC 1285, where the halo, however, appears at fainter intensities. At $z=40\arcsec$ the convolved structure is $0.56$ ($1.5$) {\magg} brighter than the model structure, the required S/N$=1.7$ ($4.0$).

\subsection{A closer look at the role of time-dependent scattered light to the analysis of stacked images}\label{sec:stack}
Galaxy halo intensities typically reach levels much below one per cent of the background sky. One can assume that haloes share fundamental properties the same way galaxies do, to overcome the difficulty in measuring such faint intensities; \citet{DoWhBr.:10} show that parameters of stacked images created from hundreds of synthetic galaxy models with random {\sersic} profiles match average values of stacked profiles well. High signal-to-noise values in halo measurements can then be achieved by stacking a large number of images of fairly similar objects. \citet{ZiWhBr:04}, who are the first to use this approach, address the role of the PSF in stacked images of 1047 nearly edge-on disc galaxies of the SDSS; they find that haloes are an almost ubiquitous phenomenon around disc galaxies. The halo of their stacked image shows abnormal red excess that they attribute a physical origin, and not scattered light. {\Jong} finds that the reason that the PSF plays no role is that \citet{ZiWhBr:04} do not consider the extended wings of the PSF properly -- at least parts of the red excess in the halo can be explained by such extended wings. Despite accounting for extended wings of PSFs, {\Jong} concludes that red haloes are still found far away from the disc (in both {\hst} and SDSS images).

{\BZC} carefully analyse a stack of nearly edge-on LSBGs, and conclude that for this kind of galaxy the PSF is unimportant. I here demonstrate, based on the study of {\BZC}, how both colours and faint stellar haloes can be induced as a consequence of temporally varying PSFs.

{\BZC} search for nearly edge-on LSBGs in $g$-, $r$-, and $i$-band SDSS images. They use the following four criteria to select galaxies: the mean surface-brightness inside the radius encompassing half of the light in the $r$ band must satisfy $23<\mu_{\text{r},50}<25\,\magg$, the $i$-band isophotal diameter $a\ge10\arcsec$, the $i$-band axial ratio $b/a<0.25$, and the redshift $z<0.2$. The halo region of interest that they discuss extends from 60 pixels and outwards on the vertical axis, which corresponds to radii $r\ge23\farcs7$. Their final sample contains 1510 galaxies that they analyse in four separate subsamples ($A$, $B$, $C$, and $T=A+B+C$), which are setup based on the $g-r$ colour. Results are presented as statistical properties of each sample, where exponential scale lengths are measured in the range $1-16\,\text{kpc}\,h^{-1}$. The authors do not provide any information about the identity or properties of individual objects. They obtain one PSF for each of the three bandpasses from stacks of bright stars.

Furthermore, {\BZC} use two approaches to correct their reduced images for scattered-light effects. In one approach, they deconvolve each image individually, before they stack them. In a second approach, they first fit a flattened-spheroidal and an exponential-disc model to a template stack of the 35 apparently largest galaxies. Thereafter, they scale the model to the same size as each galaxy in the respective sample and stack the scaled models. They provide fitted parameters for each bandpass of their template stack, as well as isophotal contour plots of the stacked images of each bandpass and subsample. In their analysis of the stacked image data, they treat each subsample individually in terms of a thin disc, a thick disc, and a halo. Their conclusions are based on $g-r$ and $r-i$ colour profiles along the major axis for the thin disc, and along the minor axis using wedges with an opening angle of $120\degr$ (thick disc) and $60\degr$ (halo). They find that the $r-i$ colour of the halo $T$ sample decreases by $r-i\la0.2\,\magg$ due to deconvolution. The remaining colour difference between more central regions and outer regions is $\Delta\left(r-i\right)\approx0.7\,\magg$. Despite an individual deconvolution of each image, they find red excess in the halo, and conclude that scattered light is not the reason for this excess. They instead explore a physical origin to explain the red excess.

The authors do not consider temporally varying PSFs; the SDSS images are sampled during long periods of time, where such variations can be significant (cf.\ \rS). Additionally, objects of different geometrical properties are differently affected by the PSF -- it is impossible to remove PSF effects exactly, without detailed knowledge of individual object properties and measurements. The exact PSFs are unknown, and no attempt was made to reproduce the observations precisely. I used the following procedure to simulate the effects of ignoring temporally varying PSFs on the stack of galaxies. I convolved \textit{input} $g$- and $i$-band surface-brightness models with the early-time PSFs ({\PSFVa} and \PSFIa), to give \textit{observed} surface-brightness structures. Thereafter, I deconvolved the result with the later-time PSFs ({\PSFVb} and \PSFIb), to give \textit{PSF-affected input} model structures. I used the Richardson--Lucy algorithm in the form of the freely available program \textsc{sgp} for the Interactive Data Language \citep[\textsc{idl};][]{PrCaZa.:12}\footnote{\textsc{sgp} is available at: \href{http://www.unife.it/prisma/software}{http://www.unife.it/prisma/software}.}. I did not add any noise to the models or the PSFs, and using the same PSF I checked that the deconvolution always reproduced the input model structure when I deconvolved any convolved input model. I compared surface-brightness structures and the $g-i$ colour of input models and PSF-affected input models with corresponding measurements.

\begin{figure*}
\sidecaption
\includegraphics[width=12cm]{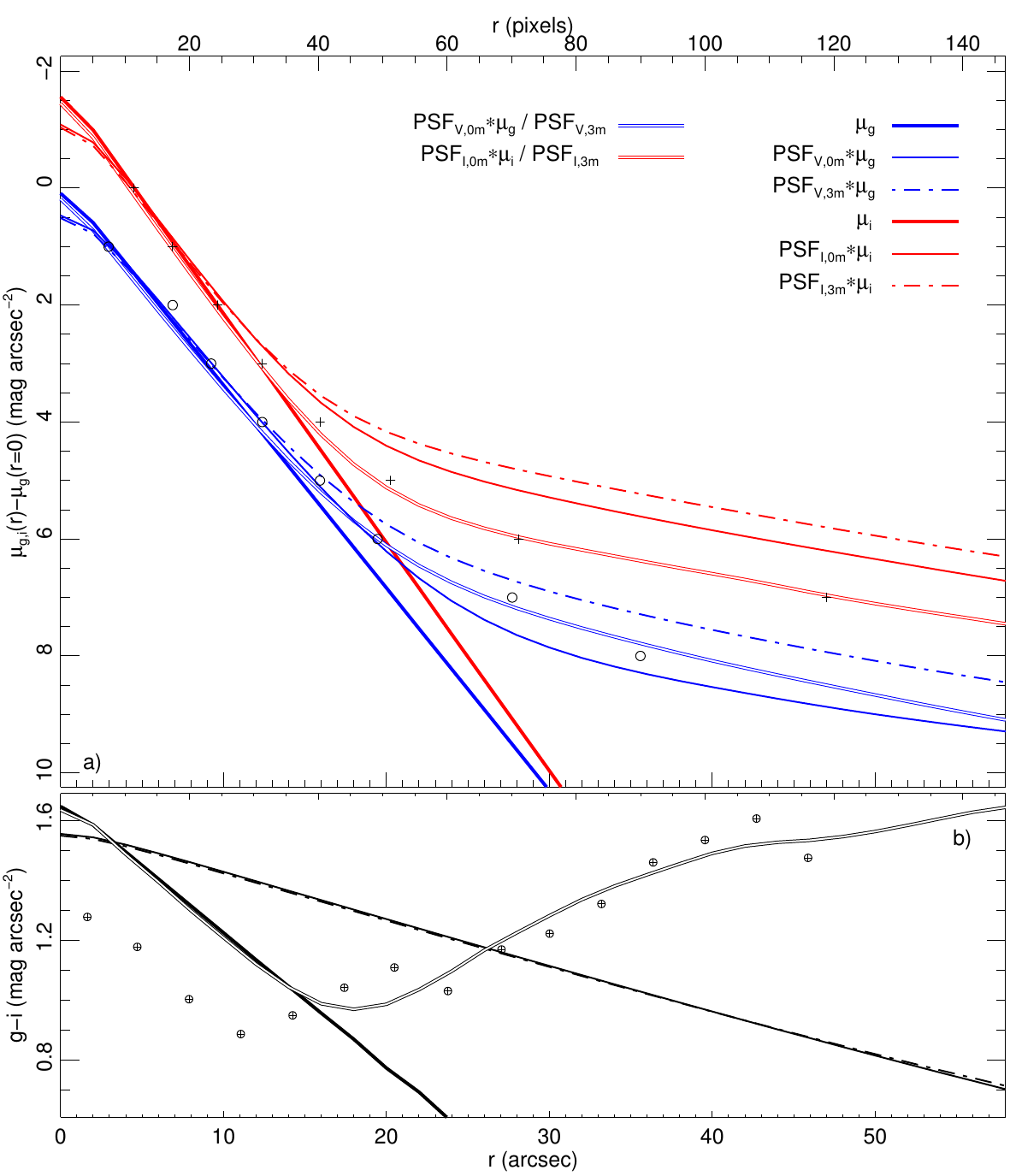}
\caption{Minor-axis $g$- and $i$-band surface-brightness profiles versus the vertical distance $z$ of the LSBG model. Panels and lines are as described in Fig.~\ref{fig:thickdisc}. Circles (crosses) indicate measured values of the $g$ band ($i$ band; \BZC). In panel \textbf{a}, the blue and the red solid lines that are overstriked with a white solid line show the PSF-affected input model. In panel \textbf{b}, the corresponding $g-i$ colour measurements are shown with crossed circles, and the thick overstriked line shows the colour of the PSF-affected model.}
\label{fig:Bergvall}
\end{figure*}

I first defined the scattered-light affected input model. I measured the isophote positions on the major and the minor axes of the $g$- and $i$-band contour plots of the $B$ subsample (fig.~7 in \BZC); the isophotes are separated by one magnitude. (I chose the $B$ subsample as it appears less noisy than the other subsamples in the same figure.) Effects are expected to be slightly larger (smaller) for objects of the smaller (larger) objects of the $A$ ($C$) subsample. I fitted the major-axis values with two separate {\sersic} profiles where $n_{g}=1.0$, $r_{\text{e},g}=26\arcsec$, $\mu_{\text{e},g}=24.0\,\magg$, $n_{i}=1.2$, $r_{\text{e},i}=21\arcsec$, and $\mu_{\text{e},i}=22.3\,\magg$. I then created two-dimensional surface-brightness structures by combining the two {\sersic} profiles with the major-to-minor axis flattened spheroidal parameters $(c/a)_{g}=0.34$ and $(c/a)_{i}=0.50$, which the authors derive from the faintest isophotes in their template model stack. To have the PSF-affected input model reproduce the measurements on the minor axis in the isophotal image, it is necessary to use lower values since scattered light affects the ellipticity at large radii (see the discussion for the BCG ESO 400-G043 in Sect.~\ref{sec:bcgeso}). I found, by trial and error, that $n_{g,i}=1.0$, $(c/a)_{g}=0.20$, and $(c/a)_{i}=0.22$ produce good results. Minor-axis plots of all models are shown in Fig.~\ref{fig:Bergvall}.

Figure~\ref{fig:Bergvall}a shows a reasonable agreement between the two PSF-affected input models and the measurements in both the $g$ and the $i$ bands; deviations are $\la0.25\,\magg$. A scattered-light halo appears in the observed structure at $r\simeq12\arcsec$ for the $g$ band, and at $r\simeq3\arcsec$ for the $i$ band. In the PSF-affected input model, this radius is $r\simeq12\arcsec$ for the $i$ band. These scattered-light haloes appear with all four PSFs that I use, and they demonstrate that galaxy surface-brightness structures that can be represented with parameters such as these are strongly affected by scattered light already at small radii.

The PSF-affected input model shows that the $i$-band scattered-light halo is more completely removed in the deconvolution than that of the $g$ band, since at most radii $(${\PSFIb}$-${\PSFIa}$)<(${\PSFVb}$-${\PSFVa}$)$. Figure~\ref{fig:Bergvall}b shows that the colour profile $g-i$ of the input model decreases with the radius. A deviation from the model colour gradient -- and a red-excess halo -- results with the PSF-affected input model, where $r\ga16\arcsec$. The agreement between the PSF-affected input model and the $g-i$ colour profile of the deconvolved halo $T$ sample is good where $r\ga15\arcsec$ (this was achieved by co-adding the two curves in fig.~14 of {\BZC}, no similar profile is presented for the $B$ subsample); deviations are $\la0.12\,\magg$. The offset of $r-i\simeq0.3\,\magg$ where $r\la12\arcsec$ could arise due to the less excellent model fit in the $g$ band (see the value at $r\simeq7\arcsec$), differences between the $B$ and the $T$ samples, or due to PSF differences in the inner region, or because of all these reasons.

My simulations cannot prove that the stacked-images halo arises only due to scattered light, or that there is no red excess in the halo. However, the results provide strong indications of overlooked systematic properties that alone can explain both these features. Additionally, \citet{TaDo:11} stack 42\,000 SDSS images of luminous red galaxies, and also use temporally averaged PSFs. These authors find support to the conclusions of \citet{ZiWhBr:04}, in terms of decreasing $r-i$ colour profiles in the inner parts of the stacked galaxy. The geometrical properties of their stacked image result in models that are less affected by PSF effects than the stack studied here. Although, the outer parts of a model that I calculated for such luminous red galaxies show significant red excess also in this case (colours are not shown for this region in their paper). My analysis suggests that the complete neglect of extended PSFs in stacked images \citep[for example,][]{ZiWhBr:04,ZiWhScBr:05,SoKaWaVe:14} affects results more than just neglecting the temporal variations of the extended PSFs.

\section{About analyses of observations of faint structures around face-on disc galaxies}\label{sec:faceon}
Surface-brightness profiles of face-on disc galaxies are categorized into three types. Type~I profiles are well described by a single exponential, whilst type~II profiles show a break outside the main region of spiral-arm activity, which is followed by a region with a steeper decline \citep{Fr:70}. A third type was added to explain galaxies where surface-brightness profiles become shallower beyond a break at a large radius; these type-III profiles show an `anti-truncation' of excess light that appears at $3.2$--$6.0$ inner scale lengths \citep{ErBePo:05}. The inflection in type-III-s profiles -- where the outer profile is believed to be part of the spheroid -- is smooth and curved, and whilst isophotes are elliptical in the inner region for inclined galaxies, they become rounder at large radii. Meanwhile, the outer isophotes are not signficantly rounder in the other subclass, type~III-d, where the outer profile is thought to be part of the disc. My selection of objects to study is made in Sect.~\ref{sec:faceono}, which also outlines the remaining parts of this section.

\subsection{Details of my selection, the observations, and the models of the five examined galaxies}\label{sec:faceono}
I selected five galaxies from {\EPB}, which all have outer parts that are classified as certain, uncertain, or possible type-III or type-III-s structures. Two galaxies are discussed in Sect.~\ref{sec:faceonsl}: NGC 7280 shows an odd structure and NGC 4102 was observed by two studies who report discrepant profiles and a possible type III-s structure. Three additional galaxies where the influence of scattered light is perhaps less clear are discussed in Appendix~\ref{sec:faceonnsl}: \object{NGC 3507} and \object{NGC 4477} were observed with SDSS and are classified with different types, and \object{NGC 2880} is classified as type III-s.

I modelled these galaxies using the same single-disc model parameters as {\EPB}. I also added a bulge to each galaxy model, which I fitted by trial and error to have convolved surface-brightness structures match the ($R$-band) measurements in the centre. Details of the observations are given in Table~\ref{tab:obs} and model parameters including galaxy types and inclination angles are collected in Table~\ref{tab:faceon}.

{\EPB} discuss how the use of free or fixed fits of ellipses affect the profiles. They clearly show that free ellipse fits result in wrong profiles in the centre regions of the galaxies, as a free fit may track bar-distorted isophotes. I tilted all disc models using the provided inclinations to achieve elliptical isophotes, and thereafter obtained the surface-brightness profiles with fixed ellipses. I never tilted the surface-brightness structure of the centre bulge.

\begin{figure*}
\sidecaption
\includegraphics{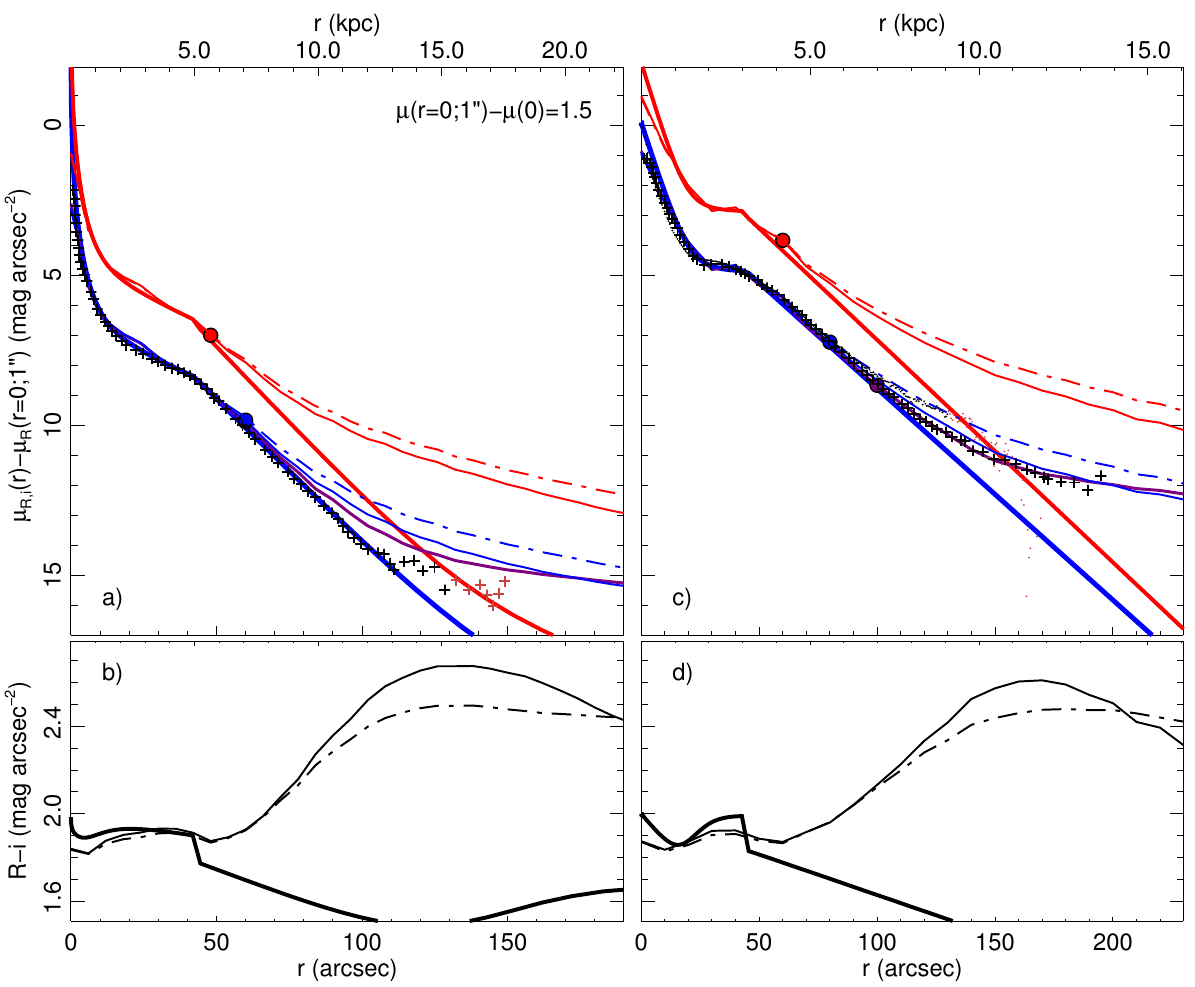}
\caption{Radial $R$-band and $i$-band surface-brightness profiles versus radius for two face-on disc galaxies: \textbf{a}) and \textbf{b}) NGC 7280, and \textbf{c}) and \textbf{d}) NGC 4102. Panels, lines, and coloured bullets are as described in Fig.~\ref{fig:thickdisc}. For NGC 7280, the measured centre magnitudes at $1\arcsec$ seeing are $1.5\,\magg$ fainter than the model magnitude at the centre; there is no additional offset here between the $i$-band and the $R$-band profiles. Measurements of {\EPB} are shown with crosses in panels \textbf{a} and \textbf{c} (for the $R$ band, their fig.~14). The $r$-band measurements of \citet{Co:96} for NGC 4102 are also shown with black dots in panel \textbf{c}. Red symbols indicate values that are fainter than the sky-uncertainty limit.}
\label{fig:faceon}
\end{figure*}

\subsection{Two galaxies where scattered light dominates faint structures: NGC 7280 and NGC 4102}\label{sec:faceonsl}
Figure~\ref{fig:faceon}a shows surface-brightness profiles of my models of NGC 7280 together with the $R$-band measurements of {\EPB}. An anti-truncation break appears at $r_{\text{abr}}\simeq100\arcsec$. The convolved profiles overlap the measurements very closely out to the scattered-light halo radius $50\la r_{110}\la60\arcsec$. Outside of both $r_{110}$ and $r_{\text{abr}}$, there is more scattered light than the measurements indicate. The suggested type III structure seems to be explained by the procedure where {\EPB} mask extended `haloes' of nearby bright stars, whereby they also remove halo light from NGC 7280 (cf.\ their sect.~5). The colour profiles in Fig.~\ref{fig:faceon}b show red excess immediately outside of the break radius at $r_{\text{br}}=42\farcs5$. For larger radii, the values are $\Delta\left(R-i\right)\simeq0.23\,\magg$ at $r=60\arcsec$ and about $0.84$--$0.96\,\magg$ at the reported anti-truncation break radius $r_{\text{abr}}=100\arcsec$. The amount of red excess decreases outside the maximum at $r=120\arcsec$. The increasingly red model profile for $r\ga135\arcsec$ originates in the outer parts of the $r^{1/4}$ bulge.

Corresponding surface-brightness profiles of the models of NGC 4102 are shown in Fig.~\ref{fig:faceon}c. They show a good match between the $R$-band measurements and the profiles of the convolved models that use either {\PSFVa} or {\PSFK}. The kink in the convolved profiles of NGC 7280 and NGC 4102, where $30\la r\la40\arcsec$, is due to the limiting radius where isophotes at larger radii were fitted with ellipses instead of circles. Using {\PSFK}, the anti-truncation radius $r_{\text{abr}}\approx100\arcsec$ agrees well with the scattered-light halo radius $r_{110}$.

The measurements of \citet{Co:96} are brighter than those of {\EPB} by up to 1\,\magg for $r\la150\arcsec$. A difference such as this could be explained by different PSFs or inaccurate sky subtraction. The figure suggests that the PSF of \citet{Co:96} could be closer to {\PSFVb}; his measurements are clearly underestimated for $r\ga150\arcsec$, where $\mu\ga26.7\,\magg$, but these data points are fainter than his reliability limit value $\mu=26\,\magg$. There is red excess immediately outside of the break radius at $r_{\text{br}}=42\farcs9$ also for this galaxy, Fig.~\ref{fig:faceon}d. At larger radii, the values are $\Delta\left(R-i\right)\simeq0.080$--$0.11\,\magg$ at $r=60\arcsec$ and about $0.48\,\magg$ at $r_{\text{abr}}$.

\section{About analyses of galaxy structures that are fitted with S{\'e}rsic-type profiles}\label{sec:sersic}
Some of the first astronomical studies that address the role of scattered light examine observations of nearby elliptical and lenticular galaxies that appear large, where these effects are found to be small and even insignificant. Galaxies included in such studies include \object{NGC 221}, NGC 3115, \object{NGC 3379}, \object{NGC 4494}, and \object{NGC 4649} \citep{Va:48,Va:53,VaCa:79,CaVa:83,CaHeNi:87} and M 31 \citep{Va:58}. Excess light in the outer regions of the cD galaxy \object{A 2029} closely follows de Vaucouleur's law, after that the diffuse scattered-light component of field stars is carefully removed \citep{UsBoKu:91}. A less dramatic influence of scattered light is found in a later study of envelopes of four additional cD and D galaxies \citep[\object{NGC 7647}, \object{NGC 7720}, \object{NGC 7728}, and \object{A 407 G1}]{Mac:92}.

\citet{Mi:99} and \citet{MiPo:00} correct colour profiles in centre regions of elliptical and lenticular galaxies from effects of variable PSFs (``differential seeing''). \citet{IdMiFr:02} study colour profiles of 36 elliptical and lenticular galaxies and correct colours for wavelength-dependent scattered-light effects; to remove artificial colours, they convolve data of one filter with the PSF of the other filter.\footnote{Studies that choose to use this method should check its validity as there is no comment about its accuracy for different sets of parameters.} {\Michard} studies the lenticular galaxy NGC 3115 as well as the four elliptical galaxies \object{NGC 4406}, \object{NGC 4473}, \object{NGC 4551}, and \object{NGC 4550}. The two latter galaxies are smaller than the two earlier elliptical galaxies, and are found to be more strongly affected by scattered light. The general opinion, however, that PSF effects are unimportant in elliptical-type galaxies, seems to originate in the early studies.

The studies I mention above provide object-specific results regarding the influence of scattered light. There is, so far, no general {\sersic}-profile parameter study that shows how strong scattered-light effects are in such {\sersic}-type galaxies of variable properties. A systematic search of scattered-light haloes in a large parameter range that includes objects that appear both small and large is provided in Sect.~\ref{sec:systematic}. Host galaxies of BCGs are typically modelled with $n=1$, and these objects are therefore relatively strongly affected by scattered light. Multiple examples of BCGs are presented in Sect.~\ref{sec:bcg}. Thereafter, an example of a galaxy with intermediate effects of scattered light is discussed in Sect.~\ref{sec:intermediate}, and two examples of large galaxies where effects are small are shown in Sect.~\ref{sec:small}.

\begin{figure*}
\centering
\includegraphics{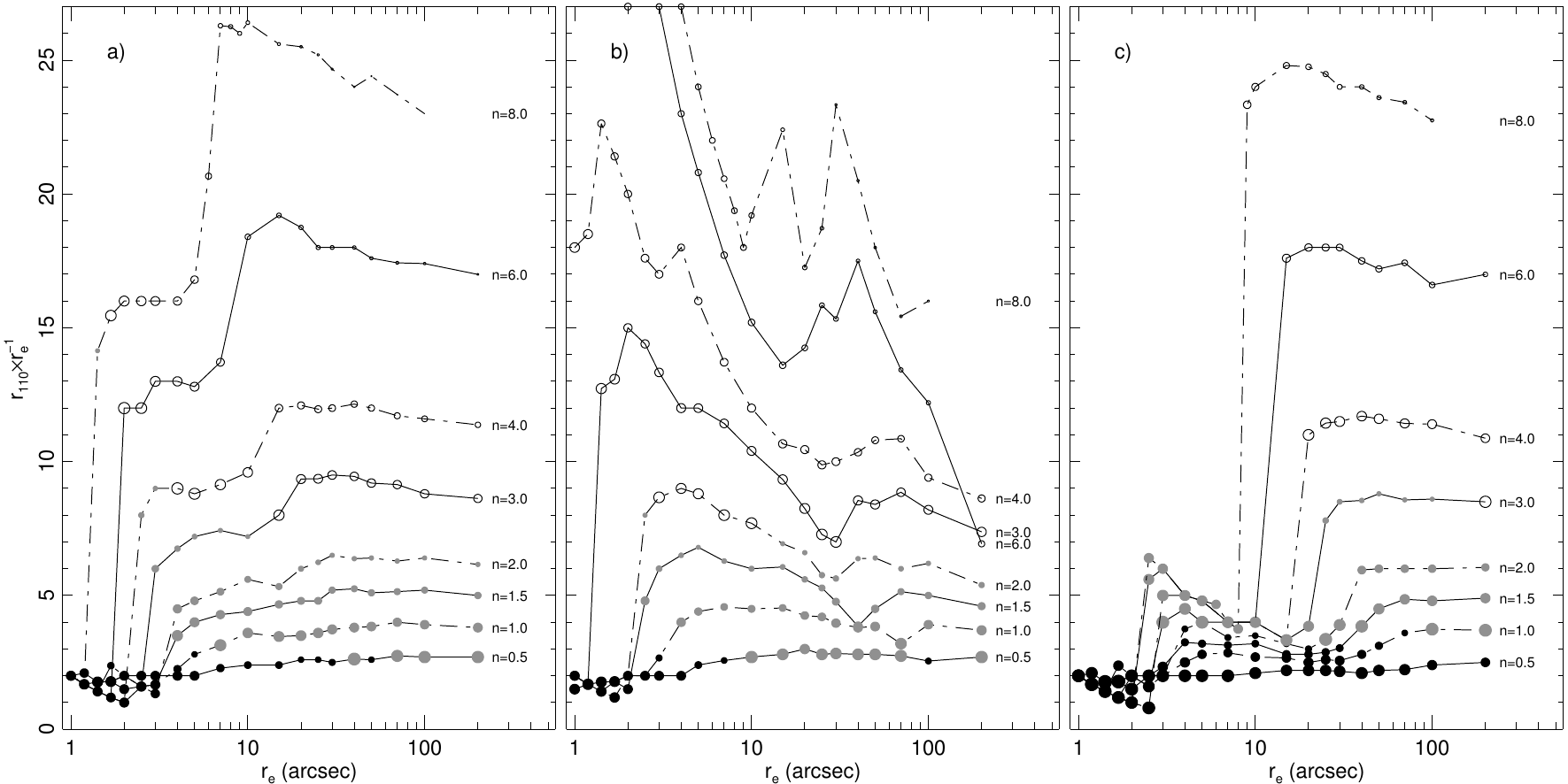}
\caption{Scattered-light halo radius $r_{110}$ in {\sersic}-type galaxies that are modelled with the {\sersic} profile. $r_{110}\ere^{-1}$ is drawn versus $r_{\text{e}}$ for $n=0.5$, $1.0$, $1.5$, $2.0$, $3.0$, $4.0$, $6.0$, and $8.0$. Each model was convolved using three different PSFs: \textbf{a}) {\PSFVa}, \textbf{b}) {\PSFK}, and \textbf{c}) {\PSFIa}. The magnitude at $r_{110}$ relative to the maximum value $0\,\magg$ at the centre is indicated with three different symbols: {\large$\bullet$}, $3<\mu\le7\,\magg$; \textcolor{gray}{\large$\bullet$}, $7<\mu\le11\,\magg$; and {\large$\circ$}, $11<\mu\le15\,\magg$. The symbol sizes are scaled linearly, where larger (smaller) symbols indicate a lower (higher) magnitude.}
\label{fig:shalo}
\end{figure*}

\subsection{A systematic search for the appearance of scattered-light haloes in {\sersic}-profile galaxy models}\label{sec:systematic}
I calculated and analysed a large number of elliptical-type galaxy models, to characterize PSF effects for a mostly complete range of the {\sersic}-profile parameters $n$ and $r_{\text{e}}$. Eight values on $n$ were used with the models: $0.5$, $1$, $1.5$, $2$, $3$, $4$, $6$, and $8$. For each value of $n$, a $V$-band and an $I$-band model were calculated using sixteen different half-light radii $r_{\text{e}}$: $1$, $2$, $3$, $4$, $5$, $7$, $10$, $15$, $20$, $25$, $30$, $40$, $50$, $70$, $100$, and $200\arcsec$. Not all parameter combinations are expected in structures of real galaxies. Larger values on $n$ and {\ere} require grids with very large pixels, and therefore low resolution in the centre. I assumed that $\mu_{\text{e},I}=\mu_{\text{e},V}-2$. Each model was convolved with {\PSFVa}, {\PSFK}, and {\PSFIa}. The scattered-light halo radius $r_{110}$ is shown versus $r_{\text{e}}$ and $n$ for all three PSFs in Fig.~\ref{fig:shalo}. As the figure shows, a scattered-light halo appears at smaller radii and lower magnitudes for lower values of $n$ and {\ere}.

With {\PSFVa} (Fig.~\ref{fig:shalo}a) and when $n\la2.0$, haloes are visible at a magnitude that is $\mbox{<10}\,\magg$ fainter than the centre surface brightness, and a radius $r_{110}\la6\ere$. The corresponding radius is $r_{110}\la4\ere$ ($6.7h$) for galaxies where $n=1.0$, which includes edge-on and face-on disc galaxies. Surface brightnesses at $r_{110}$ are for higher values on $n$ and {\ere} so faint that the scattered-light halo is unlikely to be observed in galaxies that are already faint at the centre; coarse limiting parameter values are, for example, $n=3.0$ and $\ere\simeq3\arcsec$ or $n=4.0$ and $\ere\simeq2\farcs5$.

The scattered-light halo radius $r_{110}$ moves outwards to larger radii with the lower-limit scattering {\PSFK} (Fig.~\ref{fig:shalo}b); for example, $r_{110}\la4.5\ere$ for $n=1.0$ and $r_{110}\la9\ere$ for $n=2.0$. Examples of coarse limiting parameters for models where surface brightnesses are too faint to be observed at $r_{110}$ are $\ere\ga1\farcs5$ for $n=3.0$ and $\ere\ga1\arcsec$ for $n=4.0$. In contrast, because of the red-halo effect \citep{SiClHa.:98}, $r_{110}$ moves inwards to shorter radii with the more strongly scattering {\PSFIa} (Fig.~\ref{fig:shalo}c). This is particularly noticable for values $\ere\la10$--$20\arcsec$.

\subsection{On observations and general properties of BCGs}\label{sec:bcg}
\citet{SaSe:70} conclude that compact galaxies are isolated extragalactic \ion{H}{ii} regions. Later, \citet{SeSa:72} and \citet{SeSaBa:73} find two possibilities regarding the origin of BCGs; at first, they argue that they are young galaxies that form their first generation of stars, but they settle with an explanation where star formation occurs in short bursts that are separated by long quiescent periods. Various morphological classes are introduced to explain both the star-formation region and the enclosing host galaxy \citep{LoTh:86a,KuMaVi:88,SaAlBo:89,TeMeTe:97}. Radial isophotes of faint host galaxies are found to decline as an exponential \citep[e.g.][]{Be:85,TeTe:97}, a de Vaucoleur's law \citep[i.e.\ a {\sersic} profile with $n=4$, e.g.][]{LoTh:86,DoCoPe.:97}, multi-component models \citep[hereafter \Papaderos]{PaLoThFr:96} and with {\sersic} profiles that use other values on $n$ \citep[\citealt{BeOs:02}, hereafter \BO; \citealt{CaCaAgMu:05}, hereafter {\Caon};][]{AmMuAg.:07,AmAgMuCa:09}. General physical and observable properties of BCGs are overviewed by, for example, \citet{KuOs:00}.

Comparisons of structural properties, average colours, and colour gradients between different classes of dwarf galaxies provide a mean to test evolutionary scenarios between the classes \citep{Th:85,BoMoCaGi:86,DaPh:88}. The assessment of properties of the host is important to establish the evolutionary state of the galaxy and its star-formation history. However, the host is fainter than the starburst regions, which makes it difficult to measure.

In hosts with an exponential decline, scattered-light haloes appear at relatively bright intensities and small radii $r_{110}\simeq6.7$--$7.5h$ ($4$--$4.5\ere$), see Fig.~\ref{fig:shalo}. Integrated light from the host is modulated by scattered light from the bright, compact, and possibly asymmetrically situated central starburst regions, which may complicate the interpretation of the structure. Scattered-light haloes should be visible in BCGs, but they were never reported.

Here, I discuss surface-brightness profiles of five BCGs, with a focus on the faint outer parts of the host. I selected galaxies where there are measurements for $r\ga25\arcsec$, to avoid problems with the relatively high seeing of the PSFs I use and resulting poorly resolved central regions. Most BCGs are very small and are only observed to shallow depths. I study surface-brightness profiles -- including the role of the ellipticity -- of the extreme BCG ESO 400-G043 in Sect.~\ref{sec:bcgeso}, which surface-brightness profiles are dominated by the bright starburst regions. {\Caon} find that a {\sersic} profile fits the measurements of Mrk 5 and I Zw 123 better than an exponential profile. I discuss how scattered light could be a part of the explanation of the result for Mrk 5 in Sect.~\ref{sec:bcgMrk5}, which in comparison to ESO 400-G043 clearly also shows parts of the exponential profile of the host. Three additional BCGs are discussed in Appendix~\ref{sec:abcg}: the extreme BCG ESO 350-IG038 (Haro 11) that hosts asymmetrically placed starburst regions; observations of the BCG Mrk 297, which seem to be affected by varying levels of sky subtraction in addition to scattered light; and UM 465, which is included in a recent extensive study of BCGs \citep[hereafter {\Michevab}]{MiOsZa.:13}. Parameters of observed and modelled surface-brightness profiles of the discussed BCGs are collected in Table~\ref{tab:bcg}, and details of the observations are given in Table~\ref{tab:obs}.

\subsubsection{The extreme BCG ESO 400-G043}\label{sec:bcgeso}
Out of four BCGs in {\BO}, I fitted the surface-brightness profiles of ESO 400-G043 with the smallest value on the scale length, resulting in effects of scattered light that are slightly larger than in ESO 350-IG038 and much larger than in ESO 338-IG04 or ESO 480-IG12. {\BO} present $B$-band profiles of ESO 400-G043 that they fit with both an exponential disc and a {\sersic} profile. More recent $V$-band profiles are presented by \citet[hereafter {\Michevaa}]{MiOsBe.:13}, who measure an ellipticity of $e=0.22$. The authors fit their measured surface-brightness profiles with two exponential discs in two separate magnitude ranges.

\begin{figure*}
\sidecaption
\includegraphics{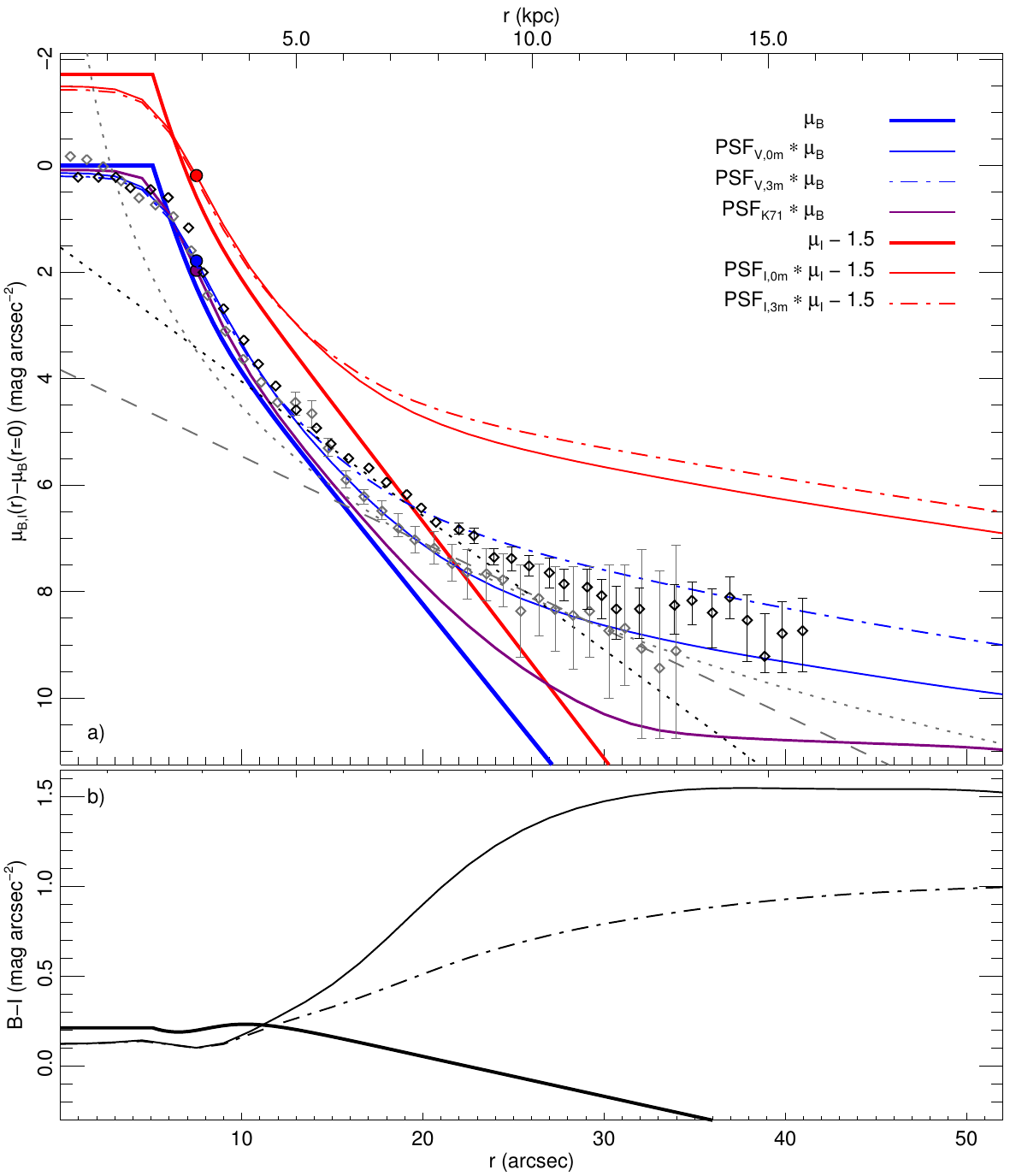}
\caption{Surface-brightness profiles versus radius for the BCG ESO 400-G043. Panels, lines, and coloured bullets are as described in Fig.~\ref{fig:thickdisc}. The $I$-band profiles are offset by $-1.5\,\magg$ from the $B$-band profiles for increased visibility. Grey and black bullets and error bars mark measured $B$-band (\BO) and $V$-band (\Michevaa) values, respectively. The grey dotted (dashed) line shows the best-fit {\sersic}-type model (exponential fit to the disc in the range $16<r<32\arcsec$) of {\BO}. The black dotted line shows the best exponential fit to the disc of {\Michevaa} (for $24\le\mu_{V}\le26\,$\magg).}
\label{fig:ESO400-G043}
\end{figure*}

I set the model parameters to achieve a reasonable agreement between convolved structures and both the $B$-band and the $V$-band measurements. My model consisted of a bulge, where $n=1.0$ and $\ere=1\farcs5$, that is superposed on an exponential-disc host where $n=1.0$ and $\ere=4\farcs3$, $\mue^{\text{host}}-\mue^{\text{bulge}}=5.75\,\magg$, and $\mu_{\text{e},B}-\mu_{\text{e},I}=0.5\,\magg$. I also set the intensity constant for $r<5\arcsec$. Major-axis surface-brightness profiles, the measurements, and all three fits are shown in Fig.~\ref{fig:ESO400-G043}a.

Outside a fairly flat core, intensities decrease steeply until they are about $\mu=5\,\magg$ fainter than at the centre. A scattered-light halo dominates at fainter intensities, but begins already near the centre; see the locations of the coloured bullets in the figure. The $B$-band measurements match the profile of the model that was convolved using {\PSFVa} well. The $V$-band measurements better match the profile that used {\PSFVb}. The dependence on scattered light with time -- as is represented by the different PSFs -- is clearly seen in the separation between the profiles of the convolved models. For example, compare the blue and the purple lines that differ by about $3\,\magg$ at $r=32\arcsec$. My parameter values are lower than those of {\BO} and {\Michevaa}; they measure their profiles in the radial range where scattered-light effects appear to dominate, $r\ga14\arcsec$.

The $B-I$ colour profiles in Fig.~\ref{fig:ESO400-G043}b clearly show more red excess with the two earlier {\PSFVa} and {\PSFIa} than with the later {\PSFVb} and {\PSFIb}. Examples of values on the red excess at $r=30\arcsec$ are $\Delta\left(B-I\right)\simeq1.6\,\magg$ with {\PSFVa} and {\PSFIa}, and about $1.0\,\magg$ with {\PSFVb} and {\PSFIb}.

In view of the results of this study, the outer parts of the surface-brightness and colour profiles of the more distant ESO 400-G043 and ESO 350-IG038 appear extreme because of the scattered light, which originates in the very bright starburst regions in the centre. The starburst regions in the two BCGs ESO 338-IG04 and ESO 480-IG12 are fainter and the profiles thereby less extreme (not shown here). My study suggests that the host is visible in the inner regions of the extended parts of these two latter BCGs as well as in Mrk 5 (see below), whilst it is completely concealed in the first two objects. Effects of scattered light are significant in all four galaxies due to the small value on {\ere}. I did not separately model the bands $B$, $V$, or $K$, but it seems reasonable that a similar conclusion applies to the remaining strange red-halo objects that are reported by \citet{BeMaPe.:05}.

\begin{figure}
\centering
\includegraphics{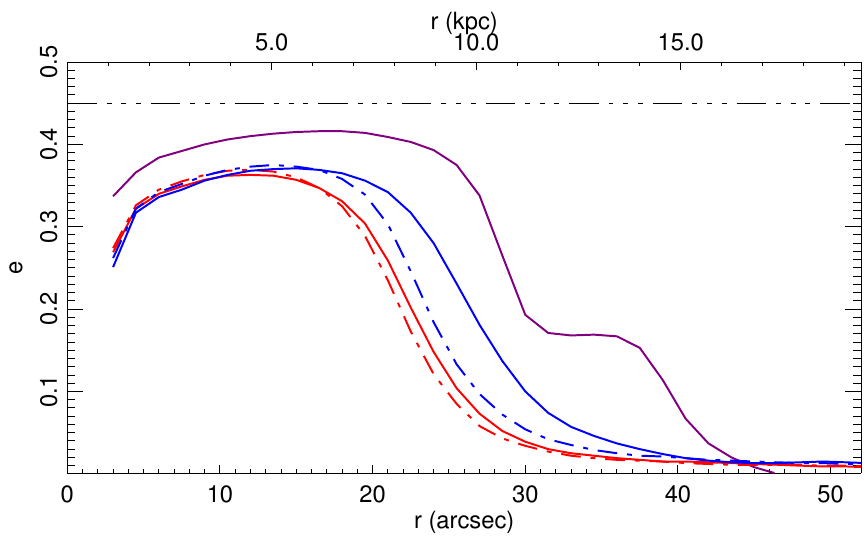}
\caption{Ellipticity versus radius $e(r)$ of convolved models of ESO 400-G043; although, using the model ellipticity $e=0.45$ (dash-tripple-dotted line). The model was convolved using five PSFs: {\PSFK} (purple line), {\PSFVa} and {\PSFVb} (blue solid and dash-dotted lines), {\PSFIa} and {\PSFIb} (red solid and dash-dotted lines).}
\label{fig:ellipticity}
\end{figure}

Besides inducing haloes and red excess, scattered light affects the flattening that is measured in elongated objects. I show one example here to illustrate how measured ellipticities of objects such as ESO 400-G043 depend on the PSF. In this context, {\Michevaa} measure the ellipticity $e=0.22$ for this BCG. I instead used $e=0.45$, to enhance the effects of scattered light. I convolved the flattened model with five PSFs and measured the ellipticity in the resulting model, see Fig.~\ref{fig:ellipticity}.

None of the measured ellipticities reach the constant model value. All measurements show more circular values, i.e.\ the ellipticity is smaller. PSFs with more scattered light show more circular structures; compare, for example, the curve of {\PSFIb} with that of {\PSFK}. In the innermost region, the ellipticity increases with radius for all PSFs, because of the bulge that was not flattened. For example, using {\PSFVa}, the ellipticity increases slightly from $e(r\simeq3\arcsec)=0.25$ to $e(15\la r\la20\arcsec)=0.37$. At larger radii, the ellipticity decreases towards zero for all PSFs. Scattered-light effects in flattened galaxies are more pronounced on the minor axis than on the major axis, but measurements along ellipses deviate minutely from major-axis measurements (not shown). It appears that the only way to measure true radial ellipticity structures is to deconvolve the measured surface-brightness structure.

\subsubsection{The BCG Mrk 5}\label{sec:bcgMrk5}
Observations of Mrk 5 were made in the $B$-band \citep[hereafter \Cairos]{CaViGo.:01}, who fit the measurements with an exponential profile. The data are reassessed by {\Caon}, who fit the host-galaxy measurements with a {\sersic} profile; these authors extend the set of published data points to $r=58\arcsec$. \citet{AmAgMuCa:09} also use the data to make a two-dimensional {\sersic}-profile fit.

\begin{figure*}
\sidecaption
\includegraphics{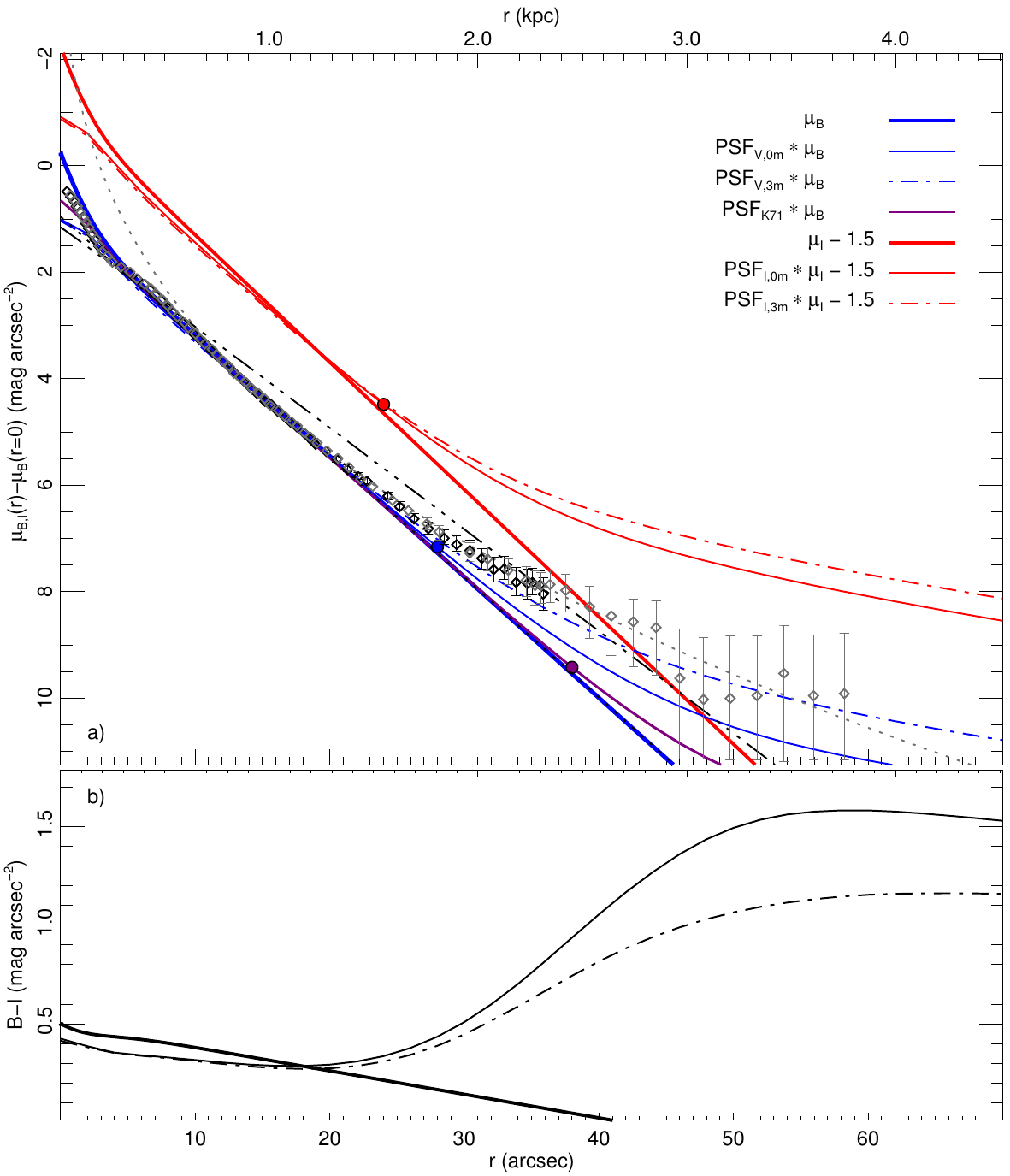}
\caption{Surface-brightness profiles versus radius for the BCG Mrk 5. Panels, lines, and coloured bullets are as described in Fig.~\ref{fig:thickdisc}. The $I$-band profiles are offset by $-1.5\,\magg$ from the $B$-band profiles for increased visibility. Black (grey) diamonds and error bars mark measured $B$-band values of {\Cairos} (\Caon). The black (grey) dotted line shows the best-fit {\sersic}-type model of {\Cairos} (\Caon); the black line is difficult to see, as it falls atop the $B$-band model line. The dash-tripple-dotted line shows the fitted profile of \citet{AmAgMuCa:09}.}
\label{fig:Mrk5}
\end{figure*}

I matched the measurements with a model using a bulge where $n=1$, $\ere=2\farcs0$, and $\mue=21.3\,\magg$, and a host where $n=1$, $\ere=8\farcs0$, and $\mue=22.0\,\magg$. I also set $\mu_{\text{e},B}-\mu_{\text{e},I}=0.5\,\magg$. It is only possibly to achieve a fair agreement between the bulge and the measurements for $r\la3\arcsec$, but there is only little intensity in this region, which effects on the larger object are small. Surface-brightness profiles of the models and the three sets of profile fits are shown in Fig.~\ref{fig:Mrk5}a.

According to {\Michevaa}, the {\PSFMich} of the NOT is close to {\PSFK} and even decreases for $r\ga30\arcsec$, see Fig.~\ref{fig:p2psf}; they conclude it is too faint to influence the observations. No PSF shows a decline that is steeper than a $r^{-2}$ power-law decline (also see figs.~1 and 2 in {\rS}) and it seems unlikely that the PSF of the NOT is any different. With my model parameters, the outer measurements of Mrk 5 appear to require a PSF that is as bright as {\PSFVb} to be explained as scattered light. There are three circumstances that make this possible. The NOT PSF could, as other PSFs, vary with time and actually be as bright as {\PSFVb}, there is not enough information about which data were used to derive {\PSFMich}. The PSF needs to be 110\% as extended as the measurements to correctly estimate its influence (Appendix~\ref{sec:toy}), considering {\PSFMich} this would affect all values for $r\ga25\arcsec$, including nearly all of the host (or halo). It is unknown how the used sky-subtraction procedure has affected the measurements, in particular where $\mu_{B}>26\,\magg$ for $r\ga24\arcsec$ (compare with the discussion for Mrk 297 in Appendix~\ref{sec:bcgMrk297}).

The scattered-light halo radius $r_{110}\simeq25\arcsec$ for {\PSFVa} and $r_{110}\simeq36\arcsec$ for {\PSFK}, whilst it is closer to $r_{110}\simeq22\arcsec$ for {\PSFVb}. The $B$-band measurements of {\Cairos} and the reassessed values of {\Caon} overlap each other and the convolved model using {\PSFVb}; the values deviate slightly for $r\la3\arcsec$, but that deviation is unimportant to the discussion here. The exponential fit of {\Cairos} agrees with the value I chose for the single-disc host component. {\Caon} fit values at all radii with a single {\sersic} profile and thereby need to use the higher value $n=2.83$; they also use an equally high value with the smaller appearing BCG I Zw 123 (not shown here). The fit of \citet{AmAgMuCa:09} agrees poorly with the measurements in large intervals (as with Mrk 297).

The $B-I$ colour profiles in Fig.~\ref{fig:Mrk5}b show dramatic amounts of red excess for $r\ga18\arcsec$. For example, at $r=30\arcsec$ the red excess is $\Delta\left(B-I\right)\simeq0.31$--$0.37\,\magg$ and at $r=40\arcsec$ the values are $\Delta\left(B-I\right)\simeq0.8$--$1.0\,\magg$.

\begin{figure}
\centering
\includegraphics{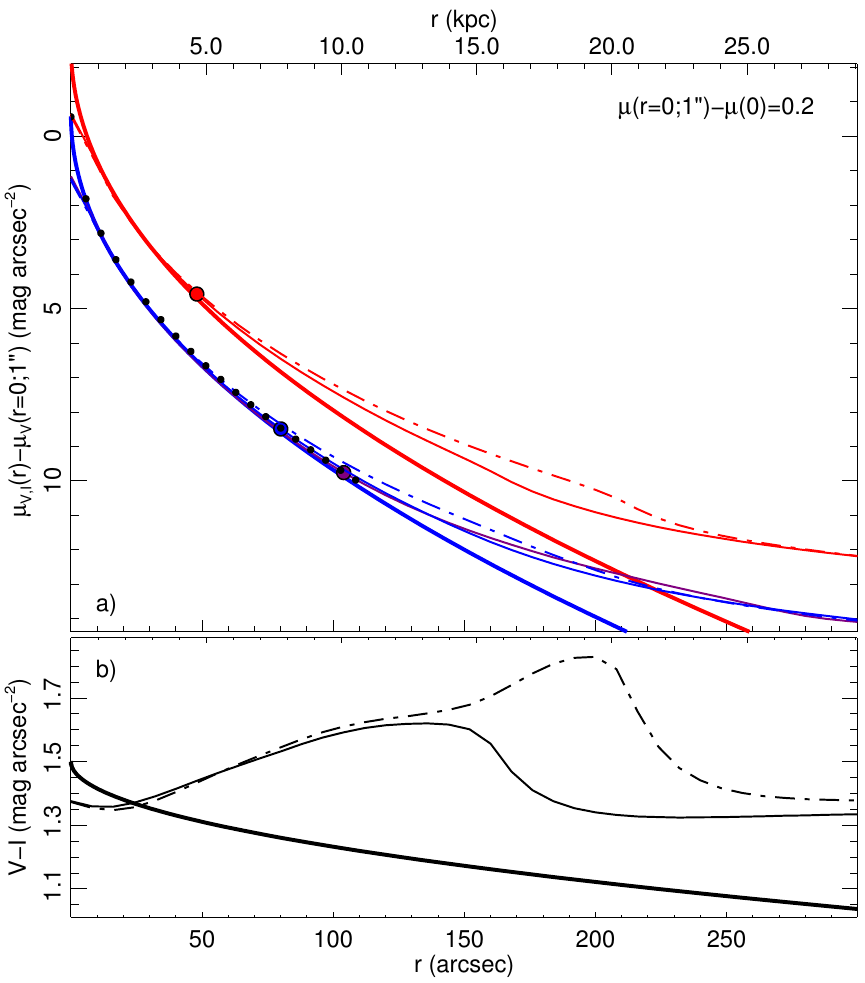}
\caption{Surface-brightness profiles verus radius for NGC 4551. Panels, lines, coloured bullets, and offsets are as described in Figs.~\ref{fig:thickdisc} and \ref{fig:faceon}. The $I$-band profile in panel \textbf{a} is offset by $-0.5\,\magg$ from the $V$-band profile. Black bullets show a fitted {\sersic} profile (the parameters are taken from {\Kormendy}).}
\label{fig:smallE}
\end{figure}

\subsection{About galaxies where scattered-light effects are intermediate}\label{sec:intermediate}
I classify intermediate effects of scattered light as those that appear in surface-brightness structures where the brighter parts are well fitted with $1<n<4$. I study one such galaxy here, NGC 4551. \citet[hereafter {\Kormendy}]{KoFiCo.:09} present observations of NGC 4551 of eight different sources in the $B$, $V$, and $R$ bands. The authors fit a $V$-band model to the data with the parameters $n=1.98$, $\ere=15\farcs5$, $\mu_{\text{e}}=20.75\,\magg$, and $D=20\,\text{Mpc}$. The profile extends out to $r=110\arcsec$ where $\mu\simeq27.3\,\magg$. {\Michard} presents observations that extend out to $r=70\arcsec$. I modelled NGC 4551 using $n=2.0$ and $\ere=15\arcsec$, neglecting the somewhat brighter region where $r<2\farcs8$. To broaden the applicability of this one study to faint structures of similar but brighter galaxies, I extended the models out to $r=300\arcsec$. Surface-brightness and colour profiles as well as the $V$-band measurements are shown in Fig.~\ref{fig:smallE}a. Other Virgo-cluster galaxies, at about the same distance, which surface-brightness profiles are fitted with similar parameters, include \object{IC 0798}, \object{IC 3509}, \object{NGC 4387}, \object{NGC 4434}, \object{NGC 4458}, \object{NGC 4478}, \object{NGC 4515}, \object{VCC 1407}, \object{VCC 1545}, \object{VCC 1828}, and \object{VCC 1910} (\Kormendy).

The scattered-light halo radii are $r_{110}\simeq80\arcsec$ (\PSFVa), $\sim105\arcsec$ (\PSFK), and $\sim45\arcsec$ (\PSFIa), also see Fig.~\ref{fig:shalo}. The first two radii are larger than (similar to) the radial range that is covered by {\Michard} (\Kormendy). Differences between profiles that were calculated using {\PSFK}, {\PSFVa}, and {\PSFVb} (or {\PSFIa} and {\PSFIb}) are small, and are more pronounced where $r>80\arcsec$. Considering that the PSFs are less accurate at large radii, say for $r\ga100\arcsec$, it is difficult to draw a conclusion on how sensitive the outer structure is to temporal variations in the PSFs. Convolved surface-brightness profiles of the separate PSFs differ at $r=80\arcsec$ by at most $0.15\,\magg$ in both bands, and at $r=150\arcsec$ by $\la0.3\,\magg$ in the $V$ band and by $\la0.5\,\magg$ in the $I$ band. The convolved $V-I$ colour gradient differs from the model gradient throughout the radial range, also where $r<r_{110}$, cf.\ Fig.~\ref{fig:smallE}b. There is red excess for $r>25\arcsec$, which reaches $V-I\simeq0.29\,\magg$ at $r=130\arcsec$ using {\PSFVa} and {\PSFIa}, and $V-I\simeq0.30\,\magg$ using {\PSFVb} and {\PSFIb}. The colour gradients are positive in the regions $20\la r\la150\arcsec$ and $r\ga250\arcsec$. The noticeable bump between the two colour profiles, where $150\la r\la250\arcsec$, appears as a result of how I extrapolated the PSFs at larger radii.

Surface-brightness profiles of galaxies fitted with models using similar {\sersic} parameters, such as those that I mention above, will show similar amounts of red excess. Whilst scattered-light effects always appear at some radius, the measured surface brightness is often too faint to reveal the scattered-light halo. However, the red excess seen in colours is detectable also at lower magnitudes and radii. Galaxy profiles fitted with smaller {\ere} have steeper colour gradients and brighter scattered-light haloes; examples of such galaxies in the Virgo cluster include \object{IC 0809}, \object{IC 3461}, \object{IC 3470}, \object{IC 3490}, \object{IC 3653}, \object{NGC 4464}, \object{NGC 4486A}, \object{NGC 4486B}, \object{VCC 1871}, \object{NGC 4467}, \object{VCC 1199}, \object{VCC 1440}, and \object{VCC 1627} (\Kormendy).

\begin{figure*}
\sidecaption
\includegraphics[width=12cm]{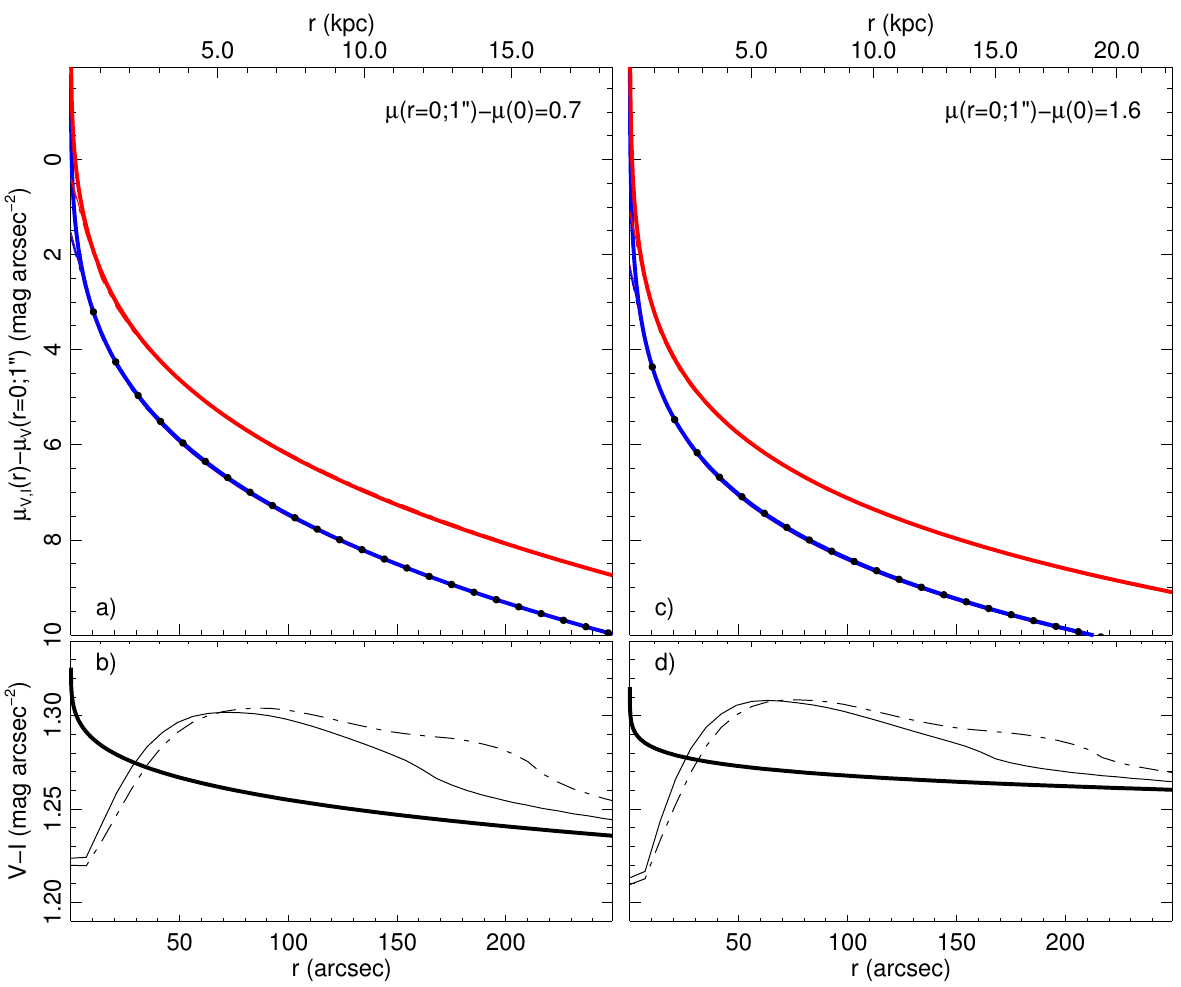}
\caption{Scattered-light effects of two Virgo-cluster galaxy models: NGC 4473 (left panels) and NGC 4374 (right panels). Panels, lines, and offsets are as described in Figs.~\ref{fig:thickdisc} and \ref{fig:faceon}. The convolved profiles are invisible as they fall atop the model profiles. Black bullets show a fitted {\sersic} profile for the respective galaxy (the parameters are taken from {\Kormendy}).}
\label{fig:largeE}
\end{figure*}

\subsection{About galaxies where scattered-light effects are smaller}\label{sec:small}
Scattered-light effects in surface-brightness structures of larger elliptical and cD galaxies ($n>4$) are typically small to very small. Instead of discussing the appearance of scattered-light haloes at large radii and faint intensities in such galaxies, I focus on colour gradients at smaller radii where $r<250\arcsec$. Surface-brightness profiles are shown in Figs.~\ref{fig:largeE}a and \ref{fig:largeE}c for two larger galaxies, NGC 4473 and \object{NGC 4374}.

{\Kormendy} fit observed surface-brightness structures of NGC 4473 with $n=4.0$, $\ere=51\farcs8$, using the distance $D=15.28\,$Mpc, and of NGC 4374 with $n=8.0$, $\ere=142\arcsec$, and $D=18.45\,$Mpc. I used the same model parameters, and I set $r_{\text{e},I}=0.97r_{\text{e},V}$ for both galaxies. With these galaxy models, a scattered-light halo appears at $r_{110}^{\text{NGC 4473}}\simeq600\arcsec$ and $r_{110}^{\text{NGC 4374}}\simeq3200\arcsec$ (cf.\ Fig.~\ref{fig:shalo} and Sect.~\ref{sec:systematic}). Both convolved models show $V-I$ colour profiles that deviate from the respective model profile where $r\la250\arcsec$, cf.\ Figs.~\ref{fig:largeE}b and \ref{fig:largeE}d.

The radial colour profile of NGC 4473 increases with radius for $r\la80\arcsec$ -- in agreement with the result of {\Michard} for the same galaxy -- and decreases for larger radii. Intensity differences due to the used temporally variable PSFs are minute; colour profiles of the convolved models differ by $<0.01\,\magg$ for $r<80\arcsec$ and by $<0.28\,\magg$ for $r<250\arcsec$. Colours are much less affected than with lower {\sersic}-profile parameter values, but there is still some blue excess in the innermost region and red excess in the remaining region. The maximum colour difference across the shown radial range is $\left(V-I\right)_{\max}\simeq0.08\,\magg$. The agreement is striking between these colour profiles and those of the much larger galaxy NGC 4374, both regarding amplitudes and shape. It seems that in this case the colour difference appears because of the convolution with a PSF, but the result is nearly insensitive to the exact properties of the model.

Compared to models of smaller galaxies with low values on $n$, it is comparatively easy to deconvolve measured data of large galaxies such as these. The accuracy requirements on the measurements and the PSF to remove the colour artefacts are small.

\section{Discussion}\label{sec:discussion}
The presence of diffuse excess light around galaxies is ubiquitous, and it is tightly linked with the existence of colour excess at intermediate to large radii. The outcome and implications of this paper are discussed in the following sections.

\subsection{Edge-on galaxies}\label{sec:disceo}
All three scrutinies of haloes around edge-on galaxies studied here, including IC 5249, NGC 4565, and NGC 4244, corroborate the result for NGC 5907 in {\rS} -- the faint parts of the observed surface-brightness profiles could be explained by scattered light. Specifically, \citet[hereafter \Fry]{FrMoHaBo:99} fit their surface-brightness profiles of NGC 4244 with a single exponential disc and argue that the galaxy does not have a halo. My scrutiny shows that their data trace the expected scattered-light profiles at large radii, where they deviate from an exponential profile. The agreement between photometry and resolved-stars measurements of this object is poor, which illustrates the need for further research to understand the differences.
      
A number of scenarios were proposed to explain the formation of currently discovered thick discs; see the overview in, for example, \citet{YoDa:08a}. It is argued that thick discs are a necessary consequence of disc formation in galaxy formation models \citep{CoElKn.:11}, and they are described as a `local missing baryons' reservoir. Another study discusses the thin-disc to thick-disc dichotomy in the Milky Way, and argues for the term `thicker disc component' instead of `thick disc' \citep{BoRiHo:12}. Both the thick disc and the halo indicate a presence of excess light, and they can be treated similarly \citep{Mo:99}. My results regarding diffuse thick discs is no different to the haloes. The two thick discs of FGC 310 and FGC 1285 bracket all the 34 objects of {\YD} (the effects of scattered light are stronger in FGC 1063 than in FGC 310 since its apparent size is smaller). All evidence points at scattered light being the ubiquitous component that conceals any real thick disc in addition to modifying the scale height of the thin disc. The geometrical shape of surface-brightness profiles of edge-on galaxies implies strong effects of scattered light in faint regions, regardless of the size and scale height of the studied object.

The analysis of the role of scattered light in stacked data can be more difficult when exact dates of the galaxy observations are unknown and their geometrical properties differ. I analysed parts of the stacked LSBG data of {\BZC}, who deconvolve their data, but neglect temporally varying PSFs. My results show that it is possible to model colour profiles with similar amounts of red excess in the outer parts as the presented deconvolved observations.

\citet{FeHaCi.:13} stack images of Ly$\alpha$ emitting galaxies at $z\sim2.1$ \citep[data from][]{GuGaPa.:10} and at $z\sim3.1$ \citep[data from][]{GrCiHi.:07,MaYaHa.:12}. They extract PSFs from bright stars in their stacks and use them to deconvolve the respective stack. The objects are very distant and therefore appear as point sources, which is why compact PSFs suffice. The $z\sim2.1$ data were observed within eleven days (2007 December 3--13), which means that one average PSF should suffice to analyse the data accurately. The data at $z\sim3.1$ were observed during a time interval of about 16 months \citep{GrCiHi.:07} and 38 months \citep{MaYaHa.:11,MaYaHa.:12}, which means that a single PSF might be an inaccurate approximation. Whilst \citet{FeHaCi.:13} explore an alternative origin of the halo for this case in other analysis-related issues, it cannot be excluded that the PSFs of the images used to find a (weakened) halo were slightly different from the mean PSF; the PSFs presented here in Fig.~\ref{fig:p2psf} show that differences are very small at the short radii of these point sources.

\subsection{Face-on disc galaxies}\label{sec:discfo}
Both {\PSFVa} and {\PSFK} appear to be close to the true PSF in the observations of NGC 4102 of {\EPB}. The surface-brightness profiles of the three galaxies observed with the SDSS and the WIYN telescope (NGC 4477, NGC 3507, and NGC 2880) also appear to be explainable with scattered light, except the outermost faintest measurements. In each case, the scattered-light halo radius $r_{110}$ agrees well with the anti-truncation radius $r_{\text{abr}}$, which is used as an indicator for type III-s profiles. It is important to remember that the models are simplified representations of the surface-brightness profiles of the real galaxies. Real variations in the intensity due to dust lanes or any azimuthal irregularities are neglected. It cannot be excluded that {\PSFVb} is closer to the true PSF, and that my models are lower limits. Besides the too large subtracted sky component, the differences of up to about $1.5\,\magg$ in the surface-brightness profiles of NGC 7280 between the profiles convolved using {\PSFVa}, {\PSFVb}, and {\PSFK}, indicate that the structures of this smaller appearing galaxy are highly sensitive to temporally variable PSFs.

There are three further complications with the observed galaxy profiles: the PSFs are unknown, especially at larger radii, which makes it impossible to accurately dismiss their influence; the accuracy of the sky subtraction seems to differ in some cases; and the adopted ellipticity affects the resulting profiles and the analysis. For example, scattered light causes a decrease of the ellipticity at larger radii in a galaxy model with a radially constant ellipticity (see the study of the BGC ESO 400-G043 in Sect.~\ref{sec:bcgeso}). Inclination angles may be underestimated when data are not first deconvolved.

The models of all five galaxies show significant amounts of red excess, which only appears outside of the inner break in those galaxies that have one. For galaxies without a break radius, the red excess begins already near the centre. Any study of radial colour gradients in the outer fainter parts of face-on disc galaxies is affected by scattered light.

Not all observations of face-on disc galaxies are deep enough to detect a type III profile, and depending on the model parameters of the inner and the outer disc the anti truncation may only appear at very large radii and faint intensities. The models presented here present good evidence that excess light at large radii around face-on disc galaxies is scattered light; the type III profile and in particular the type III-s profile could be a sign for it. Only after the scattered light is removed, it is possible to see if the excess light instead comes from the extended parts of the bulge in the centre, as is currently assumed for type III-s.

There are few identifications of haloes and anomalous colours in space-based observations. The uncertainties in the outermost parts of the \textit{HST} PSFs are undetermined, this concerns both the \textit{Hubble} Ultra Deep Field \citep[HUDF;][]{BeStKo.:06} edge-on galaxy study of \citet{ZiFe:04} that is addressed by {\Jong}, and the analysis of the haloes around UDF 3372 and UDF 5417 by \citet{TrBa:13}. Observations with space telescopes, such as the \textit{HST}, are less affected by temporally varying PSFs than ground-based observations, but it cannot be excluded that also here may there be some excess light at larger radii of the PSFs, so far, unaccounted for (cf.\ appendix~A in \rS).

\subsection{Galaxies fitted with {\sersic}-type profiles}\label{sec:discse}
All galaxy profiles modelled with a {\sersic} profile are always -- beginning at some radius -- affected by scattered light, for all reasonable values on $n$ and {\ere}. The exact values on the radius and the surface brightness where the scattered-light halo appears depend on the object properties and the PSF. With higher parameter values, the halo moves outwards to larger radii and fainter intensities, and resulting colour gradients become shallower. Whilst haloes of the largest appearing galaxies ($n\ga4$) are so faint that they are unlikely to be observed, colour gradients are still somewhat affected in the centre region, as is seen here with NGC 4473 and NGC 4374. At the other end of smaller appearing galaxies, a scattered-light halo appears at measurable magnitudes for all values of $\ere$ when $n=1$, and colours are dominated by differences between PSFs in large radial intervals.

Numerous measurements of BCG hosts are affected by scattered light. My analyses of the extreme BCGs ESO 400-G043 and ESO 350-IG038 indicate that the host-galaxy part in the observations is completely dominated by scattered light that originates in the bright starburst regions in the centre. The BCGs Mrk 297 and UM 465 were observed in different studies, they contain fainter starbursts and are less affected; here, the host galaxy appears to be visible near the centre, whilst scattered light still dominates in the fainter parts of the profiles. Current measurements of Mrk 297 differ in the outermost parts, plausibly so because of differences in the sky subtraction or the PSFs at the time of the observations -- results are ambiguous when both the sky subtraction is uncertain and the PSFs are unknown. The analysis of the BCG Mrk 5 that {\Caon} present shows values of the {\sersic} index $n\ga2.6$. The authors argue that this BCG together with I Zw 123 represents a new class of objects where indices $n>1$. My analysis suggests that a single exponential profile suffices to explain the measurements when scattered light is considered.

\subsection{Effects of scattered light in other wavelength ranges}\label{sec:discwave}
This study has focused on effects of scattered light in the visual wavelength range. It has also earlier been found and studied at radio wavelengths, where it is referred to as `stray radiation' \citep[for example,][]{KaMeRe:80}.

Several studies present surface-brightness profiles for the same types of galaxies strudied here in infrared bandpasses (see the respective overview sections). As is also noted in {\rS}, the PSFs of these bands, as measured with data of the 2MASS Large Galaxies Atlas \citep[][where observations were made at Mt.\ Hopkins in Arizona and the Cerro Tololo Inter-American Observatory]{JaChCu.:03}, are even brighter at larger radii than the $I$-band PSF \citep{Mi:07}. Extended PSFs are even less well determined in the infrared. Haloes, thick discs, and colour gradients are thereby just as questionable as those in the visual wavelength range.

A recent study reports on the ubiquity of diffuse haloes around galaxies in the ultraviolet wavelength range \citep{HoBr:14}. The study uses the Galaxy Evolution Explorer 50cm telescope and the space-based \textit{Swift} Observatory. The authors subtract some scattered light, but neglect integrated light altogether. Considering the results that are presented here in the visual wavelength range, it would seem obvious that scattered light plays an important role to explain the ubiquity also in the ultraviolet.

\subsection{Effects of scattered light are underestimated in analyses made in one dimension}\label{sec:discone}
A recent study finds that the outer discs and haloes in its observations of 698 disc galaxies are little affected by scattered light \citep{ZhThHe.:15}. Their PSFs appear similar to {\PSFK} (their figs.~28 and 29). The region they study extends out to about $10\arcsec$ (Zheng, priv.\ comm 2014). The authors present a one dimensional galaxy model convolved with their one dimensional PSFs that is very similar to the input model, and they conclude that scattered light has minor effects on their results.

\begin{figure}
\centering
\includegraphics{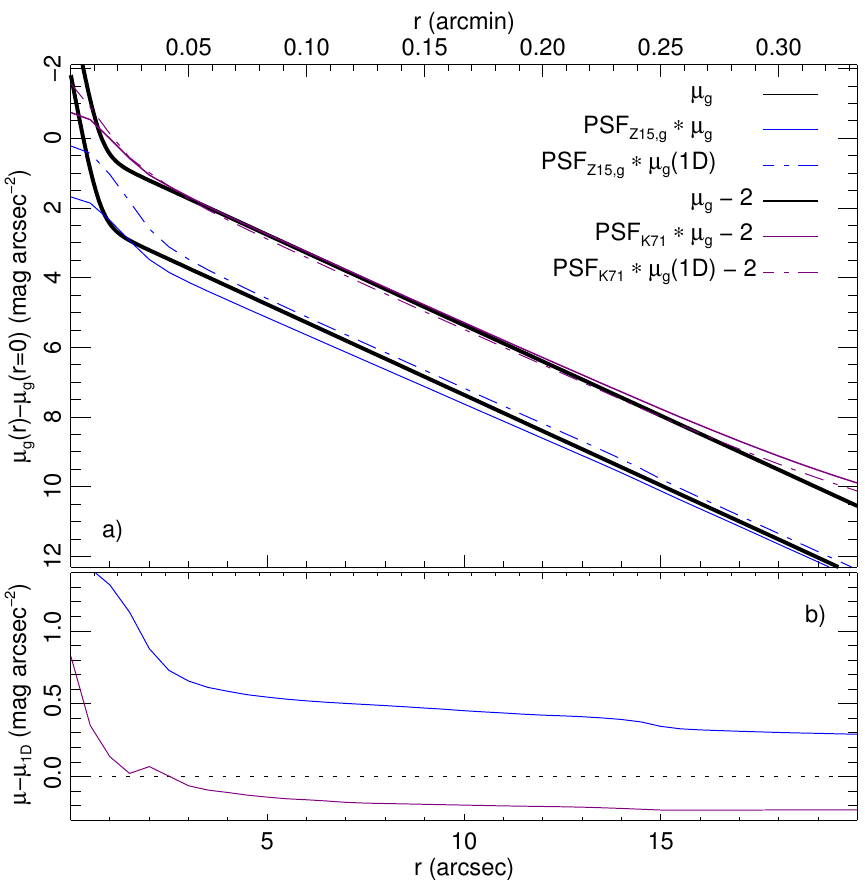}
\caption{Radial surface-brightness profiles. Panel \textbf{a}) shows the model of \citet{ZhThHe.:15} convolved with PSF$_{\text{Z15},g}$ (blue lines) and with \PSFK\ (purple lines), respectively. The regular (one-dimensional) model convolved with the regular (one-dimensional) PSF is shown with a solid (dash-dotted) line. Panel \textbf{b}) shows the difference between the convolved regular and one-dimensional model.}
\label{fig:Zheng}
\end{figure}

I modelled the disc galaxy in the $g$ band both in one and two dimensions, using $\ere=3\farcs5$ and a bulge where $n=1$, $\ere=0\farcs5$, and $\mu_{0}-\mu_{\text{e}}=4.0\,\magg$; I matched their convolved model with my one-dimensional convolved model. The results depend on if the model is convolved with their PSF$_{\text{Z15},g}$ or with, for example, {\PSFK}, in particular at the seeing-dependent centre, see Fig.~\ref{fig:Zheng}. Differences between the one and two-dimensional treatments are larger with the better-seeing PSF$_{\text{Z15},g}$ than with {\PSFK}. At larger radii, $r\ga13\arcsec$, the comparison with the profiles of the lower-limit {\PSFK} suggests that their $g$-band PSF underestimates the scattered light. Notably, the difference between a two-dimensional and one-dimensional treatment increases at larger radii for larger galaxies (compare the lines of the one-dimensional example models in Fig.~\ref{fig:toy} with the other models); the reason is that more points contribute to the intensity at any one point in a two-dimensional approach compared with a one-dimensional line. The amount of scattered light is in general underestimated at larger radii when the analysis is made in one dimension, and the analysis therefore needs to be made in two dimensions.

\subsection{Suggestions for future observations of surface-brightness profiles that, in particular, include haloes and thick discs}\label{sec:discsug}
My analysis has shown that we know little if anything about haloes and thick discs observed in diffuse emission. Time- and instrument-specific scattered-light effects need to be removed at first, before the existence of any real (red) halo is studied, measurements of host-galaxies are reassessed, or properties of stellar populations and amounts of non-baryonic dark matter and other phenomena are scrutinized \citep[for example, \BO;][\Michevaa; \Michevab; \citealt{MaTrKn.:14}]{ZaBeOs.:06,ZaFl:08,ZaJoMi:12,HeJiAl:13}. It also appears premature to discuss relations between thin-disc and thick-disc properties in edge-on disc galaxies \citep[for example,][]{DaBe:02,YoDa:05} before the observations are corrected for scattered light; the same conclusion applies to kinematic studies of thick discs \citep{YoDa:05,YoDa:08a}.

The radius and intensity where haloes and thick discs appear depend on the object properties, the brightness of the sky, and the instrument and the telescope (as well as the atmosphere) that give rise to the PSF. The following points should be taken into account in future observations of haloes and thick discs to correctly address effects of scattered light:

\begin{itemize}
\item The level of accuracy of the measurements must be determined for each image or data set individually. Images that need to be stacked could be sorted in groups that span a time interval that is short enough to ensure that the PSF is invariant -- say, two weeks.\footnote{I suggest a value this small to be on the safe side since the period and amplitude of actual temporal variations are unknown.} It is particularly important to determine the accuracy of the sky background if parts of the object structure are fainter than the sky. The accuracy can be tested by adding an exponential profile in different parts of the raw image, which is then reduced and analysed to find the level where the profile deviates from the exponential profile. My example models indicate that the accuracy of the background level of the sky is about $2.5\,\magg$ higher than the surface brightness at the radius where the reduced profile deviates by more than $\delta=0.10\,\magg$ from an exponential profile.
\item In case of {\sersic}-type and face-on disc (edge-on) galaxy profiles, it is necessary to use PSFs that are accurately determined out to radii that are 1.1 (1.5) times the maximum considered radius (vertical distance). It must be remembered that the PSF varies with both the wavelength and the time. Guidelines on how the PSF can be measured accurately out to large radii are provided in {\rS}.
\item A model of the galaxy surface-brightness structure must account for both the accuracy of the sky background and the contribution of scattered light, which define a maximum useful radius. Measurements must be brighter than the level where a convolved profile deviates from the modelled exponential profile by less than $\delta=0.05$--$0.15\,\magg$, and they must be inside of the maximum useful radius. Nothing can be said about measurements at larger radii. The analysis needs to be done more accurately if the measurements and the scattered-light model are similar and the intention is to use the data for scientific conclusions.
\item Points for further improvement of the analysis include accounting for possible flaws of the model, such as asymmetries and other data-reduction issues. It may also be necessary to use a two-dimensional PSF that accounts for spatial variations across the observed field.
\end{itemize}

The ultimate solution is to deconvolve the data, but this approach requires that both the measurements and the PSF are sufficiently accurate; it still remains to be determined exactly how accurate the data must be to do this. The required accuracy of the measurements, the sky level, and the PSF becomes prohibitively high already at intermediate radii for many galaxies. It is then difficult and even impossible to show that there is anything but scattered light at those radii. Instead, it appears more fruitful to analyse the data at smaller radii and then progress outwards as far as the errors allow. Except features such as much fainter tidal streams and tails, there should be little excess at larger radii.

\section{Conclusions}\label{sec:conclusions}
I have presented multiple examples of models and measurements of surface-brightness structures of different types of galaxies. The results show that scattered light produces structures that are so similar to the faint measurements at large radii that they can explain their origin. Specifically, the affected parts of the structures include haloes and thick discs around edge-on galaxies, haloes and type III profiles around face-on disc galaxies, halo-like hosts around BCGs, haloes around elliptical galaxies, and increased scale lengths in the centre regions for smaller appearing galaxies of all kinds.

The influence of scattered light has been modelled in terms of the minimum and larger amounts. As in {\rS}, {\PSFK} \citep{Ki:71} has been used as a lower limit and the four brighter {\PSFVa}, {\PSFVb}, {\PSFIa}, and {\PSFIb} of {\Michard} as intermediate-to-higher amount PSFs -- in particular, {\rS} shows that {\PSFK} can be used as a generally applicable lower limit with all instruments and (reflective) telescopes. With these models, it has been possible to study effects of temporally variable scattered light using radially extended PSFs in two bandpasses and one colour. All differences have been neglected between PSFs in the visual wavelength range, except in comparison with the $I$ band. The surface-brightness structures have been modelled using standard galaxy-specific profiles. Each of the models has been convolved individually with all five PSFs, and the resulting surface-brightness profiles have been compared with measurements of different studies in different bandpasses. The combined influence of scattered light and the accuracy of the sky subtraction has also been analysed. Together, these two components provide strict limits on how far out it is possible to measure surface-brightness structures that can be accurately corrected.

The colour profiles of all modelled galaxies show significant to drastic amounts of red excess across large radial intervals regardless of the galaxy type, and in particular in the faint halo regions. I have only studied one colour, such as $R-i$, but the results are indicative of PSF-dependent colours in the outer regions. Any radial differences between PSFs of different wavelengths or bandpasses will induce artificial colour profiles that could be either red or blue. Notably, artificial colours are also seen at smaller radii than where scattered-light haloes appear.

The selected sample of galaxy observations demonstrates that scattered light plays an integral part to explain faint surface-brightness and colour profiles around various types of galaxies. Currently, diffuse data only show ubiquitous scattered-light haloes. To measure physical haloes it is necessary to first remove the scattered-light component, which requires accurate measurements of extended PSFs (cf.\ \rS) and accurate analysis as shown here. For as long as scattered light is not excluded as the cause of the excess light, it cannot be expected that the faintest parts of the modelled profiles are anything but scattered light, as it is a more likely solution than other more exotic explanations. Notably, the increasing number of deep surveys have so far rarely measured extended PSFs. There are hints that PSFs of these surveys may be brighter than the ones used in the analysis here \citep{DuCuKa.:15}. Such bright PSFs would further amplify the effects and make the results even more alarming. Meanwhile, it seems that the only way to measure haloes are through observations of asymmetric tidal streams and tails as well as of resolved stars of nearby galaxies. I have limited this study to galaxies in the visual wavelength range, but the result applies to other wavelength ranges, haloes, excess light, and colour profiles around all kinds of objects on the sky.

\begin{acknowledgements}
I thank A.\ Partl, D.\ Streich, C.\ Vocks, H.\ Enke, and L.\ Strandberg for comments on the manuscript. C.S.\ was supported by funds of PTDESY-05A12BA1 and 05A14BA1, and the BMBF VIP program 03V0843. This research has made use of the NASA/IPAC Extragalactic Database (NED) which is operated by the Jet Propulsion Laboratory, California Institute of Technology, under contract with the National Aeronautics and Space Administration.
\end{acknowledgements}

\bibliographystyle{aa}

\appendix

\section{Complementary measurements of the PSF and details of the observations}\label{sec:apsf}
\begin{figure*}
\sidecaption
\includegraphics{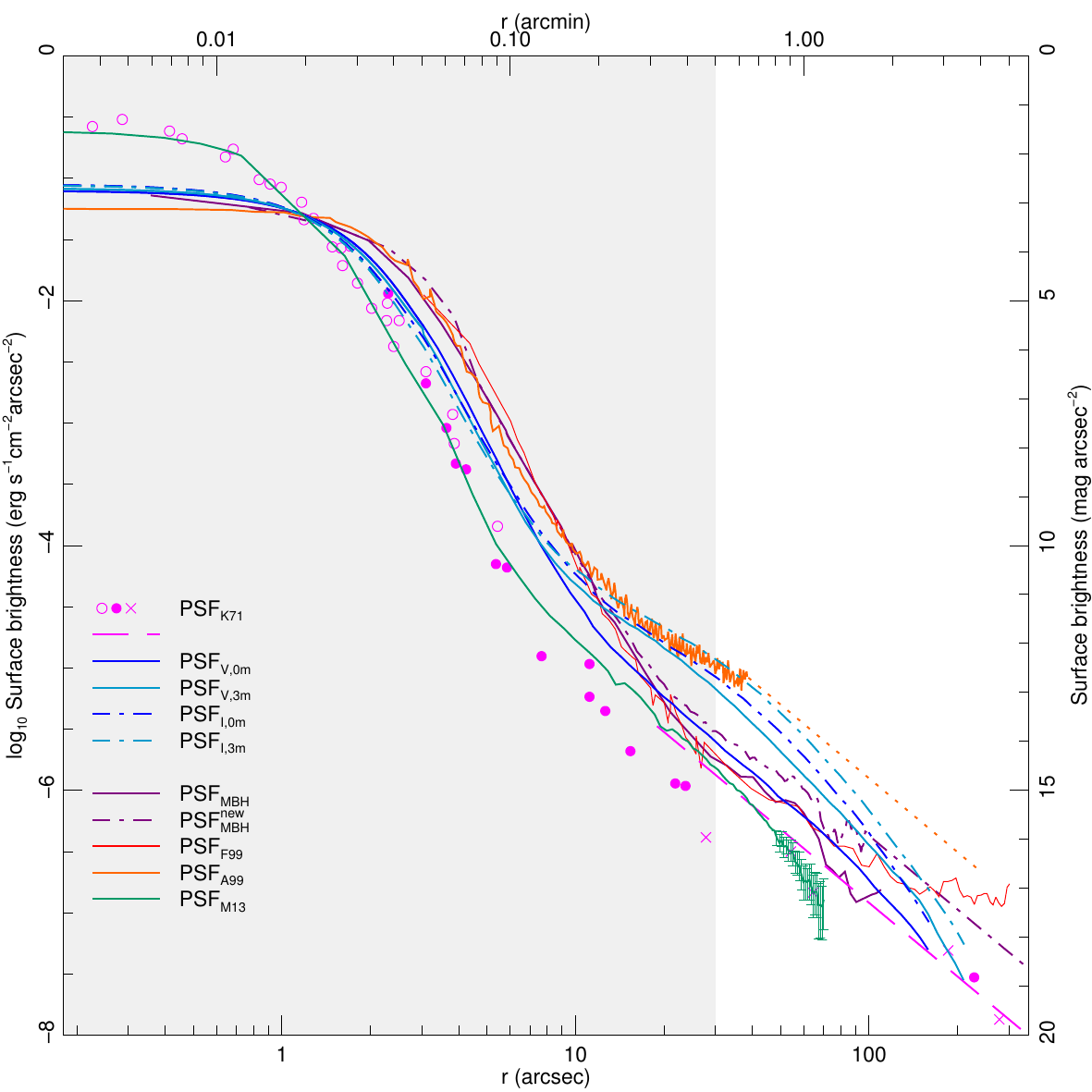}
\caption{PSF surface-brightness profiles versus the radius $r$ for a 0\,mag point source. Individual PSFs are drawn with coloured lines and symbols as indicated in the figure. {\PSFK}, {\PSFVa}, {\PSFIa}, {\PSFIb}, {\PSFMBH}, and {\PSFMBHn} are shown in this figure as references. The light grey region indicates the PSF core, and the white region the aureole, as defined in \rS. The extrapolated part of {\PSFA} is shown with a dotted line.}
\label{fig:p2psf}
\end{figure*}

The PSFs discussed in this paper are shown in Fig.~\ref{fig:p2psf} that is complementary to figs.~1 and 2 in {\rS}. The plot includes {\PSFK}, {\PSFVa}, {\PSFVb}, {\PSFIa}, {\PSFIb}, {\PSFMBH} \citep{MoBoHa:94}, and {\PSFMBHn} (\rS) that are all introduced in {\rS}. The following three PSFs are studied here: {\PSFF} (\Fry), {\PSFA} \citep[hereafter {\Abe}]{AbBoCa.:99}, and {\PSFMich} (\Michevaa).

Details of all observations related to the measurements scrutinized in this paper are collected in Table~\ref{tab:obs}; the information allows a straightforward comparison of the measurements in the context of scattered light. The table also contains the details of the observations of NGC 5907 discussed in {\rS}.

\begin{table*}
\caption{Observations.}
\centering
\label{tab:obs}
\tabcolsep=3.1pt
\begin{tabular}{llllllll}\hline\hline\\[-1.8ex]
Galaxy & Sect. & Ref. & Band & Telescope / Observatory & Date & Exp.time & Comments \\\hline\\[-1.8ex]
NGC 5907      & \rS                    & (1)  & $R$ Harris           & Kitt Peak Nat.\ Obs.\ No.\ 1 0.9m & 1990-04-29, 30 & 30min$\times6$ & $3\farcs5$, $0\farcs77$/px\\
              &                        & (2)  & $V$                  & Canada-France-Hawaii Tel. & 1995-06-03-- & 30min$\times5$ & $<1\arcsec$, $0\farcs206$/px\\
              &                        &      &                      & \quad (CFHT)              & \quad 1995-06-07\\
              &                        & (3)  & $6660$ {\AA}\tablefootmark{a} & Beijing Astr.\ Obs.\ 0.6/0.9m Tel.\ & 1995-01-31-- & 10, 20min & $4\arcsec$, $1\farcs71$/px\\
              &                        &      &                      &                           & \quad 1995-06-27      & \quad $\times137$  & \quad 23 nights\\
IC 5249       & \ref{sec:ic5249}       & (4)  & $R_{\text{C}}$\tablefootmark{b} & 0.6m \textit{Boller Chivens} Tel.\ & 1997-07-04, & 20min$\times7$ & $4\arcsec$ , $0\farcs645$/px\\
              &                        &      &                      & \quad Mount John Univ.\ Obs. & \quad 1997-07-06\\
NGC 4565      & \ref{sec:ngc4565}      & (5)  & $J$\tablefootmark{c} & 48" Schmidt, Palomar Obs.\ & 1976-03-30-- & 55min$\times2$ & $1\arcsec$, $2.013\arcsec$/px\\
              &                        &      &                      &                            & \quad 1976-04-02\\
              & \ref{sec:ngc4565}      & (6)  & $B$, $r$             & 48" Schmidt, Palomar Obs.\ & 1975-03-31-- & 190min &\\
              &                        &      &                      &                            & \quad 1977-03-13\\
              &                        & (7)  & $V$                  & 2.52m Nordic Optical Tel.\ & 1994-02-10, 11 & 25min & $1\arcsec$, $0\farcs462$/px\\
              &                        & (8)  & $6660$ {\AA}\tablefootmark{a} & 60/90cm Schmidt Tel., Xinlong & 1995-01-28-- & 42.8h & $2\farcs9$--$4\farcs4$, $1\farcs7$/px\\
              &                        &      &                      & \quad Stat., Nat.\ Astr.\ Obs.\ China & \quad 1997-05-28\\
NGC 4244      & \ref{sec:ngc4244}      & (5)  & $J$\tablefootmark{c} & 48" Schmidt, Palomar Obs.\            & 1976-03-30-- & 55min & $1\arcsec$, $2.013\arcsec$/px\\
              &                        &      &                      &                                       & \quad 1976-04-02\\
              &                        & (9)  & $R$ Harris           & 24/36" \textit{Burrell Schmidt} Tel.\ & 1997-03 (5n), & 20min & $2\farcs03$/px\\
              &                        &      &                      &                                       & \quad 1997-04 (6n) & \quad $\times66$\\
FGC 310,      & \ref{sec:fgc}          & (10) & $R$                  & \textit{du Pont} 2.5m, & 1997-09-21 & 360s & $0\farcs9$, $0\farcs259$/px\\
FGC 1285      & \ref{sec:fgc}          & (10) & $R$                  & \quad Las Campanas Obs.\ & 1998-04-15 & 900s & $1\farcs1$, $0\farcs259$/px\\
LSBGs         & \ref{sec:stack}        & (11) & $g$, $r$, $i$        & SDSS, Apache Point 2.5m & & &\\
NGC 7280      & \ref{sec:faceonsl}     & (12) & $R$                  & 2.5m \textit{Isaac Newton} Tel.\ & 2003-09-20 & 10,200s & $1\farcs0$--$1\farcs5$\\
NGC 4102      & \ref{sec:faceonsl}     & (12) & $R$                  & \quad Wide-Field Camera & 2004-03-17 & 1200s & $0\farcs7$--$3\farcs4$\\
NGC 4477      & \ref{sec:faceonnsl}    & (12) & $r$                  & SDSS, Apache Point 2.5m & & &\\
NGC 3507      & \ref{sec:faceonnsl}    & (12) & $r$                  & SDSS, Apache Point 2.5m & & &\\
NGC 2880      & \ref{sec:faceonnsl}    & (12) & $R$                  & Wisconsin Indiana Yale & 1997-03-02 & 300s &\\
              &                        &      &                      & \quad NOAO (WIYN) Tel.\ \\
ESO 400-G043  & \ref{sec:bcgeso}       & (13) & $B$ Cousins          & ESO 2.2m Tel.\ & 1989 & 30min & $<1\arcsec$\\
              &                        & (14) & $V$                  & ESO Multi-M.\ Ins.\ (EMMI) & 2005 & 77min & \\
              &                        &      &                      & \quad New Tech.\ Tel.\ (NTT)\\
Mrk 5         & \ref{sec:bcgMrk5}      & (15) & $B$                  & NOT & 1999-01 & 3100s &\\
ESO 350-IG038 & \ref{sec:ESO350-IG038} & (13) & $B$ Cousins          & ESO 2.2m Tel.\ & 1984 & 25min & $<1\arcsec$\\
              &                        & (16) & $V$                  & NOT, MOSaic CAmera & 2008 & 40min & $0\farcs217$/px\\
Mrk 297       & \ref{sec:bcgMrk297}    & (17) & $B$ Mould            & Kitt Peak No.1\ 0.9m Tel.\ & 1983-05-15--17 & 50min & $1\farcs1$\\
              &                        & (15) & $B$                  & Calar Alto 2.2m Tel.\ & 1998-08 & 2000s &\\
UM 465        & \ref{sec:abcgUM465}    & (18) & $B$                  & NOT, Andalucia Faint Obj.\ & 2004 & 40min & \\
              &                        &      &                      & \quad Spectrograph (ALFOSC)\\
VCC 0001      & \ref{sec:abcgUM465}    & (19) & $g$, $r$, $i$        & SDSS, Apache Point 2.5m\\
NGC 4551      & \ref{sec:intermediate} & (20) & $V$                  & Various telescopes\\
NGC 4473      & \ref{sec:small}        & (20) & $V$                  & Various telescopes\\
NGC 4374      & \ref{sec:small}        & (20) & $V$                  & Various telescopes\\
\hline
\end{tabular}
\tablefoot{Column 1, galaxy name; Col.~2, the galaxy is referred to in this section or appendix; Col.~3, reference; Col.~4, considered bandpass out of the used ones; Col.~5, the telescope and observatory of the measurements; Col.~6, dates of the measurements; Col.~7, total exposure time of the measurements; Col.~8, additional comments such as seeing and spatial resolution, where this information is available.\\
\tablefoottext{a}{The BATC filter is centred on 6660 {\AA} and has a bandwidth of 480{\AA}.}
\tablefoottext{b}{The observations were made with a `step function' filter that includes 60 per cent of the $R_{\text{C}}$ passband; the measurements were converted to the $r$ band.}
\tablefoottext{c}{The photographic exposures were on IIIa-J emulsion with a Wratten-2C filter, which gives a bandpass of about 4000--5400{\AA}; the resulting magnitude scale is denoted $J$.}
}
\tablebib{(1)~\citet{MoBoHa:94}; (2)~\citet{LeFoDa.:96}; (3)~\citet{ZhShSu.:99}; (4)~\Abe; (5)~\citet{Kr:79}; (6)~\citet{JeTh:82}; (7)~\citet{NaJo:97}; (8)~{\Wu}; (9)~{\Fry}; (10)~\DB; (11)~\BZC; (12)~\EPB; (13)~\BO; (14)~\Michevaa; (15)~\Cairos; (16)~\Michevac; (17)~\Papaderos; (18)~\Michevab; (19)~\citet{MeLiJaPa:14}; (20)~\Kormendy.}
\end{table*}

\begin{table}
\caption{Example model parameters.}
\centering
\label{tab:toy}
\begin{tabular}{lllll}\hline\hline\\[-1.8ex]
Galaxy model & \multicolumn{1}{l}{$n$, $\ere$ / $h_{\text{i}}$, $h_{\text{o}}$, $\Delta\mu_{0}$} & \multicolumn{1}{c}{$r_{\text{tr},1}$} & \multicolumn{1}{c}{$r_{\text{tr},2}$} & \multicolumn{1}{l}{Fig.}\\\hline\\[-1.8ex]
small {\sersic} & $1$, $\phantom{0}1.5$ & $4.0$ & $\phantom{0}10.0$ & \ref{fig:toy}a\\
medium {\sersic} & $1$, $\phantom{0}4.0$ & $10$ & $\phantom{0}25.0$ & \ref{fig:toy}b\\
large {\sersic} & $1$, $40$  & $10$ & $250$ & \ref{fig:toy}c\\[0.3ex]
small face-on  & $\phantom{0}13.5$, $\phantom{0}0.897$, $3.0$ & $\phantom{0}4.0$ & $\phantom{0}10.0$ & \ref{fig:toy}d\\[0.3ex]
medium face-on & $\phantom{0}35.9$, $\phantom{0}2.39$, $\phantom{0}3.0$ & $10$ & $\phantom{0}25.0$ & \ref{fig:toy}e\\[0.3ex]
large face-on  & $359$, $\phantom{.0}23.9$, \phantom{00}$3.0$ & $10$ & $250$ & \ref{fig:toy}f\\[0.3ex]
\hline
\end{tabular}
\tablefoot{Column 1, example galaxy model; Col.~2, {\sersic}-profile index and half-light radius (arcsec), or the face-on galaxy inner and outer scale lengths (arcsec), and the intensity difference $\Delta\mu_{0}=\mu_{0,\text{i}}-\mu_{0,\text{o}}$ (\magg); Col.~3, first truncation radius (arcsec); Col.~4, second truncation radius (arcsec); and Col.~5, results are shown in this figure.}
\end{table}

\begin{figure*}
\centering
\includegraphics{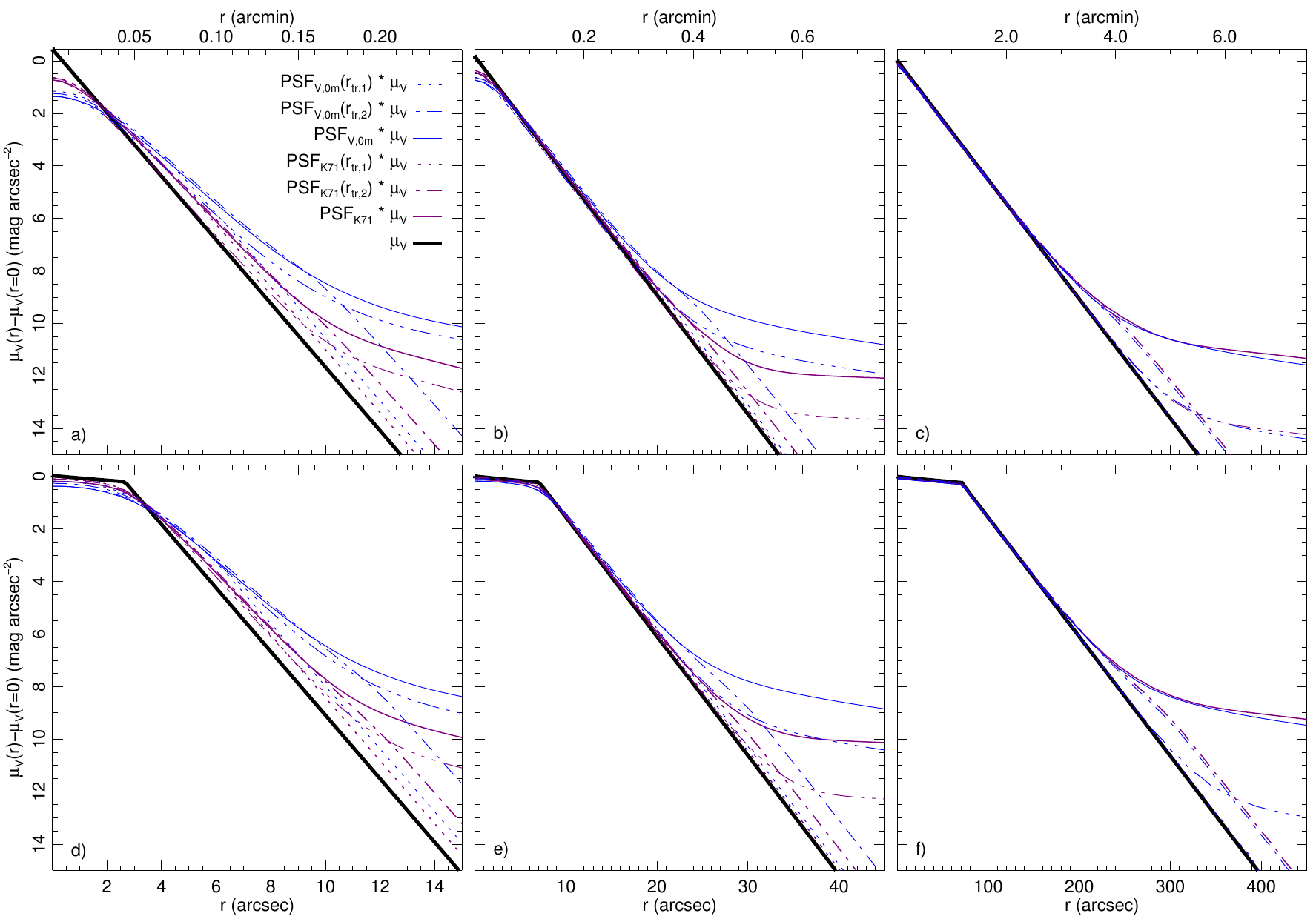}
\caption{Surface-brightness profiles of example models of three {\sersic}-type galaxies (the upper panels \textbf{a}--\textbf{c}) and three face-on disc galaxies (the lower panels \textbf{d}--\textbf{f}) that illustrate effects of a radial truncation of {\PSFK} and {\PSFVa}. The three columns show: \textbf{a}) and \textbf{d}) two small galaxies, \textbf{b}) and \textbf{e}) two intermediate-size galaxies, and \textbf{c}) and \textbf{f}) two large galaxies. In each panel, the galaxy model profile is drawn with a thick black solid line, and the model profiles that were convolved with {\PSFK} (\PSFVa) are drawn with thin purple (blue) lines. The profile of the model that was convolved with the complete PSF is drawn with a thin solid line. Profiles are also shown where each PSF was truncated at $r_{\text{tr},1}$ ($r_{\text{tr},2}$) with a dotted (dash-dotted) line in each panel; \textbf{a}) and \textbf{d}) $r_{\text{tr},1}=4\arcsec$ and $r_{\text{tr},2}=10\arcsec$; \textbf{b}) and \textbf{e}) $r_{\text{tr},1}=10\arcsec$ and $r_{\text{tr},2}=25\arcsec$; \textbf{c}) and \textbf{f}) $r_{\text{tr},1}=10\arcsec$ (lines fall atop the model lines and are not visible) and $r_{\text{tr},2}=250\arcsec$ (Table~\ref{tab:toy}). The dash-tripple-dotted lines show the result of one-dimensional models that were convolved with the respective one-dimensional PSF.}
\label{fig:toy}
\end{figure*}

\section{Application of the method on example models of {\sersic}-type and face-on disc galaxies}\label{sec:toy}
The example models of surface-brightness profiles of edge-on galaxies in {\rS} are complemented here with models of {\sersic}-type and face-on disc galaxies. I delimit the former type galaxy examples to $n=1$, which results in exponential slopes; galaxy profiles of this type are less affected when $n\gg1$ (cf.\ Sect.~\ref{sec:sersic}). Galaxy structures well represented by {\sersic}-type profiles are less affected by scattered light than galaxies viewed edge on; simply viewed, the brightest intensities are concentrated to a single point instead of a line. Surface-brightness profiles of face-on disc galaxies are also modelled with exponential slopes. However, effects of scattered light in the outer regions can be enhanced compared to my {\sersic}-type profiles when intensity declines in the central region are shallower inside a break radius; due to, for example, spiral arms or a bar.

I calculated three example models of a small, an intermediate, and a large object for both the {\sersic}-type galaxy and the face-on disc galaxy. I varied the half-light radius $\ere$ and set the {\sersic} index $n=1$ in the {\sersic}-type galaxy models. In the face-on disc galaxy models, I varied the outer scale length $h_{\text{o}}$, set the inner scale length $h_{\text{i}}=15h_{\text{o}}$, $\mu_{0,\text{i}}=\mu_{0,\text{o}}+3\,\magg$, and the break radius $r_{\text{br}}=3h_{\text{o}}$; I did not model an additional bulge. Each model was convolved with {\PSFK} and a radially extrapolated {\PSFVa} -- where both PSFs are more than two times as extended as the respective galaxy model. Additionally, each model was convolved with either PSF, after the PSF was truncated at the smaller radius $r_{\text{tr},1}$ or the larger radius $r_{\text{tr},2}$. Model parameters and truncation radii $r_{\text{tr}}$ are given for each of the six models in Table~\ref{tab:toy}. The resulting surface-brightness profiles are shown in Fig.~\ref{fig:toy}. (Models were also convolved in one dimension with the respective PSF, the corresponding profiles are discussed in Sect.~\ref{sec:discsug}.) Values quoted in parentheses below were modeled with {\PSFK}, and the other values were modeled with {\PSFVa}. I describe each of the three {\sersic}-type galaxy models separately, where I compare the resulting properties with the results of the respective face-on and edge-on disc galaxy model.

\subsection{The two small galaxies}
The convolved profiles of the small galaxy show excess light at all distances $r\ga2\farcs0$, Fig.~\ref{fig:toy}a. Less light is seen within this radius; this decrease is largely caused by the centre part of the poor-seeing PSFs I used (FWHM $\simeq2$--$4\arcsec$). The scattered-light halo of the convolved model is too faint for $r\ga4\farcs7$ ($r\ga5\farcs3$) when the PSF is truncated at $r_{\text{tr}}=4\arcsec$. The corresponding value for $r_{\text{tr}}=10\arcsec$ is $r\ga9\farcs3$ ($r\ga9\farcs3$). The surface-brightnesses that were calculated using the three versions of the truncated {\PSFVa} differ by up to about $0.7\,\magg$ at the centre, due to the variations of the individual normalization of the PSFs. The correct amount of scattered light at $r=15\arcsec$ is only achieved when the PSF is not truncated within, say, $r\approx16\arcsec$ ($15\arcsec\times10\arcsec/9\farcs3$). Compared to the centre, both {\PSFK} and {\PSFVa} are about $10\,\magg$ fainter at $r=15\arcsec$ (see Fig.~\ref{fig:p2psf}). The model surface brightness is about $4.7$ ($3.3$) {\magg} fainter than the convolved structure at $r=12\arcsec$, where the surface brightness is $\Delta\mu=14\,\magg$ fainter than at the centre; the signal-to-noise (S/N) value that would be required to extract the model structure from the observed (convolved) structure is about 76 (21). At $r=7\farcs5$ ($\Delta\mu\simeq8.7\,\magg$), the corresponding values are $2.0\,\magg$ and S/N$\approx$6.2 ($0.95\,\magg$ and S/N$\approx$2.4).

The face-on galaxy shows excess light for $r\ga3\farcs7$, which is just outside of the break radius $r_{\text{br}}=2\farcs7$, Fig.~\ref{fig:toy}d. The centre plateau results in brighter profiles at larger radii. The model surface brightness is about $5.8$ ($4.3$) {\magg} fainter than the convolved structure at $r=14\arcsec$ ($\Delta\mu\simeq14\,\magg$); the S/N value that would be required to extract the model structure from the observed structure is about 200 (53). At $r=9\farcs7$ ($\Delta\mu\simeq8.7\,\magg$), the corresponding values are $2.4\,\magg$ and S/N$\approx$9.2 ($1.2\,\magg$ and S/N$\approx$3.0). These values are $0.25$--$1.1\,\magg$ brighter than for the {\sersic}-type galaxy.

For the small edge-on galaxy model in {\rS}, the model surface brightness is about $6.7$ ($5.1$) {\magg} fainter than the convolved structure at $z=10\farcs7$ ($\Delta\mu\simeq14\,\magg$). At $z=7\farcs1$ ($\Delta\mu\simeq8.7\,\magg$), the corresponding value is $3.3$ ($1.9$) {\magg}. Compared to the {\sersic}-type galaxy, these values are $0.95$--$2.0\,\magg$ brighter.

\subsection{The two intermediate-size galaxies}
There appears to be less excess light for the intermediate-size galaxy, at least near the centre, Fig.~\ref{fig:toy}b. The convolved profiles show excess light at all radii $r\ga5\arcsec$. When the PSF is truncated at $r_{\text{tr}}=10\arcsec$, there is an excess due to scattered light of up to about $0.35$ ($0.17$) $\magg$ for $r\ga10\arcsec$. When the PSF is instead truncated at $r_{\text{tr}}=25\arcsec$, the corresponding convolved model becomes significantly fainter than the model convolved using the full PSF for $r\ga23\arcsec$ ($r\ga24\arcsec$). The correct amount of scattered light at, say, $r=40\arcsec$ is only achieved when the PSF is not truncated within, say, $r=44\arcsec$ ($40\arcsec\times25\arcsec/23\arcsec$). Compared to the centre, the PSF is about $12$ ($14$) $\magg$ fainter at $r=40\arcsec$. The model surface brightness is about $4.1$ ($2.4$) $\magg$ fainter than the convolved structure at $r=31\farcs2$ ($\Delta\mu\simeq14\,\magg$), and the required S/N$\approx$44 (S/N$\approx$9.1). At $r=19\farcs5$ ($\Delta\mu\simeq8.7\,\magg$), the corresponding values are about $0.78\,\magg$ and S/N$\approx$2.1 ($0.26\,\magg$ and S/N$\approx$1.3).

The face-on galaxy shows excess light for $r\ga12\arcsec$, where the break radius is at $r_{\text{br}}=7\farcs2$, Fig.~\ref{fig:toy}e. The model surface brightness is about $5.6$ ($4.1$) $\magg$ fainter than the convolved structure at $r=37\farcs5$ ($\Delta\mu\simeq14\,\magg$); the S/N value that would be required to extract the intensity structure of the model structure is about 170 (44). At $r=26\arcsec$ ($\Delta\mu\simeq8.7\,\magg$), the corresponding values are $1.6\,\magg$ and S/N$\approx$4.4 ($0.6\,\magg$ and S/N$\approx$1.7). These values are $0.34$--$1.7\,\magg$ brighter than in the case of the {\sersic}-type galaxy.

For the intermediate-size edge-on galaxy model in {\rS}, the model surface brightness is about $6.4$ ($4.7$) {\magg} fainter than the convolved structure at $z=28\farcs5$ ($\Delta\mu\simeq14\,\magg$). At $z=19\arcsec$ ($\Delta\mu\simeq8.7\,\magg$), the corresponding value is $2.1$ ($0.94$) $\magg$. Compared to the {\sersic}-type galaxy model, these values are $0.68$--$2.3\,\magg$ brighter.

\subsection{The two large galaxies}
Using either one of the two  PSFs, the convolved model shows significant excess due to scattered light, beginning at $r\simeq160\arcsec$, Fig.~\ref{fig:toy}c. The halo is not reproduced at all when the PSF is truncated at $r_{\text{tr}}=10\arcsec$. The scattered-light halo of the convolved model is too faint for $r\ga230\arcsec$ when the PSF is truncated at $r=250\arcsec$. The PSFs are about $20\,\magg$ fainter at $r=330\arcsec$ than at the centre. To measure the modelled intensity at $r\simeq250\arcsec$, which is about $1.4\,\magg$ fainter than the convolved model, the required S/N $\approx3.6$, the PSF likely needs to be more accurate. The convolved profiles of the two PSFs nearly overlap since they are the same for $r>200\arcsec$. As in the case of the edge-on galaxy in {\rS}, the surface brightness of the model convolved with {\PSFK} in Fig.~\ref{fig:toy}c is brighter at large radii than it is in Figs.~\ref{fig:toy}a and \ref{fig:toy}b; this is expected, because, compared to where $r\la30\arcsec$, the PSF slope is shallower at larger radii.

The face-on galaxy shows excess light for $r\ga180\arcsec$, where the break radius is at $r_{\text{br}}=72\arcsec$, Fig.~\ref{fig:toy}f. The model surface brightness is about $5.0\,\magg$ fainter than the convolved structure at $r=375\arcsec$ ($\Delta\mu\simeq14\,\magg$), the required S/N$\approx100$. At $r=255\arcsec$ ($\Delta\mu\simeq8.7\,\magg$), the corresponding values are $1.1\,\magg$ and S/N$\approx$2.8. These values are $0.58$--$1.7\,\magg$ brighter than in the case of the {\sersic}-type galaxy.

For the large edge-on galaxy model in {\rS}, the model surface brightness is about $5.8\,\magg$ fainter than the convolved structure at $z=285\arcsec$ ($\Delta\mu\simeq14\,\magg$). At $z=190\arcsec$ ($\Delta\mu\simeq8.7\,\magg$), the corresponding value is $1.7\,\magg$. Compared to the {\sersic}-type galaxy model, these values are $1.2$--$2.5\,\magg$ brighter.

\section{More on edge-on disc galaxies}\label{sec:aedgeon}
The three galaxies discussed here are: IC 5249, Appendix~\ref{sec:ic5249}; NGC 4565, Appendix~\ref{sec:ngc4565}; and NGC 4244, Appendix~\ref{sec:ngc4244}. All considered existing and derived model parameters of these galaxies as well as the two thick-disc galaxies FGC 310 and FGC 1285 (Sect.~\ref{sec:fgc}) are given in Table~\ref{tab:edgeon}.

\begin{table*}
\caption{Model parameters of the 7 edge-on disc galaxies considered and referred to here. The distances are in this work only used to scale the upper x-axis in the surface-brightness profile plots. Details of the original observations are given in Table~\ref{tab:obs}.}
\centering
\label{tab:edgeon}
\begin{tabular}{llcclll@{\,}llll@{\,}llllll}\hline\hline\\[-1.8ex]
       &                       & & & \multicolumn{3}{l}{single/thin disc} && \multicolumn{3}{l}{thick disc} && \multicolumn{3}{l}{bulge}\\\cline{5-7}\cline{9-11}\cline{13-15}\\[-1.8ex]
Galaxy & \multicolumn{1}{c}{D} & \multicolumn{1}{c}{Band} & \multicolumn{1}{c}{$n_{\text{s}}$} & \multicolumn{1}{c}{$h_{\text{r}}$} & \multicolumn{1}{c}{$z_{0}$} & \multicolumn{1}{c}{$\mu_{0,0}$} && \multicolumn{1}{c}{$h_{\text{r}}$} & \multicolumn{1}{c}{$z_{0}$} & \multicolumn{1}{c}{$\mu_{0,0}$} && \multicolumn{1}{c}{$n$} & \multicolumn{1}{c}{$r_{\text{e}}$} & \multicolumn{1}{c}{$\Delta\mu$} & Ref. & Fig.\\\hline\\[-1.8ex]
NGC 5907 & 11     & $R$ Harris       & 2    & \pz90 & 16 & \sd && \ssd & - & - && - & - & \sdash-& 1\\[0.3ex]
         & 11.7   &                  & 2    & \pz90 & 15 & \sd && \ssd & - & - && - & - & \sdash-& 2\\[0.3ex]
IC 5249  & 36     & $R_{\text{C}}$ & $\infty$ & \pz63 & \pz3.3 & \sd && \pz34 & \pz9.7 & - && -&- & \sdash- & 3 & \ref{fig:ic5249}\\[0.3ex]
         & 36     &                  & 1    & \pz63 & \pz4.7 & \sd && \ssd & - & - && - & - & \sdash- & & \ref{fig:ic5249}\\[0.3ex]
NGC 4565 & 10     & $J$              & 1    &   110 & 16 & 20.1 && \ssd & -& -&& -& -& \sdash-& 4, 5 & \ref{fig:ngc4565}\\[0.3ex]
         & 10     & $B$              & 1    &   124 & 17 & \sd && 120 & 24 & -&& -& -& \sdash-& 6, 7\tablefootmark{a} & \ref{fig:ngc4565}\\[0.3ex]
         &        & $r$              & 1    &   100 & 17 & \sd && 110 & 26 & -&& -& -& \sdash-& 6, 7\tablefootmark{a} & \ref{fig:ngc4565}\\[0.3ex]
         & 10     & $V$              & 1    &   120 & 12 & 20.2 && 120 & 27 & 21.8 && -& -& \sdash-& 8\tablefootmark{b} & \ref{fig:ngc4565}\\[0.3ex]
         & 14.5   & 6660{\AA}        & 1    &   115 & 17 & 22.32 && 160 & 36 & 25.52 && -& -& \sdash-& 9 & -\\[0.3ex]
         & 10     &                  & 1    &   113 & 20.6 & \sd && \ssd & - & - && 1 & 3.0 & $-0.7$ & & \ref{fig:ngc4565}\\[0.3ex]
NGC 4244 & \pz5   & $J$              & 1    &   110 & 24   & 20.6 && \ssd & -& -&& -& -& \sdash-& 4, 5 & \ref{fig:ngc4244}\\[0.3ex]
         & \pz3.6 & $R$        & $\infty$ & 105.4 & 14.1 & 20.47 && \ssd & -& -&& -& -& \sdash-& 10 & \ref{fig:ngc4244}\\[0.3ex]
         & \pz3.6 &                  & 2    &   105 & 13 & \sd && \ssd & - & - && - & - & \sdash- & & \ref{fig:ngc4244}\\[0.3ex]
\object{FGC 310}\tablefootmark{c} & 80.6 & $R$ & 1 & \pzz8.7 & \pz1.95 & 21.19 && \ssd &-&-&&-&-&\sdash-& 11 & \ref{fig:thickdisc}a \& \ref{fig:thickdisc}b\\[0.3ex]
         & 80.6   & $R$              & 1    & \pzz8.4 & \pz1.6 & 21.3 && \pzz9.6 & \pz2.9 & 22.5 && - &-&\sdash-& 11 & \ref{fig:thickdisc}a \& \ref{fig:thickdisc}b\\[0.3ex]
         & 80.6   &                  & 1    & \pzz8.4 & \pz1.6 & 21.3 && \ssd & - & -&& - & - & \sdash- && \ref{fig:thickdisc}a \& \ref{fig:thickdisc}b\\[0.3ex]
\object{FGC 1285}\tablefootmark{c} & 18.8 & $R$ & 1 & \pz19.7 & \pz6.63 & 20.99 && \ssd &-&-&&-&-&\sdash-& 11 & \ref{fig:thickdisc}e \& \ref{fig:thickdisc}f\\[0.3ex]
         & 18.8   & $R$              & 1    & \pz17.3 & \pz4.4 & 21.2 && \pz23.6 & 10.1 & 22.2 &&-&-&\sdash-& 11 & \ref{fig:thickdisc}e \& \ref{fig:thickdisc}f\\[0.3ex]
         & 18.8   &                  & 1    & \pz17.3 & \pz9.6 & 19.5 && \ssd & - & - && 1 & 4.4 & $-1.6$ & & \ref{fig:thickdisc}e \& \ref{fig:thickdisc}f\\[0.3ex]
\hline
\end{tabular}
\tablefoot{Column 1, galaxy name; Col.\ 2, distance (Mpc); Col.\ 3, bandpass used in the observations; Col.\ 4, the $n_{\text{s}}$-parameter in Eq.~(\ref{eq:edgeon}); Cols.\ 5--7, single-disc or thin-disc scale length (arcsec), scale height (arcsec), and intensity at the object centre (\magg); Cols.\ 8--10, thick-disc scale length (arcsec), scale height (arcsec), and intensity at the object centre (\magg); Cols.\ 11--13, bulge {\sersic} index, half-light radius (arcsec), and magnitude difference $\Delta\mu=\mu_{0,0}-\mu_{\text{e}}$; Col.\ 14, references, rows without a number refer to this study; and Col.\ 15, results are shown in this figure.\\
\tablefoottext{a}{The values are taken from the $\text{sech}^2$-$\text{sech}^2$-$\exp$-model in table~6 of (9).}
\tablefoottext{b}{The authors fit the galaxy profiles with three discs where the parameters of the third disc are $h_{\text{r}}=200\arcsec$ and $z_{0}=7\arcsec$.}
\tablefoottext{c}{Parameter errors are omitted, considering the uncertainty due to scattered light that appears to affect the models of (13).}}
\tablebib{(1) \citet{MoBoHa:94}; (2) {\rS}; (3) \Abe; (4) \citet{JeTh:82}; (5) \citet{ShGi:89}; (6) \citet{NaJo:97}; (7) {\Wu}; (8) \Fry; (9) {\YD}.}
\end{table*}

\subsection{The halo of IC 5249}\label{sec:ic5249}
The outcome of $R$-band observations of the edge-on Sc/Sd-type galaxy IC 5249 are presented by {\Abe}, who measure {\PSFA} that extends to $r=39\arcsec$. They derive the PSF following the approach of \citet{MoBoHa:94}; this $R$-band PSF is brighter than other typical PSFs, see Fig.~\ref{fig:p2psf}. In the outer region, the PSF agrees reasonably well with both {\PSFIb} and {\PSFIa}. {\Abe} study the contribution of the PSF to the halo by convolving it with an image of the galaxy truncated at $z=5\farcs7$. The resulting surface-brightness profiles are significantly fainter than the measurements (see their fig.~6), and they conclude that the halo is real.

\begin{figure*}
\sidecaption
\includegraphics{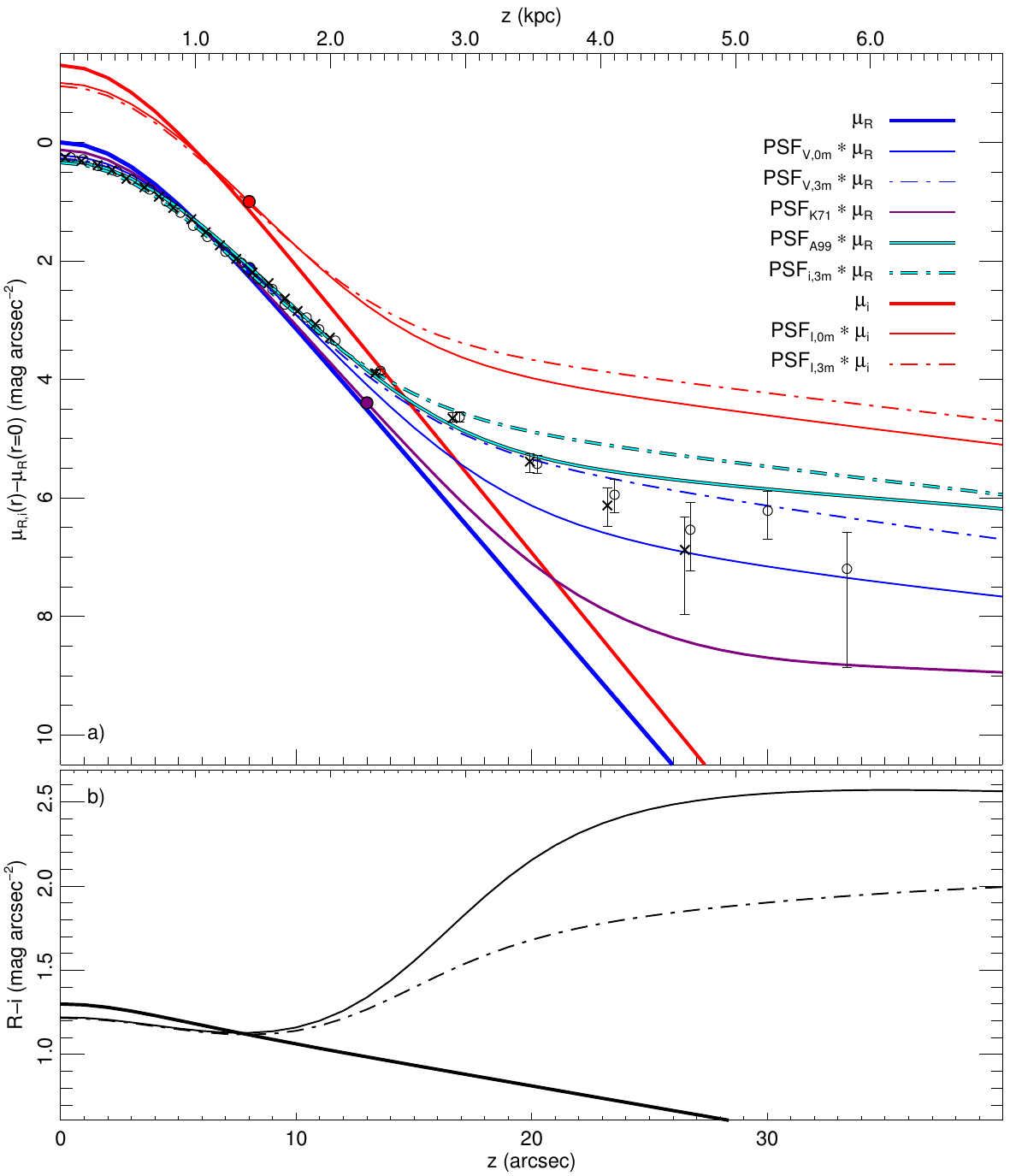}
\caption{Minor-axis $R$-band and $i$-band surface-brightness profiles versus the vertical distance $z$ of models and measurements of the edge-on galaxy IC 5249. Panels, lines, and coloured bullets are as described in Fig.~\ref{fig:thickdisc}. The $R$-band model was additionally convolved with {\PSFA} (\PSFIb) to produce the cyan-coloured solid (dash-dotted) line with a black border in panel \textbf{a}). Crosses, circles, and error bars in the same panel show region-B measurements for cuts on either side of the disc of \Abe.}
\label{fig:ic5249}
\end{figure*}

The authors model the surface-brightness profiles of IC 5249 by superposing a thick disc on a thin disc. I modelled IC 5249 with one single thin disc with similar, but not identical parameters as {\Abe}, where $z_{0}=4\farcs7$, $h_{\text{r}}=63\arcsec$, and $n_{\text{s}}=1$. I convolved the $R$-band model with the usual three PSFs that I use in this paper, but also with {\PSFA} and {\PSFIb}. Before the convolution, I extrapolated {\PSFA} with an $r^{-2}$ power-law profile where $r>40\arcsec$, cf.\ Fig.~\ref{fig:p2psf}. Surface-brightness profiles of both models and measurements are shown in Fig.~\ref{fig:ic5249}a, along vertical-axis profiles of the part of the galaxy that {\Abe} refer to as region B, which is the region most closely centred on the minor axis.

Because of the larger distance to IC 5249, the surface-brightness profiles are steeper than in NGC 5907 (where $z_{0}=15\arcsec$). Intensity and colour profiles are thereby also more sensitive to varying PSFs in IC 5249. With \PSFVa, the scattered-light halo begins already at $z\simeq8\arcsec$. The $R$-band model convolved with {\PSFA} agrees slightly better with the measurements than the model convolved with {\PSFIb}; compare the two measurements at $r\simeq17\arcsec$ and $r\simeq20\arcsec$. {\PSFA} extends to $38\farcs6$, which is enough to analyse measurements out to about $r=25\arcsec$ ($38\farcs6/1.5$). The six measurements for $r\ga22\arcsec$ are fainter than either model, they are too low for either PSF and have too small upper error limits. The halo is about $0.5\,\magg$ too bright with {\PSFIb}. Notably, the required accuracy at the location of the three outermost measurements, which would allow the data to be deconvolved to an exponential profile (ignoring the achievable accuracy of the sky subtraction), are about S/N=$40$, $300$, and $500$, respectively.

The colour profiles show red excess for $z>r_{110}=8\arcsec$, Fig.~\ref{fig:ic5249}b. The excess at $z=20\arcsec$ is $\Delta\left(R-i\right)=0.85$--$1.35\,\magg$, compared to the modelled profile, which grows to $\Delta\left(R-i\right)=1.35$--$2.00\,{\magg}$ at $z=30\arcsec$. The smaller (larger) values used {\PSFVb} and {\PSFIb} ({\PSFVa} and {\PSFIa}). As is the case with the separate surface-brightness profiles, the large range of magnitudes between the two colour profiles indicates a strong sensitivity to the PSF.

IC 5249 is a good example where it is clear that the observed halo is strongly affected by scattered light. It is possible to draw this conclusion because of the availability of a PSF with unusually bright wings observed together with the object. NGC 3957 is another galaxy with surface-brightness structures that are fitted with a somewhat similar scale height as IC 5249 \citep{JaTaCoFe:10}. Whilst similar models also appear to explain its halo as scattered light (not shown here), a bar-like region at intermediate radii \citep{PoBaLuDe:04} complicate the analysis.

\begin{figure*}
\sidecaption
\includegraphics{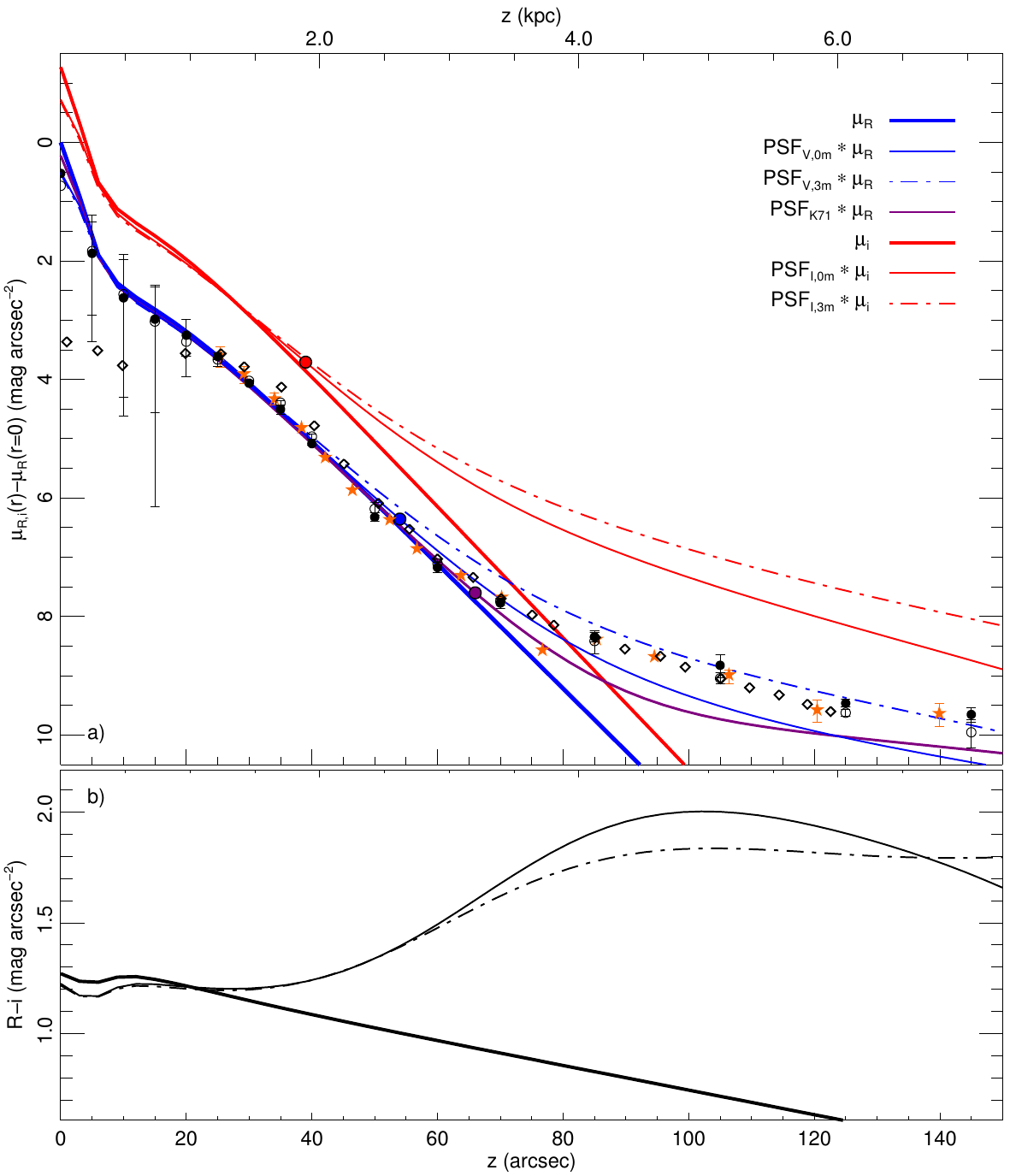}
\caption{Minor-axis surface-brightness profiles versus the vertical distance $z$ for the edge-on disc galaxy NGC 4565. Panels, lines, and coloured bullets are as described in Fig.~\ref{fig:thickdisc}. The bullets and circles (diamonds; orange stars) in panel \textbf{a} show measured $r$-band and $B$-band ($J$-band; $V$-band) values of \citet[\KrSe; \citealt{NaJo:97}]{JeTh:82}.}
\label{fig:ngc4565}
\end{figure*}

\subsection{The thick disc and the halo of NGC 4565}\label{sec:ngc4565}
The galaxy NGC 4565 is (along with NGC 891) an outstanding example of a bright edge-on galaxy similar to the Milky Way \citep{Kr:84}. \citet{Kr:79} observed NGC 4565 in the $J$ band, and the resulting plates were analysed by {\KrSe}, who smooth the data and find a thick disc. Later, \citet{JeTh:82} present profiles in the $B$ band and the $r$ band of new observations (their table~10 and fig.~18), which they compare with previous measurements of the same galaxy (their section~IIIe). The two studies present surface-brightness profiles that differ in the centre parts, this is likely a consequence of how they smooth their data. Additional profiles are presented by \citet[$V$ band]{NaJo:97} and \citep[hereafter {\Wu}]{WuBuDe.:02} (6660{\AA} band), where the latter authors find both a thick disc and a halo in their deep observations. {\Wu}, furthermore, claim they measure a PSF out to $r=1700\arcsec$, but it is not shown, and it is used to convolve a model instead of deconvolving it. The PSF is said to be derived from separate observations of bright stars, whilst the galaxy observations were made in a time span of more than two years; there is no information on when those observations were made, and the PSF data are meanwhile lost (Wu, priv.\ comm.\ 2014).

A study of near-infrared (NIR) observations of NGC 4565 (3.5--5$\mu$m) from outside of the Earth atmosphere finds no proof for a halo \citep{UeBoKa.:98}. Unlike other studies, these authors measure a PSF with a full-width-at-half-maximum of about 25{\arcsec} out to $r=10\arcmin$, where it is $12.5\,\magg$ fainter than at the centre (the PSF is not shown). They use the PSF to show that their halo is scattered light. A quick check in fig.~1 in {\rS} shows that the corresponding value of optical wavelength-range PSFs at the same radius is some $18\,\magg$ fainter than at the centre; because of the steeper decline of the PSF in the optical wavelength range, one might be led to conclude that scattered light is of little concern in this case.

{\KrSe} fit the brightest parts of the spatially smoothed surface-brightness structure. \citet{ShGi:89} make multiple two- and three-component fits of the surface-brightness structure \citep[using the data of][]{JeTh:82}. The final $r$-band and $B$-band models (their table~6) contain a thin disc and a thick disc, and a halo with an exponential fall-off ($h=53\arcsec$). \citet{NaJo:97} fit their $V$-band galaxy profile with three discs. {\Wu} fit a thin disc, a thick disc and a power-law halo to their surface-brightness profile. They compare their fit with nine previous studies.

I fitted the measurements of \citet{JeTh:82} with a small bulge superposed on a single thin disc, where $h_{\text{r}}=113\arcsec$, $z_{0}=20\farcs6$, $n_{\text{s}}=1$, $n=1$, $\ere=3\arcsec$, and $\mu_{0,0}-\mu_{\text{e}}=-0.7\,\magg$. The resulting minor-axis surface-brightness profiles and the measurements of {\KrSe}, \citet{JeTh:82}, and \citet{NaJo:97} are shown in Fig.~\ref{fig:ngc4565}a (the measurements of {\Wu} are left out as they would overfill the plot). The inclusion of the bulge has minor effects on the outer parts of the convolved profiles.

Despite that the apparent size of NGC 4565 is much larger than IC 5249 and NGC 3957, the colour profiles are similarly affected by scattered light, Fig.~\ref{fig:ngc4565}b. The red excess amounts to $\Delta\left(R-i\right)\simeq0.85$--$0.96\,\magg$ at $z=80\arcsec$ and about $1.2$--$1.3\,\magg$ at $z=120\arcsec$, where the larger (smaller) values were calculated using {\PSFVa} and {\PSFIa} ({\PSFVb} and {\PSFIb}).

The presented profiles of NGC 4565 provide strong hints that scattered light can induce a structure that appears as both a thick disc and a halo around this galaxy. The four measured profiles are suspiciously similar (as is the fifth one of {\Wu}); in comparison, the measured profiles of NGC 5907 show larger variation. The studies do not provide enough information to exclude that they have matched their sky subtraction to achieve the very similar profiles of NGC 4565. To correctly deconvolve observations of this galaxy out to $r=100\arcsec$, it is necessary to use a PSF that extends out to, at least, $r=150\arcsec$ in each considered bandpass. In this context, it is worth to mention that the PSF is poorly known at large radii, and at the time of the observations it could very well be brighter than the lower-limit {\PSFK}, and even as bright as {\PSFVb} for $r\ga100\arcsec$. At $r=100\arcsec$, the assumed exponential model structure is about $2\,\magg$ fainter than the observed structure. If the data would be deconvolved to recover the exponential profile, the required signal-to-noise is about S/N$=6.5$ at the same radius.

\begin{figure*}
\sidecaption
\includegraphics{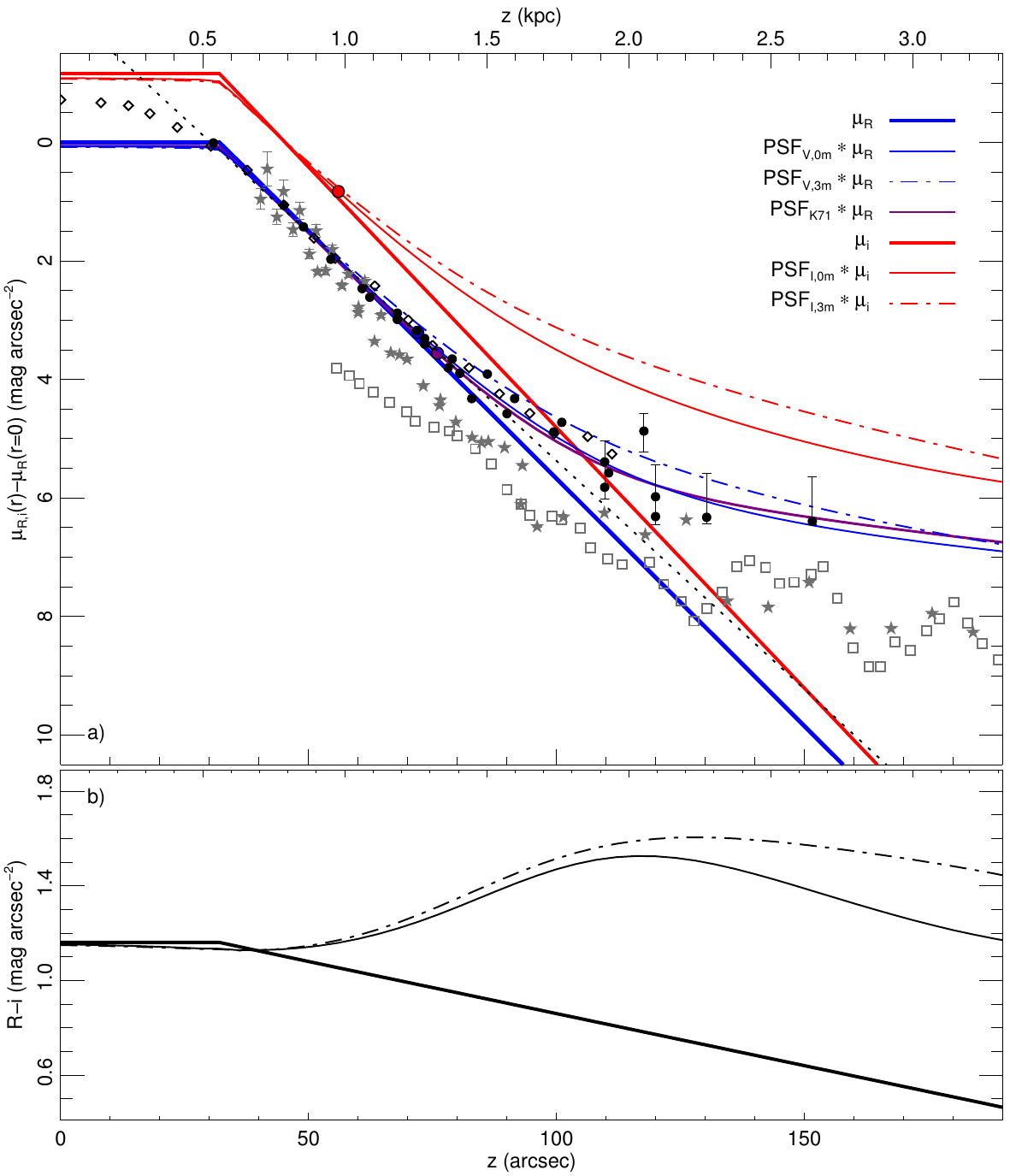}
\caption{Minor-axis surface-brightness profiles versus the vertical distance $z$ for the edge-on disc galaxy NGC 4244. Panels, lines, and coloured bullets are as described in Fig.~\ref{fig:thickdisc}. The black bullets and error bars (diamonds) in panel \textbf{a} show measured $R$-band ($J$-band) values of {\Fry} (\KrSe). The grey boxes and stars show scaled single-star measurements derived from \textit{HST} star counts of \citet{TiGa:05} and Streich et al.\ (in prep.), respectively; the values of the latter reference are only shown for $z>40\arcsec$. The best-fit profile of {\Fry} ($z_{0}=14\farcs1$) is shown with a black dotted line.}
\label{fig:ngc4244}
\end{figure*}

\subsection{The existence of a thick disc and a halo in visual wavelength-range observations of NGC 4244}\label{sec:ngc4244}
Using the photographic observations of NGC 4244 made by \citet{Kr:79}, {\KrSe} analyse and smooth the data, but do not find a thick disc. Surface-brightness profiles of more recent observations in the $R$ band are presented by {\Fry}. The authors follow the approach of \citet{MoBoHa:94} and measure {\PSFF} that extends out to $r=305\arcsec$; no information is provided regarding the saturated stars used to derive the outer profile for $r\ga20\arcsec$. {\PSFF} is shown in Fig.~\ref{fig:p2psf} where it is seen to be similar to {\PSFMBH} and {\PSFMBHn}. Any influence of scattered light is dismissed; the PSF is merely scaled to the measurements of the centre intensity of NGC 4244 to show that the PSF is fainter at large radii.

{\Fry} fit a single exponential disc to the data. I also fitted the measurements with a single disc, where $h_{\text{r}}=105\arcsec$, $z_{\text{0}}=13\arcsec$, and $n_{\text{s}}=2$. I set the model intensity constant for $z<29\arcsec$. Minor-axis surface-brightness profiles and the measurements of both {\KrSe} and {\Fry} are shown in Fig.~\ref{fig:ngc4244}a; only five error bars were extracted from fig.~9 in the latter reference since the figure is so small \citep[some of the data points are also reproduced in a larger figure in][]{Mo:99}. \citet{TiGa:05} and \citet{SeDaJo:05} use \textit{HST} data and count resolved red-giant stars out to large vertical distances. The re-analysed and scaled data points of the second data set (Streich et al.\ in prep.) are shown together with the photometry data in Fig.~\ref{fig:ngc4244}a. The data of \citet{TiGa:05} were scaled to match the second data set where $r\ga80\arcsec$.

The colour profiles in Fig.~\ref{fig:ngc4244}b show red excess for $z\ga45\arcsec$, which amounts to $\Delta\left(R-i\right)\simeq0.36$--$0.39\,\magg$ at $z=80\arcsec$ and about $0.75$--$0.83\,\magg$ at $z=120\arcsec$. This galaxy is intermediate in apparent size between NGC 3957 and NGC 4565, and despite a reported lack of a thick disc or halo, the models reveal significant amounts of red excess.

Both sets of photometry measurements of NGC 4244 show good agreement with the surface-brightness profiles of the convolved models. As with the other edge-on galaxies, photometry of this object also reveals a scattered-light halo, which deviates from an exponential profile for $r\ga80\arcsec$. This galaxy is somewhat fainter than NGC 5907, which could explain why there are fewer measurements of the scattered-light halo at large radii. The star-count profiles deviate from the photometry profiles, possibly due to how stellar populations are calculated from the red-giant star counts. The important attribute is that the single-star counts result in a halo that appears to be some $2\,\magg$ fainter than the photometry measurements for $r\ga100\arcsec$.

\section{More on face-on disc galaxies}\label{sec:afaceon}

\begin{table*}
\caption{Fitted $R$-band model parameters of the 7 face-on disc galaxies considered and referred to here. The galaxy type, distance, inclination, and the disc parameters are identical to the values of {\EPB} (their tables~1 and 4 and fig.~14). A bulge was also added to each model. Details of the original observations are given in Table~\ref{tab:obs}.}
\label{tab:faceon}
\tabcolsep=4.6pt
\begin{tabular}{lllrllllcccllrl}\hline\hline\\[-1.8ex]
 & & & & \multicolumn{5}{l}{Inner and outer discs} & & \multicolumn{3}{l}{Bulge} & &\\\cline{5-9}\cline{11-13}\\[-1.8ex]
Galaxy & \multicolumn{1}{l}{Type} & \multicolumn{1}{c}{$D$} & \multicolumn{1}{c}{$\hat{\textit{\i}}$} & \multicolumn{1}{c}{$h_{\text{i}}$} & \multicolumn{1}{c}{$h_{\text{o}}$} & \multicolumn{1}{c}{$\mu_{0,\text{i}}$} & \multicolumn{1}{c}{$\mu_{0,\text{o}}$} & \multicolumn{1}{c}{$r_{\text{br}}$} && \multicolumn{1}{c}{$n$} & \multicolumn{1}{c}{\ere} & \multicolumn{1}{c}{$\mu_{\text{e}}$} & \multicolumn{1}{c}{$r_{\text{abr}}$} & \multicolumn{1}{l}{Fig.}\\\hline\\[-1.8ex]
NGC 7280 & II.o-OLR(?) [+ III?]   & 24.3 & 50 & $\phantom{0}25.4$ & $11.2$ & $20.20$ & $17.90$ & $42.5$ && $4.0$ & $2.5$ & $17$    & 100 & \ref{fig:faceon}a, b\\
NGC 4102 & II.o-OLR + III-s(?)    & 14.4 & 55 & $344.1$           & $15.4$ & $20.58$ & $17.59$ & $42.9$ && $1.0$ & $7.5$ & $17.6$  & 100 & \ref{fig:faceon}c, d\\
NGC 4477 & I [or III?]            & 16.5 & 33 & $\phantom{0}35.7$ & \sd    & $19.87$ & \sd     & -      && $4.0$ & $7.0$ & $17.5$  & 200 & \ref{fig:nfaceon}a, b\\[0.3ex]
NGC 3507 & II.o-OLR(?) [+ III?]   & 14.2 & 27 & $\phantom{0}36.7$ & $18.9$ & $20.31$ & $18.22$ & $75.1$ && $4.0$ & $1.5$ & $17.3$  & 130 & \ref{fig:nfaceon}c, d\\
NGC 2880\tablefootmark{c} & III-s & 21.9 & 52 & $\phantom{0}20.0$ & \sd    & $19.77$ & \sd     & -      && $2.0$ & $3.0$ & $15.85$ & 120 & \ref{fig:nfaceon}e, f\\
\hline
\end{tabular}

\medskip
\tablefoot{Column~1, galaxy name; Col.~2, galaxy type; Col.~3, distance (Mpc); Col.~4, disc inclination ($\degr$); Cols.~5--9, disc parameters of {\EPB}, the scale lengths of the inner and outer discs (arcsec), the inner and outer disc centre intensities (\magg), and the break radius between the inner and the outer discs (arcsec); Cols.~10--12, bulge {\sersic}-profile index, half-light radius (arcsec), and centre intensity (\magg); Col.~13, approximate anti-truncation break radius (arcsec); and Col.~14, galaxy model is shown in these figures.\\
\tablefoottext{a}{This value results in a large offset between the inner and the outer discs; I used the value $\mu_{0,\text{o}}=17.27\,\magg$ instead.}}
\end{table*}

\subsection{Model parameters}
The model parameters of all six face-on disc galaxies considered and referred to here and in Sect.~\ref{sec:faceon} are given in Table~\ref{tab:faceon}.

\begin{figure*}
\centering
\includegraphics{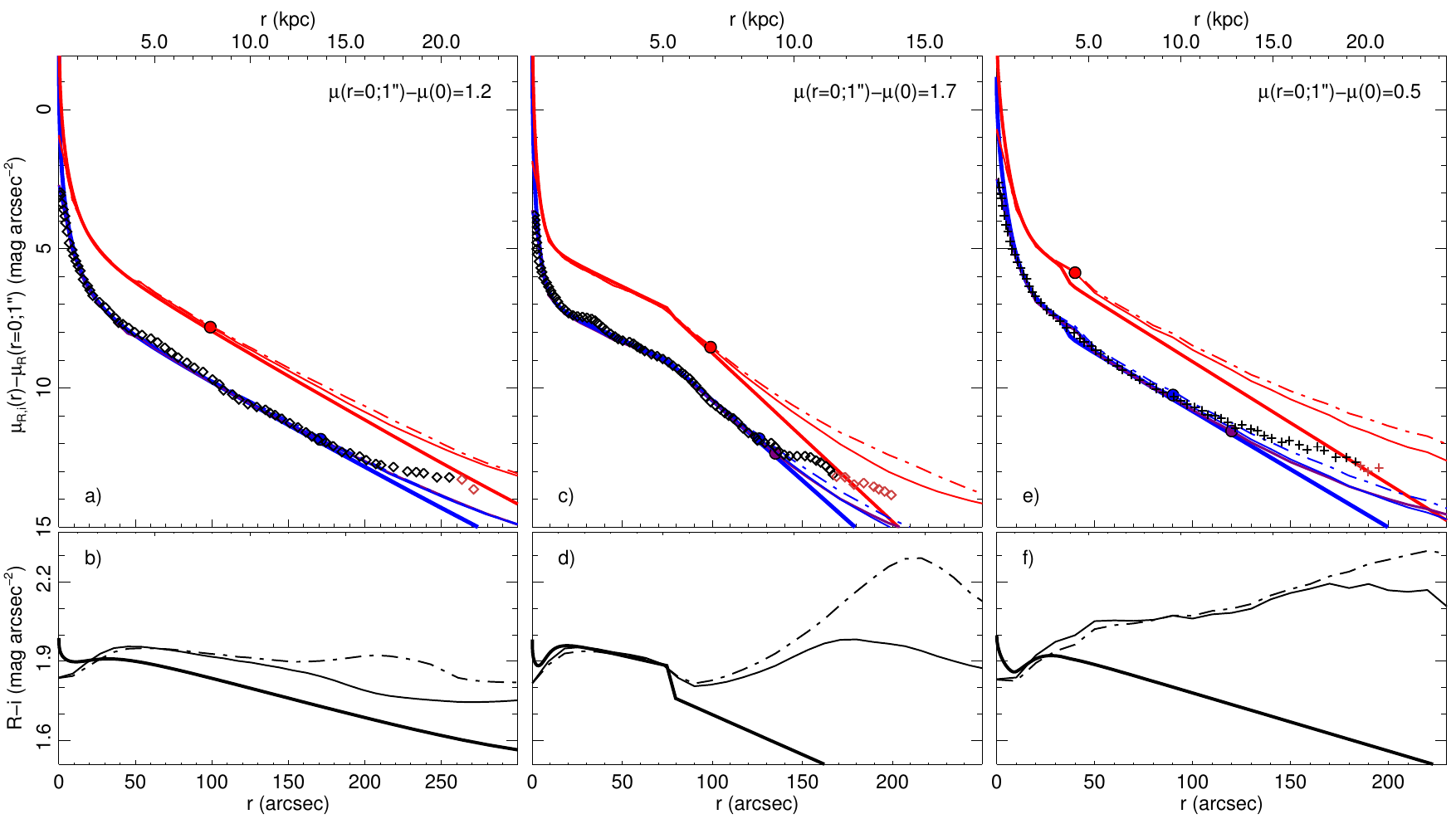}
\caption{Radial $R$-band and $i$-band surface-brightness profiles versus radius for three face-on disc galaxies: \textbf{a}) and \textbf{b}) NGC 4477, \textbf{c}) and \textbf{d}) NGC 3507, and \textbf{e}) and \textbf{f}) NGC 2880. The panels, lines, coloured bullets, and offsets are as described in Figs.~\ref{fig:thickdisc} and \ref{fig:faceon}. Diamonds (crosses) show $R$-band SDSS (INT) measurements (\EPB, their fig.~14); red symbols indicate values that are fainter than their sky-uncertainty limit.}
\label{fig:nfaceon}
\end{figure*}

\subsection{Three examples where the PSF may explain faint structures: NGC 4477, NGC 3507, and NGC 2880}\label{sec:faceonnsl}
Surface-brightness profiles of models of NGC 4477 are shown in Fig.~\ref{fig:nfaceon}a, together with the SDSS $r$-band measurements of {\EPB}. The profile is classified as type I or possible type III. An anti-truncation break appears at $r_{\text{abr}}\simeq200\arcsec$. The scattered-light halo radii are $r_{110,\text{\PSFK}}=r_{110,\text{\PSFVa}}\approx170\arcsec$ and $r_{110,\text{\PSFIa}}\approx100\arcsec$. This galaxy is so large that the convolved $R$-band profiles overlap as the outermost parts of the PSFs all share the same values. The measurements are brighter than the convolved model profiles for radii $r\ga210\arcsec$, which is only slightly larger than the anti-truncation radius $r_{\text{abr}}\simeq200\arcsec$. The colour profiles in Fig.~\ref{fig:nfaceon}b show smaller amounts of red excess than in the other face-on galaxies of this study, because of the large scale length; shortwards of $r\simeq25\arcsec$ there is also some blue excess. The values are $\Delta\left(R-i\right)\simeq0.086$--$0.095\,\magg$ at $r=100\arcsec$ and about $0.10$--$0.24\,\magg$ at $r_{\text{abr}}$.

Model surface-brightness profiles of NGC 3507 are shown in Fig.~\ref{fig:nfaceon}c, also in this case with SDSS $r$-band measurements of {\EPB}. The anti-truncation radius $r_\text{abr}\approx135\arcsec$ agrees with the scattered-light halo radius $r_{110}\simeq125$--$135\arcsec$. This galaxy is not as extended as NGC 4477, but the convolved profiles that were calculated using {\PSFK} and {\PSFVa} overlap for $r>r_{110}$. The measurements are brighter than the profiles for $r>r_{\text{abr}}\simeq130\arcsec$ also here. The colour profiles in Fig.~\ref{fig:nfaceon}d are reminicent of those for NGC 4102 in Fig.~\ref{fig:faceon}d. There is red excess at all radii outside the break radius $r_{\text{br}}=75\farcs1$, as with the other galaxies with such a break. At larger radii, the colour excess values are $\Delta\left(R-i\right)\simeq0.11$--$0.13\,\magg$ at $r=100\arcsec$ and about $0.25$--$0.30\,\magg$ at $r_{\text{abr}}$. The two radii $r_{110}$ and $r_{\text{abr}}$ are very similar for both NGC 4477 and NGC 3507, and it cannot be excluded that more extended and brighter SDSS PSFs could explain the suggested type III profile in the outermost parts of these two galaxies.

The final galaxy considered is NGC 2880, which surface-brightness profiles are shown in Fig.~\ref{fig:nfaceon}e, together with the measurements of the WIYN telescope. The parameters of this galaxy are similar to those of NGC 3507, but no inner break radius is used with this galaxy. The scattered-light halo radii $r_{110}$ are again similar to the anti-truncation radius $r_{\text{abr}}=120\arcsec$. The PSFs, and {\PSFVb} in particular, seem to be able to explain the measured profile out to $r=150\arcsec$. There are no known measurements of the PSF of the WIYN telescope at larger radii. The colour profiles in Fig.~\ref{fig:nfaceon}f show red excess outside of $r\simeq25\arcsec$. The values are $\Delta\left(R-i\right)\simeq0.13$--$0.17\,\magg$ at $r=50\arcsec$ and about $0.34$--$0.36\,\magg$ at $r_{\text{abr}}$.

\section{More on BCGs}\label{sec:abcg}
Three BCGs are studied in the following subsections: ESO 350-IG038, Appendix~\ref{sec:ESO350-IG038}; Mrk 297, Appendix~\ref{sec:bcgMrk297}; and UM 465, Appendix~\ref{sec:abcgUM465}. The model parameters of all nine BCGs considered and referred to here and in Sect.~\ref{sec:bcg} are given in Table~\ref{tab:bcg}.

\begin{table*}
\caption{Model parameters of the 9 BCGs that are considered here. Only those values are shown where a value is provided. The distances are in this work only used to scale the upper x-axis in the surface-brightness profile plots. Details of the original observations are given in Table~\ref{tab:obs}.}
\centering
\label{tab:bcg}
\begin{tabular}{llcllll@{\,}clllllll}\hline\hline\\[-1.8ex]
 & & & \multicolumn{4}{l}{Low surface-brightness component} & & \multicolumn{4}{l}{Starburst component}\\\cline{4-7}\cline{9-12}\\[-1.8ex]
Galaxy & \multicolumn{1}{c}{D} & \multicolumn{1}{c}{Band} &
   \multicolumn{1}{c}{$n$} & \multicolumn{1}{c}{\ere} & \multicolumn{1}{c}{$h$} & \multicolumn{1}{c}{$\mu$} & &
   \multicolumn{1}{c}{$n$} & \multicolumn{1}{c}{\ere} &
   \multicolumn{1}{c}{$h$} & \multicolumn{1}{c}{$\Delta\mu$} & \multicolumn{1}{c}{$r_{\text{X}}$} &
   Ref. & Fig.\\\hline\\[-1.8ex]
\object{ESO 400-G043}  & 77    & $B$ & $17.1$      & \snul$1.13\,$mas  & \sd         & $-38.85$      && - & - & - & - & - & 1\tablefootmark{a} & \ref{fig:ESO400-G043}\\[0.3ex]
                       & 77    & $B$ & \snul$1$    & $11$              & \snul$6.7$  & \sdash$23.1$  && - & - & - & - & - & 1\tablefootmark{a} & \ref{fig:ESO400-G043}\\[0.3ex]
                       & 79.2  & $V$ & \snul$1$    & \snul$7.21$       & \snul$4.31$ & \sdash$20.64$ && - & - & - & - & - & 2\tablefootmark{b} & \ref{fig:ESO400-G043}\\[0.3ex]
                       & 79.2  & $V$ & \snul$1$    & $15.7$            & \snul$9.39$ & \sdash$23.60$ && - & - & - & - & - & 2\tablefootmark{c} & \ref{fig:ESO400-G043}\\[0.3ex]
                       & 79.2  &     & \snul$1$    & \snul$4.3$        & \snul$2.6$  & \sdashx-      && $1$ & $1.5$ & $0.90$ & $4.75$ & $5.0$ &  & \ref{fig:ESO400-G043}\\[0.3ex]
\object{Mrk 5}         & 13.3  & $B$ & \snul$1$    & \snul$8.02$       & \snul$4.81$ & \sdash$21.09$ && - & - & - & - & - & 3\tablefootmark{d} & \ref{fig:Mrk5}\\[0.3ex]
\ \ UGCA 130           & 13.3  & $B$ & \snul$2.83$ & \snul$3.35$       & \sd         & \sdash$20.23$ && - & - & - & - & - & 4 & \ref{fig:Mrk5}\\[0.3ex]
                       & 13.96 & $B$ & \snul$0.99$ & \snul$9.6$        & \sd         & \sdash$21.18$ && - & - & - & - & - & 5 & \ref{fig:Mrk5}\\[0.3ex]
                       & 13.3  &     & \snul$1$    & \snul$8.0$        & \snul$4.8$  & \sdashx-      && $1$ & $2.0$ & $1.2$ & $0.7$ & - &   & \ref{fig:Mrk5}\\[0.3ex]
\object{ESO 350-IG038} & 80    & $B$ & $19.8$      & \snul$0.080\,$mas & \sd         & $-53.95$      && - & - & - & - & - & 1\tablefootmark{e} & \ref{fig:haro11}\\[0.3ex]
\ \ Haro 11            & 80    & $B$ & \snul$1$    & $14$              & \snul$8.3$  & \sdash$23.5$  && - & - & - & - & - & 1\tablefootmark{e} & \ref{fig:haro11}\\[0.3ex]
                       & 82    &     & \snul$1$    & \snul$5.0$        & \snul$3.0$  & \sdashx-      && $1$ & $0.7$ & $0.42$ & $8.50$ & $2.5$ &   & \ref{fig:haro11}\\[0.3ex]
                       & 82    &     & $11.0$      & \snul$8.25\,$mas  & \sd         & $-41.20$      && - & - & - & - & - & & \ref{fig:haro11_x}\\[0.3ex]
\object{Mrk 297}\tablefootmark{f} & 63 & $B$ & \snul$1$ & $16.3$       & \snul$9.72$ & \sdash$22.43$ && - & - & - & - & - & 6 & \ref{fig:Mrk297}\\[0.3ex]
\ \ NGC 6052/54        & 64.3  & $B$ & \snul$1$    & $13.3$            & \snul$7.96$ & \sdash$22.30$ && - & - & - & - & - & 3\tablefootmark{g} & \ref{fig:Mrk297}\\[0.3ex]
\ \ UGC 10182          & 65.10 & $B$ & \snul$0.84$ & \snul$9.25$       & \sd         & \sdash$21.35$ && - & - & - & - & - & 5 & \ref{fig:Mrk297}\\[0.3ex]
                       & 64.3  &     & \multicolumn{4}{l}{$\snul5.5$, $3.5$, $18.48$, $16.6$, $16$, $22$\tablefootmark{h}} && $1$ & $2.0$ & $1.2$ & - & - & & \ref{fig:Mrk297}\\[0.3ex]
\object{UM 465}        & 20.7  & $B$ & \snul$1$    & $10.8$            & \snul$6.48$ & \sdash$19.97$ && - & - & - & - & - & 7\tablefootmark{b} & \ref{fig:UM465}\\[0.3ex]
\ \ UGC 6877           & 20.7  & $B$ & \snul$1$    & $15.7$            & \snul$9.37$ & \sdash$21.82$ && - & - & - & - & - & 7\tablefootmark{c} & \ref{fig:UM465}\\[0.3ex]
\ \ Mrk 1308           & 20.7  &     & \snul$1$    & \snul$9.53$       & \snul$5.70$ & \sdashx- && $1$ & $2.5$ & $1.5$ & $2.4$ & - & & \ref{fig:UM465}\\[0.3ex]
\object{VCC 0001}      & 32    & $r$ & \snul$1$    & $2.6$             & \snul$1.5$  & \sdash$20.3$ && - & - & - & - & - & 8 & \ref{fig:VCC0001}\\[0.3ex]
                       & 32    &     & \snul$1$    & $4.9$             & \snul$2.9$  & \sdashx- && - & - & - & - & - &  & \ref{fig:VCC0001}\\[0.3ex]
\hline
\end{tabular}
\tablefoot{Column~1, galaxy name, some alternative names are provided indented on subsequent lines; Col.~2, distance (Mpc); Col.~3, bandpass used in the observations, Cols.~4--7, {\sersic} parameters of the low surface-brightness component in the form of the index, the half-light radius (arcsec), the corresponding scale length when $n\equiv1$ (arcsec), and the intensity at $r=0$ ($n=1$) or {\ere} (\magg, $n\ne1$); Cols.~8--11, {\sersic} parameters of the starburst component in the form of the index, the half-light radius (arcsec), the corresponding scale length when $n\equiv1$ (arcsec), and the magnitude difference $\Delta\mu=\mu-\mu_{\text{e}}$ (\magg); Col.~12, inner truncation radius (arcsec); Col.~13, references, rows without a number refer to this study; and Col.~14, results are shown in this figure.\\
\tablefoottext{a}{The fit is based on measurements in the radial range $16\le r\le32\arcsec$.}
\tablefoottext{b}{The scale length and the central surface brightness are based on measurements in the radial range where $24\le\mu\le26\,\magg$.}
\tablefoottext{c}{The scale length and the central surface brightness are based on measurements in the radial range where $26\le\mu\le28\,\magg$.}
\tablefoottext{d}{The fit is based on measurements in the radial range $13\le r\le32\arcsec$.}
\tablefoottext{e}{The fit is based on measurements in the radial range $20\farcs6\le r\le41\farcs3$ (8--16\,kpc).}
\tablefoottext{f}{Besides the low surface-brightness component, the parameters of the two inner components are given in Sect.~\ref{sec:bcgMrk297}.}
\tablefoottext{g}{The fit is based on measurements in the radial range $30\farcs3\le r\le36\farcs5$.}
\tablefoottext{h}{The six values specify the five face-on galaxy parameters $h_{\text{i}}$ (arcsec), $h_{\text{o}}$ (arcsec), $\mu_{0,\text{i}}$ (\magg), $\mu_{0,\text{o}}$ (\magg), and $r_{\text{br}}$ (arcsec), as well as $\mue$ of the bulge (\magg).}
}
\tablebib{(1) {\BO}; (2) {\Michevaa}; (3) {\Cairos}; (4) {\Caon}; (5) \citet{AmAgMuCa:09}; (6) {\Papaderos}; (7) {\Michevab}; (8) \citet{MeLiJaPa:14}.}
\end{table*}

\subsection{The extreme BCG ESO 350-IG038 (Haro 11)}\label{sec:ESO350-IG038}
{\BO} present surface-brightness profiles of observations in the $B$ band. The observed surface-brightness profiles were fitted using ellipses with the ellipticity $e=0.20$ (as \citealt{MiZaOs.:10} report later, hereafter \Michevac). The latter authors present more recent profiles of observations in the $V$ band. In this case, the measured ellipticity is $e=0.20$--$0.32$.

\begin{figure*}
\sidecaption
\includegraphics{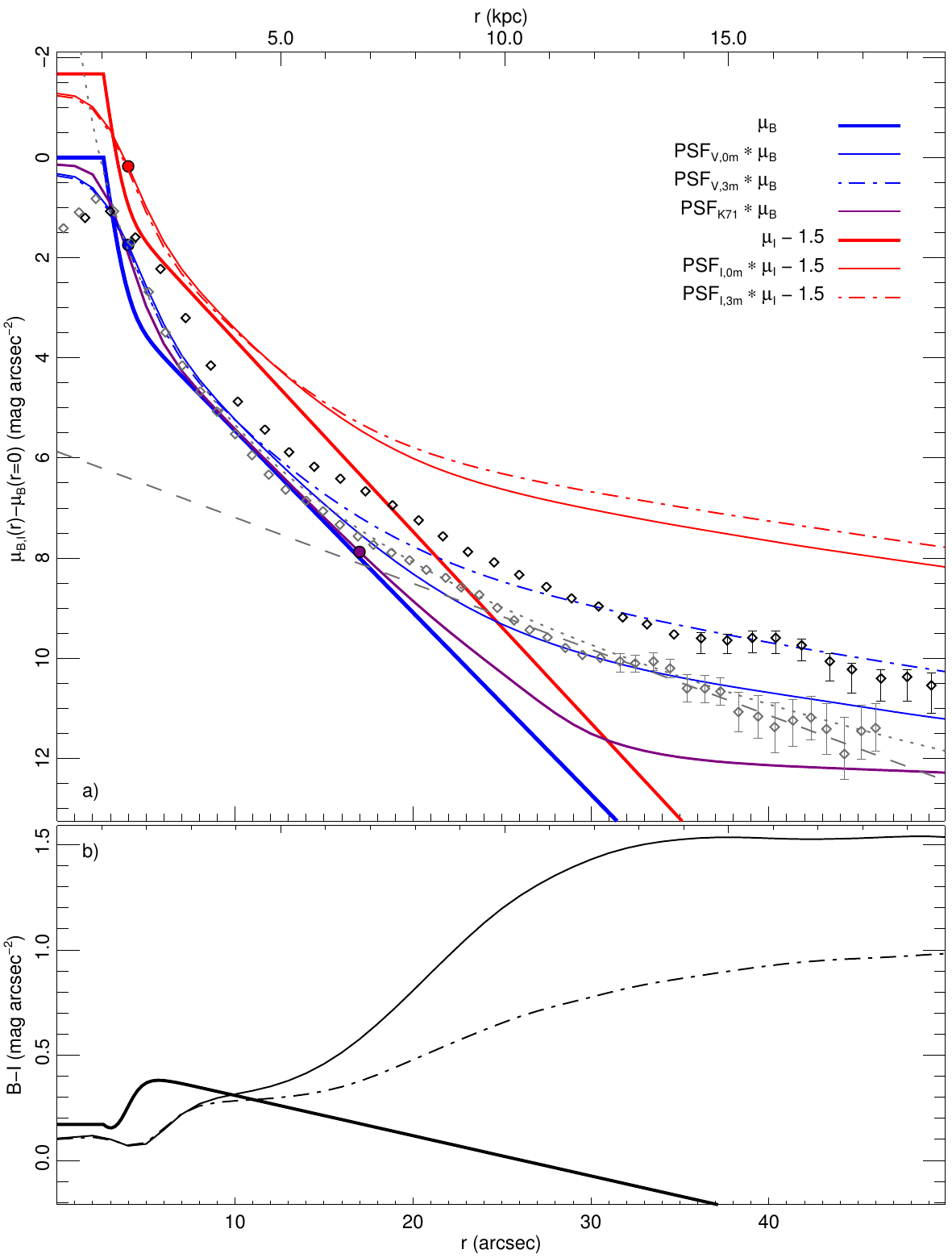}
\caption{Surface-brightness profiles versus radius for the BCG ESO 350-IG038 (Haro 11). Panels, lines, and coloured bullets are as described in Fig.~\ref{fig:thickdisc}. The $I$-band profiles are offset by $-1.5\,\magg$ from the $B$-band profiles for increased visibility. Grey and black bullets and error bars mark measured $B$-band (\BO) and $V$-band (\Michevac) values, respectively. The grey dotted (dashed) line shows the best-fit {\sersic}-type model (exponential fit to the disc in the range $20<r<41\arcsec$) of {\BO}.}
\label{fig:haro11}
\end{figure*}

The first authors fit the surface-brightness profile for a part of the radial range with a {\sersic} profile. They also fit the measurements in the same radial range with an exponential disc. I matched the observed $B$-band measurements with a convolved model where I used two {\sersic} profiles: a bulge where $n=1.0$, $\ere=0\farcs7$, and $\mue=13.5\,\magg$, and a host where $n=1.0$, $\ere=5\farcs0$, and $\mue=22.0\,\magg$. Additionally, $\mu_{\text{e,B}}-\mu_{\text{e,I}}=0.5\,\magg$. To match the measurements, the profile is truncated for $r\la2\farcs6$. The surface-brightness profiles and the two sets of measurements are shown in Fig.~\ref{fig:haro11}a. It is difficult to fit the measurements well at all radii, both the innermost and the outermost measurements are poorly fitted in either case. The convolved profiles are up to $1.0\,\magg$ too bright for $r<2\farcs0$. The measurements appear to be fairly well matched with a real PSF that is similar to {\PSFVa}. The more recent $V$-band measurements were fitted with ellipses of different ellipticity, and appear radially offset at the centre of the object -- they are poorly fitted in any case.

The three $B-I$ colour profiles are shown in Fig.~\ref{fig:haro11}b. The red excess at $r=20\arcsec$ is $\Delta(B-I)\simeq0.70\,\magg$ with {\PSFVa} and {\PSFIa}, and about $0.35\,\magg$ with {\PSFVb} and {\PSFIb}. The corresponding values at $r=35\arcsec$ are $\Delta(B-I)\simeq1.7\,\magg$ and about $1.0\,\magg$, respectively, which are very similar to the corresponding values for ESO 400-G043 at $r=30\arcsec$.

As in the case of ESO 400-G043, which was modelled using similar small-valued parameters, the scattered-light halo of ESO 350-IG038 begins already at small radii. The amount of red excess at larger radii is therefore determined by the shape of the $B$-band and the $I$-band PSFs. As with the other studied galaxies in this paper, a hypothetically temporally varying PSF can itself explain why the more recent $V$-band observations of {\Michevac} simultaneously show more scattered light and less red excess in the halo (they use the colour $V-K$).

There are several separated bright starburst regions at the centre of this BCG; their placement suggests the inapplicability of a symmetric one-dimensional function to describe the co-added structure of scattered light. The measurements of {\Michevac} are brighter than the $B$-band measurements, suggesting possible differences between how the asymmetric surface-brightness structure is sampled, and the PSF could also be different. The goal here was not to reproduce the true model profile, but to illustrate that the geometrical properties of this galaxy suggest that scattered-light effects dominate the measured halo.

To demonstrate the importance of an accurately determined PSF, I calculated a set of alternative models where I set $n=11.0$, $\ere=8.2\,$mas, and $\mue=-41.2\,\magg$. The profiles are again truncated for $r\la2\farcs6$. The resulting surface-brightness profiles are shown in (the online-only) Fig.~\ref{fig:haro11_x}a. The convolved profiles of these alternative models and the models in Fig.~\ref{fig:haro11}a agree fairly well with the $B$-band measurements. There is, in this case, slightly less red excess than before, compare Figs.~\ref{fig:haro11_x}b and \ref{fig:haro11}b. Here, the red excess at $r=30\arcsec$ is $\Delta(B-I)\simeq1.0\,\magg$ with {\PSFVa} and {\PSFIa}, and about $0.60\,\magg$ with {\PSFVb} and {\PSFIb}. Since the exact shape of the PSF is unknown and because of the asymmetry due to the brightest (starburst) regions of the object, it is difficult to deconvolve the provided surface-brightness structure to get the real physical structure.

There seem to be three possible scenarios regarding the host of ESO 350-IG038: there is no host (unlikely, but not excluded with the current data), the host is of disc type, and the host is more massive ($n\ga10$). It is impossible to say which one is correct before the scattered light of the centre regions is first removed. There is no strong physical basis for either of the two profiles that {\BO} measure in the outer region. The galaxy should be analysed after the observations are deconvolved -- accounting for the maximum possible useful radius when considering the sky subtraction and the PSF.

\subsection{The BCG Mrk 297}\label{sec:bcgMrk297}
{\Papaderos} present $B$-band profiles of Mrk 297. The authors model the surface-brightness profile with three components: a low surface-brightness region $I_{\text{E}}$, a plateau $I_{\text{P}}$, and an inner part $I_{\text{G}}$. These components are defined as follows
\begin{eqnarray}
I_{\text{E}}&=&I_{\text{E},0}\exp\left[-\frac{r}{\alpha}\right],\quad
I_{\text{P}}=I_{\text{P},0}\exp\left[-\left(\frac{r}{\beta}\right)^{\eta}\right],\quad\text{and}\\
I_{\text{G}}&=&I_{\text{G},0}\exp\left[-0.5\left(\frac{r}{\gamma}\right)^2\right]
\end{eqnarray}
where the surface-brightness profile of Mrk 297 is fitted with $\alpha=9\farcs72$, $\beta=13\farcs0$, $\eta=2.46$, $\gamma=3\farcs37$, $I_{\text{E},0}=22.43\,\magg$, $I_{\text{P},0}=20.67\,\magg$, and $I_{\text{G},0}=20.21\,\magg$. With the exception of the low surface-brightness component $I_{\text{E}}$, the values of this fit are omitted in Table~\ref{tab:bcg}. {\Cairos} present more recent observations in the $B$ band and also fit an exponential profile in a small part of the radial extent. \citet{AmAgMuCa:09} use the same data and make a two-dimensional {\sersic}-profile fit of the stellar host.

\begin{figure*}
\sidecaption
\includegraphics{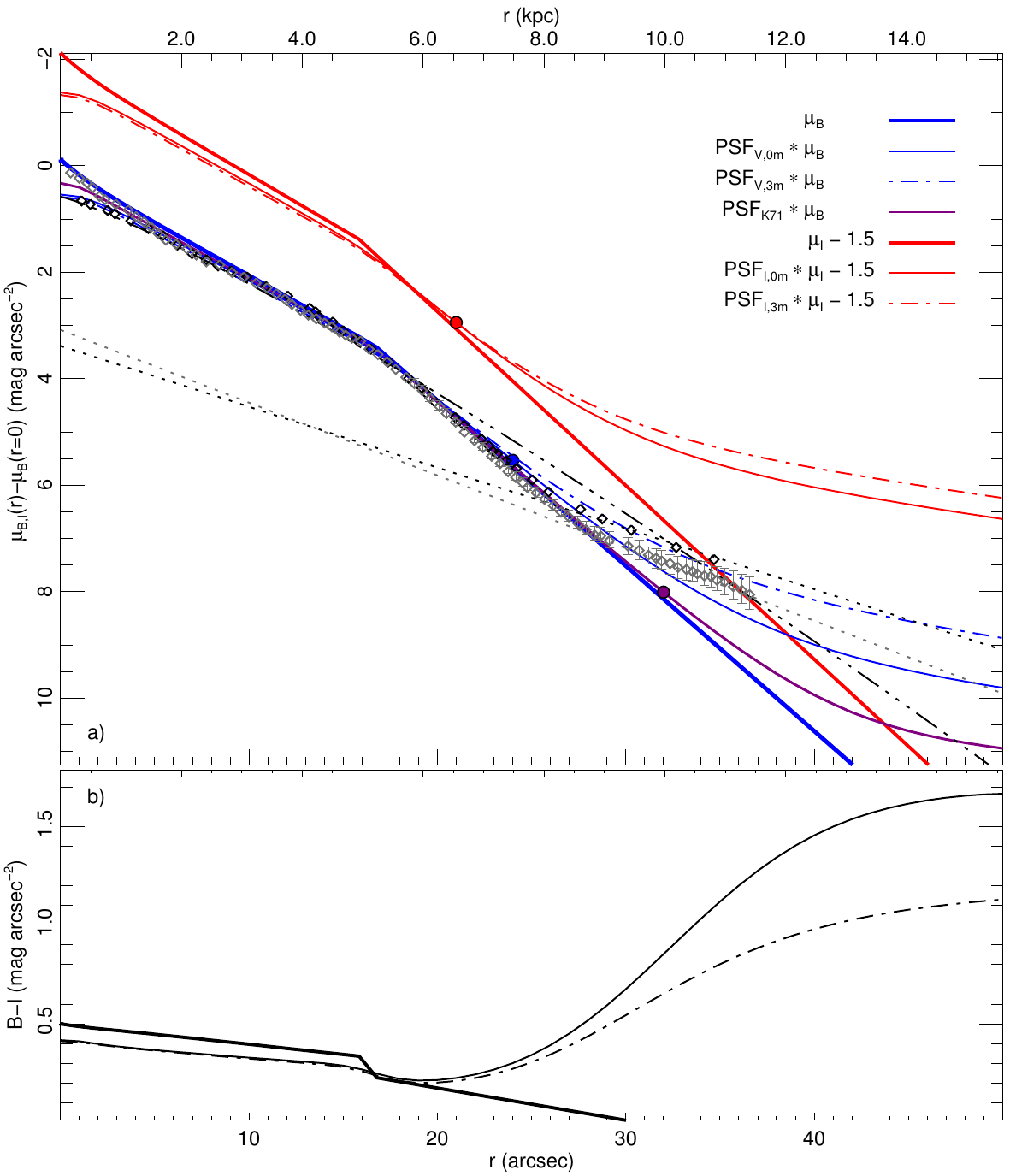}
\caption{Surface-brightness profiles versus radius for the BCG Mrk297. Panels, lines, and coloured bullets are as described in Fig.~\ref{fig:thickdisc}. The $I$-band profiles are offset by $-1.5\,\magg$ from the $B$-band profiles for increased visibility. Grey (black) diamonds and error bars mark measured $B$-band values of {\Cairos} (\Papaderos). The grey (black) dotted line shows the best-fit {\sersic}-type model of {\Cairos} (\Papaderos). The dash-tripple-dotted line shows the fitted profile of \citet{AmAgMuCa:09}.}
\label{fig:Mrk297}
\end{figure*}

I defined a bulge and face-on disc galaxy model to get a reasonable agreement between convolved structures and the measurements of {\Papaderos} and {\Cairos}. The parameters of the bulge were $n=1$, $\ere=2\farcs0$, and $\mue=22\,\magg$, and those of the face-on galaxy were $h_{\text{i}}=5\farcs5$, $h_{\text{o}}=3\farcs5$, $\mu_{0,\text{i}}=18.48\,\magg$, $\mu_{0,\text{o}}=16.6\,\magg$, and $r_{\text{br}}=16\arcsec$. I also set $\mu_{\text{e},B}-\mu_{\text{e},I}=0.5\,\magg$. Surface-brightness profiles of the models and the three sets of profile fits are shown in Fig.~\ref{fig:Mrk297}a.

The scattered-light halo radius $r_{110}\simeq22\arcsec$ for {\PSFVa} and about $30\arcsec$ for {\PSFK}. The measurements of both {\Papaderos} and {\Cairos} overlap the models well in all radial regions; the values deviate slightly for $r\la5\arcsec$, but that deviation is unimportant to the discussion here. Each study calculated a fit to the host component using the outermost measurements. Their scale lengths differ slightly as the outermost measurements that are used in the fit differ. The profiles of this study suggest that the difference in the outermost region occurs as the PSF of the two separate observations were different, and both fits merely measure the scattered-light structure for $r\ga20\arcsec$. Another possibility is that the accuracy of the sky subtraction differs (see below). The PSF at the time of the observations seems to be closer to {\PSFVa} with the measurements of {\Cairos}, whilst that of the measurements of {\Papaderos} was slightly brighter, but not as bright as {\PSFVb}. The profile is reminiscent of a type III-s structure as is seen in face-on disc galaxies. The fit of \citet{AmAgMuCa:09} is more difficult to interpret as it differs from both sets of measurements outwards of $r\simeq20\arcsec$.

Despite the small value on $\mu_{\text{e},B}-\mu_{\text{e},I}$, the $B-I$ colour profiles in Fig.~\ref{fig:Mrk297}b show dramatic amounts of red excess for $r\ga18\arcsec$. For example, at $r=30\arcsec$ the red excess is $\Delta\left(B-I\right)\simeq0.75\,\magg$ with {\PSFVa} and {\PSFIa}, and about $0.59\,\magg$ with {\PSFVb} and {\PSFIb}. The values increase dramatically at larger radii.

It cannot be excluded that the measurements of {\Papaderos} and {\Cairos} differ because of differences in the sky subtraction. The same result is achieved with models of Mrk 35 and Mrk 314 (not shown here). The measurements are ambiguous when both sky subtraction and scattered light are considered.

\subsection{The BCG UM 465}\label{sec:abcgUM465}
{\Michevab} present deep surface-brightness profiles for 21 BCGs. Several of the BCGs are too small, too large, show little intensity excess, or are severely asymmetric. These are all factors that complicate an analysis of the influence of scattered light as presented here. I considered the following five BCGs to be more easy to analyse with the high-seeing PSFs I use, as they neither seem to require a deconvolution of the images nor extend to large radii were the intensity of the PSF is unknown: UM 465, UM 483, UM 491, UM 538, and UM 559. Out of these BCGs, I chose to scrutinize the surface-brightness profiles of the most symmetric galaxy of these, UM 465. The authors point out that the $B$-band PSF for this galaxy is different from the other filters in the centre; there is no other information on this issue. The ellipticity is measured as $e=0.16$. The authors fit their measured surface-brightness profiles with two exponential discs in two separate magnitude ranges.

\begin{figure*}
\sidecaption
\includegraphics{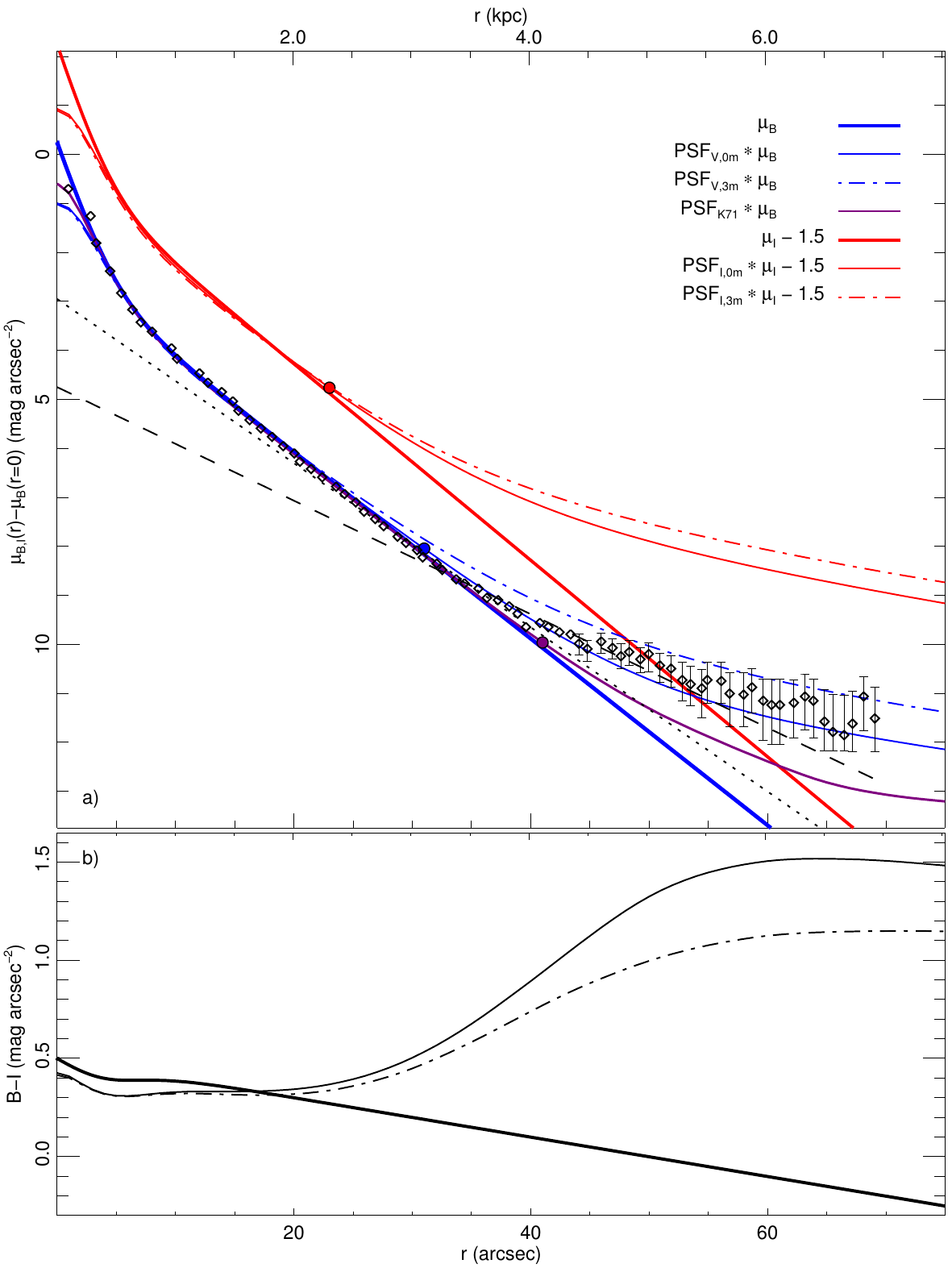}
\caption{Surface-brightness profiles versus radius for the BCG UM 465. Panels, lines, and coloured bullets are as described in Fig.~\ref{fig:thickdisc}. The $I$-band profiles are offset by $-1.5\,\magg$ from the $B$-band profiles for increased visibility. Diamonds and error bars mark measured $B$-band values of {\Michevab}. The dotted (dashed) line shows the best-fit exponential model fit of {\Michevab} for $24\le\mu_B\le26$ ($26\le\mu_B\le28$) \magg.}
\label{fig:UM465}
\end{figure*}

I chose the model parameters of a bulge where $n=1.0$ and $\ere=2\farcs5$, which is superposed on an exponential-disc host where $n=1.0$, $\ere=9\farcs53$, $\mue^{\text{host}}-\mue^{\text{bulge}}=2.4\,\magg$, and $\mu_{\text{e},B}-\mu_{\text{e},I}=0.5\,\magg$. The resulting surface-brightness profiles and the measurements are shown in Fig.~\ref{fig:UM465}a.

The measurements agree well with the model that was convolved with {\PSFVa}, throughout the full radial range. The scattered-light halo radius $r_{110}\simeq31\arcsec$ ($r_{110}\simeq41\arcsec$) with {\PSFVa} (\PSFK). {\Michevaa} measure {\PSFMich} for the same telescope that is even fainter than {\PSFK}, but this PSF only seems to be correctly measured out to $r\simeq30\arcsec$ and the times of the measurements that were used to derive {\PSFMich} are unknown (see Fig.~\ref{fig:p2psf} and Sect.~\ref{sec:bcgMrk5}). The measurements extend out to $r=70\arcsec$, which requires a PSF that extends to $r=77\arcsec$ (Sect.~\ref{sec:toy}). As with Mrk 297, my study suggests that the measurements are scattered light, more accurate measurements of the PSF would reveal to which extent. Notably, the authors make an effort to carefully subtract the sky, which is why it should be possible to exclude this element as an explanation.

I achieve the same conclusion with the surface-brightness profiles of BCG \object{VCC 0001} that are presented by \citet{MeLiJaPa:14}. These authors have chosen to only present the measurements of the host and not of the (possibly) bright starburst region, which makes an analysis of the origin of the excess light impossible (profiles are shown in the online-only Fig.~\ref{fig:VCC0001}).

The $B-I$ colour profiles in Fig.~\ref{fig:UM465}b show red excess for radii $r\ga18\arcsec$. The red excess at $r=30\arcsec$ is $\Delta\left(B-I\right)\simeq0.30\,\magg$ with {\PSFVa} and {\PSFIa} and about $0.24\,\magg$ with {\PSFVb} and {\PSFIb}. The same values at $r=40\arcsec$ are $0.79\,\magg$ and $0.63\,\magg$, respectively.

\onlfig{
  \begin{figure*}
    \sidecaption
    \includegraphics{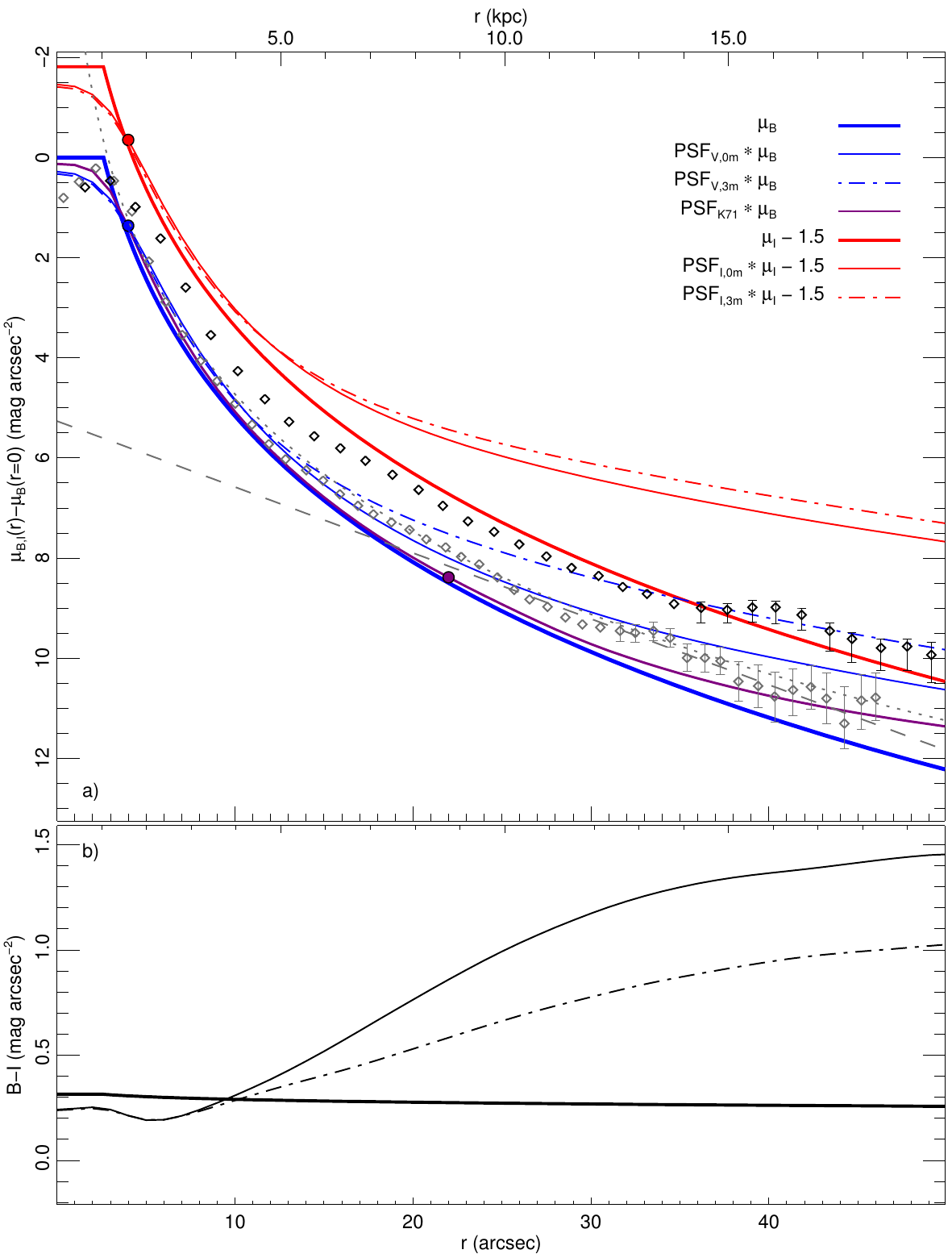}
    \caption{Surface-brightness profiles versus radius for the alternative models of ESO 350-IG038 (Haro 11), compare with Fig.~\ref{fig:haro11}. Panels, lines, and coloured bullets are as described in Fig.~\ref{fig:thickdisc}. The $I$-band profiles are offset by $-1.5\,\magg$ from the $B$-band profiles for increased visibility. Grey and black bullets and error bars mark measured $B$-band (\BO) and $V$-band (\Michevac) values, respectively. The grey dotted (dashed) line shows the best-fit {\sersic}-type model (exponential fit to the disc in the range $20<r<41\arcsec$) of {\BO}.}
    \label{fig:haro11_x}
  \end{figure*}
}

\onlfig{
\begin{figure*}
\sidecaption
\includegraphics{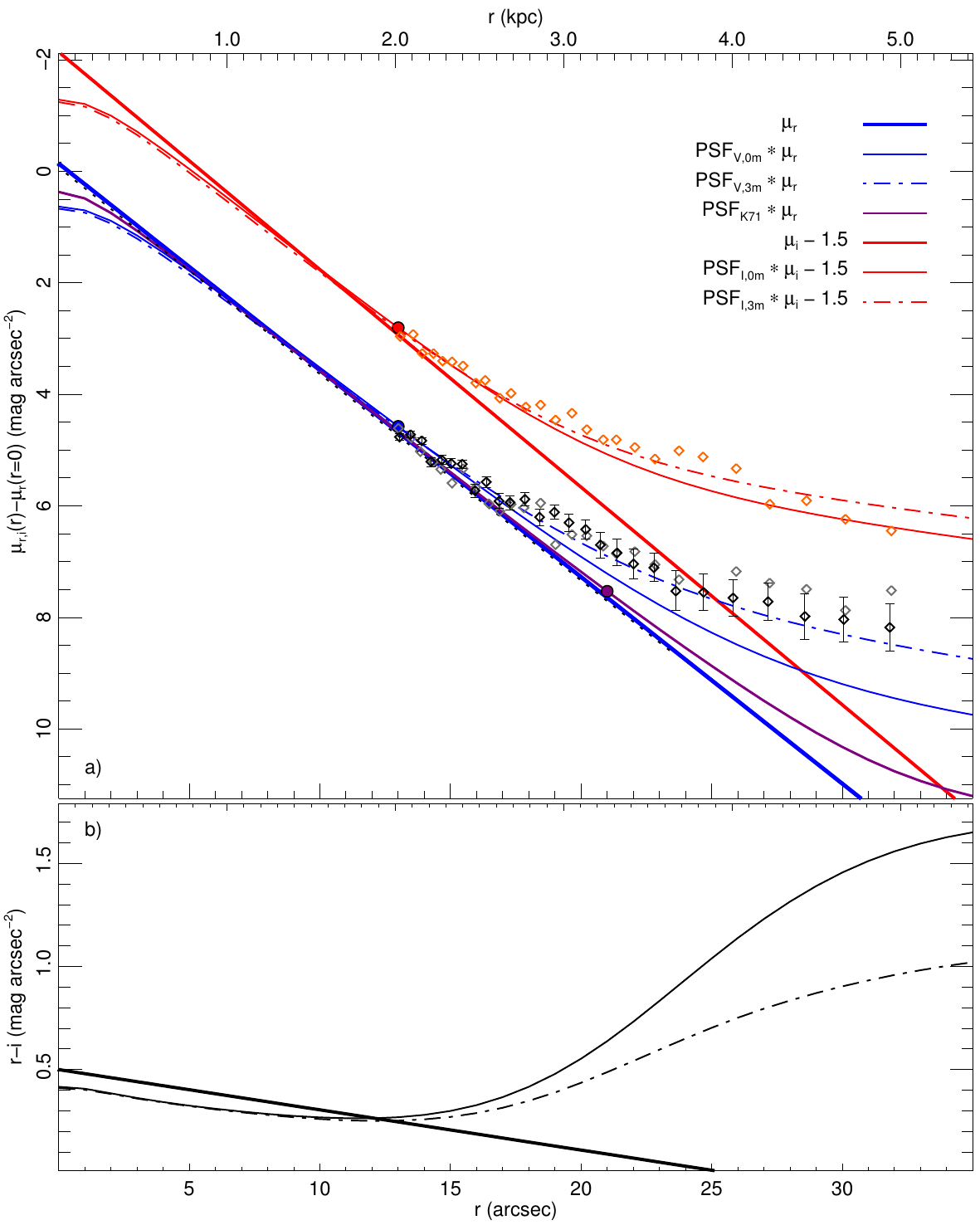}
\caption{Surface-brightness profiles versus radius for the BCG VCC 0001. Panels, lines, and coloured bullets are as described in Fig.~\ref{fig:thickdisc}. The $i$-band profiles are offset by $-1.5\,\magg$ from the $r$-band profiles for increased visibility. Black diamonds (grey; orange) diamonds, and error bars mark measured $r$-band ($g$-band; $i$-band) values of \citet{MeLiJaPa:14}.}
\label{fig:VCC0001}
\end{figure*}
}

\end{document}